\documentclass[preprint,trackchanges]{aastex631}

\newcommand{\lat}{$Fermi$-LAT}

\hypersetup{linkcolor=red,citecolor=cyan,filecolor=blue,urlcolor=magenta}
\usepackage{subfigure}
\shorttitle{GB6 J2113+1121: a possible neutrino emitter}

\begin{document}

\title{GB6 J2113+1121: A multi-wavelength flaring $\gamma$-ray blazar temporally and spatially coincident with the neutrino event IceCube-191001A}

\correspondingauthor{Neng-Hui Liao; Ting-Gui Wang}
\email{nhliao@gzu.edu.cn;twang@ustc.edu.cn}

\author[0000-0001-6614-3344]{Neng-Hui Liao}
\affiliation{Department of Physics and Astronomy, College of Physics, Guizhou University, Guiyang 550025, People’s Republic of China}

\author{Zhen-Feng Sheng}
\affiliation{Key laboratory for Research in Galaxies and Cosmology, Department of Astronomy, The University of Science and Technology of China, Chinese Academy of Sciences, Hefei, Anhui 230026, People’s Republic of China}

\author{Ning Jiang}
\affiliation{Key laboratory for Research in Galaxies and Cosmology, Department of Astronomy, The University of Science and Technology of China, Chinese Academy of Sciences, Hefei, Anhui 230026, People’s Republic of China}

\author{Yu-Ling Chang}
\affiliation{Tsung-Dao Lee Institute, Shanghai Jiao Tong University, Shanghai 200240, People’s Republic of China}

\author{Yi-Bo Wang}
\affiliation{Key laboratory for Research in Galaxies and Cosmology, Department of Astronomy, The University of Science and Technology of China, Chinese Academy of Sciences, Hefei, Anhui 230026, People’s Republic of China}

\author{Dong-Lian Xu}
\affiliation{Tsung-Dao Lee Institute, Shanghai Jiao Tong University, Shanghai 200240, People’s Republic of China}

\author{Xin-Wen Shu}
\affiliation{Department of Physics, Anhui Normal University, Wuhu, Anhui, 241000, People’s Republic of China}

\author{Yi-Zhong Fan}
\affiliation{Key Laboratory of Dark Matter and Space Astronomy, Purple Mountain Observatory, Chinese Academy of Sciences, Nanjing 210034, People’s Republic of China}

\author[0000-0002-1517-6792]{Ting-Gui Wang}
\affiliation{Key laboratory for Research in Galaxies and Cosmology, Department of Astronomy, The University of Science and Technology of China, Chinese Academy of Sciences, Hefei, Anhui 230026, People’s Republic of China}

\begin{abstract}
A radio-emitting tidal disruption event (AT2019dsg) is proposed as a likely counterpart of the IceCube neutrino event IC-191001A. In this work we have revisited the {\it Fermi}-LAT data in the direction of the neutrino and confirmed no signal at the site of AT2019dsg. Instead, at the edge of the 90\% confidential level error region of this neutrino there is a $\gamma$-ray transient source associated with a blazar GB6 J2113+1121. In May 2019, GB6 J2113+1121 was undergoing an unprecedented $\gamma$-ray flare since the start of the {\it Fermi}-LAT operation, with a variability amplitude about 20-fold. Similar violent flares of GB6 J2113+1121, unobserved before, have been also detected observed in optical bands. Moreover, the blazar remained in a high flux state in the infrared bands when IceCube-191001A arrived, though its $\gamma$-ray and optical activities has temporally ceased. Motivated by the spatial and temporal coincidence, we suggest that GB6 J2113+1121 is a candidate of the counterpart of IC-191001A. The jet properties of GB6 J2113+1121 are investigated, which are found to be comparable with that of the neutrino-emitting blazars (candidates). A specific analysis of archival IceCube data in this direction and future observations would put a further constraint on the origin of the neutrino.
\end{abstract}


\section{INTRODUCTION} \label{sec:intro}
The IceCube Neutrino Observatory \citep{2017JInst..12P3012A} at the south Pole\footnote{\url{https://icecube.wisc.edu/}} has detected high-energy neutrinos of astrophysical origin \citep{2013Sci...342E...1I}, which opened a new window onto the non-thermal universe. Unlike the charged cosmic rays (CRs) which could be affected by ambient matters and the radiation fields as well as the deflection due to the magnetic fields during the propagation, or the $\gamma$-ray photons might suffer significant attenuation, the neutrinos travel undisturbed, allowing us to directly probe a large number of extreme cosmic environments that are otherwise opaque \citep{2018AdSpR..62.2902A}. The arrival directions of the IceCube neutrinos are isotropically distributed, which indicates a predominantly extragalactic origin. Extragalactic cosmic accelerators, for instance, active galactic nuclei (AGNs) jets \citep[e.g.,][]{2013ApJ...768...54B,2014PhRvD..89l3005B}, $\gamma$-ray bursts \citep[e.g.,][]{1995PhRvL..75..386W}, starburst galaxies \citep[e.g.,][]{2006JCAP...05..003L} as well as galaxy clusters \citep[e.g.,][]{2008ApJ...689L.105M}, are hence reasonable neutrino contributors (for a review see \citealt{2015RPPh...78l6901A}). However, no compelling evidence of significant clusters of IceCube neutrinos in either space or time has been found \citep{2015ApJ...807...46A,2017ApJ...835..151A,2020PhRvL.124e1103A,2021arXiv210109836I}. Interestingly, a 2.9$\sigma$ significance excess against the background is found 0.35$\degr$ from a Seyfert II galaxy NGC 1068 by targeting a predefined list of 110 sources \citep{2020PhRvL.124e1103A}. Therefore, a substantial fraction of the observed diffuse neutrinos is suggested to be from the contribution of weak sources that are individually below the point source sensitivity \citep{2016PhRvD..94j3006M}. On the other hand, neutrino transients have been proposed to be detectable \citep[e.g.,][]{2016ApJ...831...12H}.

A notable case is the IceCube detection of a 0.3~PeV muon neutrino (IC-170922A) from a position compatible blazar TXS 0506+056 in coincidence with a multi-bands flaring period at 3$\sigma$ significance \citep{2018Sci...361.1378I}. A 160-day long neutrino flare (3.5$\sigma$ significance) has been also found in that direction, though no accompanying electromagnetic activities of the blazar exhibited \citep{2018Sci...361..147I}. Blazars, including flat-spectrum radio quasars (FSRQs) and BL Lacertae objects (BL Lacs), are an extreme subclass of AGNs that their strong relativistic jets are well aligned with our line of sight \citep{1978bllo.conf..328B,2019ARA&A..57..467B}. The highly beamed jet emissions are overwhelming and hence the emissions of blazars are highly variable \citep{1995ARA&A..33..163W,1997ARA&A..35..445U}. The universal spectral energy distributions (SEDs) of blazars are featured as a two-bump shape in log$\nu$F$\nu$-log$\nu$ plot, where one is from synchrotron emission while the other one extends to the $\gamma$-ray domain. In the hadronic radiation scenarios, $\gamma$-ray emissions of blazars could be from the decay of neutral pions generated from the interaction between cosmic ray (CR) nuclei and ambient matters or the radiation fields, meanwhile, charged pions are produced at the same time which lead to the generation of the neutrinos \citep{1991PhRvL..66.2697S,1993A&A...269...67M,2001PhRvL..87v1102A}. Therefore, the observed $\gamma$-ray emissions and the neutrinos are thought to be tightly connected. Furthermore, in addition to these two messengers, the hadronic cascades also produce electromagnetic emissions at lower energies, and hence the multiwavelength observations play a key role in identifying neutrino emitters \citep[e.g.,][]{2018ApJ...865..124M,2019ApJ...880..103G,2020ApJ...893..162F,2020A&A...640L...4G}. Since the neutrinos are solely produced by the hadronic processes but $\gamma$-ray photons can also be from the leptonic processes (i.e. inverse Compton scattering of soft photons, \citealt{1992ApJ...397L...5M,1993ApJ...416..458D,1994ApJ...421..153S,2000ApJ...545..107B}), detections of neutrinos from blazars provide a unique insight on the AGN jets \citep{2018ApJ...863L..10A}

On 1st October 2019 at 20:09:18.17 UT, IceCube detected a track-like $\sim$ 0.2~PeV neutrino event (IC-191001A) with a probability 58.9\% (i.e. {\tt Gold} alert streams) of being of astrophysical origin, with an arrival direction of R.A. $\rm 314.08^{+6.56}_{-2.26}\degr$ and Decl. $\rm 12.94^{+1.50}_{-1.47}\degr$ \citep{2019GCN.25913....1I}. Within such a region there are two known $\gamma$-ray sources categorized in the fourth {\it Fermi} $\gamma$-ray source catalog (4FGL, \citealt{2020ApJS..247...33A}) \citep{2019GCN.25932....1G}. In addition, a spatially coincident radio-emitting Tidal Disruption Event (TDE) AT2019dsg was discovered about 175 days earlier of the detection of IC-191001A \citep{2019GCN.25929....1S}. After investigations of multi-wavelength variability properties of the TDE, it was proposed to be a likely association for the neutrino, especially considering the continuously increasing radio emission \citep{2021NatAs...5..510S}. By analyzing roughly one-year long {\it Fermi}-LAT data around the arrival time of IC-191001A, no significant $\gamma$-ray emissions of the two 4FGL sources have been detected. Anyhow, a new $\gamma$-ray source (labeled as Fermi-J2113.8+1120 towards to a flat spectral source GB6 J2113+1121) emerged at the edge of the 90\% C. L. localization error box of the neutrino  \citep{2021NatAs...5..510S}, also see Figure \ref{tsmap}.  A similar temporal investigation of Fermi-J2113.8+1120 also yielded no clear flux enhancement at the exact arrival time (i.e. within one month) of the neutrino, and hence this source was also suggested to be unlikely related to the neutrino \citep{2021NatAs...5..510S}. 

In this Letter, we carry out thorough investigations of multi-wavelength properties of GB6 J2113+1121, based on 12.6-year {\it Fermi}-LAT data since the start of its operation as well as archival data from {\it Swift}, Zwicky Transient Facility (ZTF) and {\it WISE}, along with an optical spectroscopic observation carried out by Hale 5 meter telescope (Section \ref{sec:data}). Discussions of its broadband variability behaviors and the potential to be a neutrino counterpart are shown in Section \ref{sec:diss}. We adopt a ${\Lambda}$CDM cosmology with $\Omega_{M}$\,=\,0.32, $\Omega_{\Lambda}$\,=\,0.68,  and a Hubble  constant of $H_{0}$\,=\,67\,km$^{-1}$\,s$^{-1}$\,Mpc$^{-1}$ \citep{2014A&A...571A..16P}.

\section{DATA ANALYSIS AND RESULTS} \label{sec:data}
GB6 J2113+1121 is known as a flat spectral radio source \citep{1986ApJS...61....1B,2003MNRAS.341....1M,2011AJ....142..108K}, with a spectroscopic redshift measurement of 1.316 \citep{2013ApJ...767...14P}. Based on the SDSS $g$-band magnitude \citep{2007ApJS..172..634A} and the NVSS 1.4~GHz flux density \citep{1998AJ....115.1693C}, the radio loudness is estimated as high as $\simeq$ 3000.
\subsection{{\it Fermi}-LAT Data}
The first 12.6-year (MJD 54683$-$59309) {\tt SOURCE} type ({\tt evclass} = 128 and {\tt evtype} = 3) \lat~ {\tt Pass} 8 data with energy range between 0.1 and 500~GeV are collected. The analysis is carried out with the {\tt Fermitools} software of version {\tt 2.0.8} and the accompanying {\tt Fermitools-data} of version {\tt 0.18}. The zenith angle cut (i.e. $< 90^{\circ}$) together with the recommended quality-filter cuts (i.e. {\tt DATA\_QUAL==1 \&\& LAT\_CONFIG==1}) are adopted to filter the entire photon data. {\tt Unbinned} likelihood analyses implemented in the {\tt gtlike} task are used to extract $\gamma$-ray flux and spectrum. The significance level of detecting a $\gamma$-ray source is quantified by the test statistic ($\rm TS = -2{\rm ln}({L_{0}/L})$, \citealt{1996ApJ...461..396M}), where $\rm {L}$ and $\rm L_{0}$ are the maximum likelihood values for the model with and without the target source, respectively. During the likelihood analysis, 4FGL sources within 15$\degr$ of the radio position of GB6 J2113+1121, as well as the diffuse $\gamma$-ray emission templates (i.e. {\tt gll\_iem\_v07.fits} and {\tt iso\_P8R3\_SOURCE\_V3\_v1.txt}), are embraced. Parameters of the background sources within 10$\degr$ region of interest (ROI) centered at the target location as well as normalizations of the two diffuse templates are left free, whereas others are frozen as the 4FGL values. If new $\gamma$-ray sources (i.e TS $>$ 25) are found from the residual TS maps, the initial background model is updated and the likelihood fitting is then re-performed. In the temporal analysis, weak background sources (i.e. TS $<$ 10) are removed from the analysis model. For sources with TS $<$ 10, the {\tt pyLikelihood UpperLimits} tool is adopted to calculate a 95\% confidential level (C. L.) upper limit instead of a flux estimation.

The analysis of the entire data set suggests an existence of a significant $\gamma$-ray source (TS = 217.4) towards GB6 J2113+1121, along with other three $\gamma$-ray sources (TS $\leq$ 50, not in 4FGL-DR2\footnote{\url{https://fermi.gsfc.nasa.gov/ssc/data/access/lat/10yr\_catalog/}}) at the edge of the ROI among which one may associate with CGRaBS J2051+1743 \citep{2008ApJS..175...97H}. The optimized location of the central $\gamma$-ray source is at R.A. 318.4855$\degr$, decl. 11.3261$\degr$ with a 95\% C. L. error radius of 4.3$\arcmin$. Considering that the separation between the $\gamma$-ray location and radio position of GB6 J2113+1121 is 1.9$\arcmin$, and no other known radio source is found within the error radius, GB6 J2113+1121 is probably the low-energy counterpart of the $\gamma$-ray source. With an assumption of the single power-law spectral distribution (i.e. $dN/dE \propto E^{-\Gamma}$, where $\Gamma$ is the photon index), an averaged $\gamma$-ray flux is given, (1.30 $\pm$ 0.15)$\rm \times 10^{-8}$ ph $\rm cm^{-2}$ $\rm s^{-1}$, as well as a relatively soft spectrum (i.e. $\Gamma_{\gamma}$ = 2.58 $\pm$ 0.07).

\begin{figure}
\centering
\includegraphics[scale=0.45]{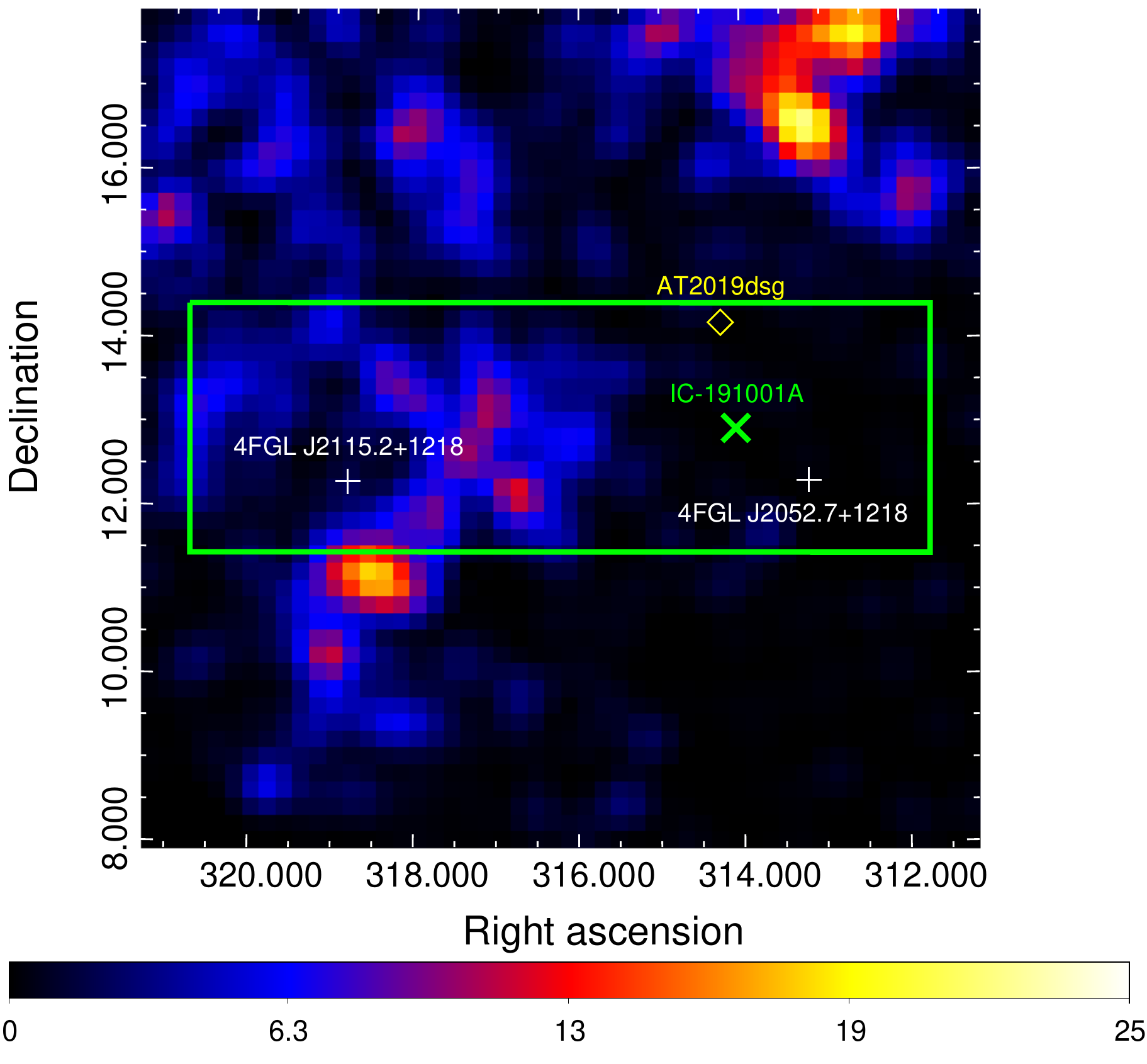}
\includegraphics[scale=0.45]{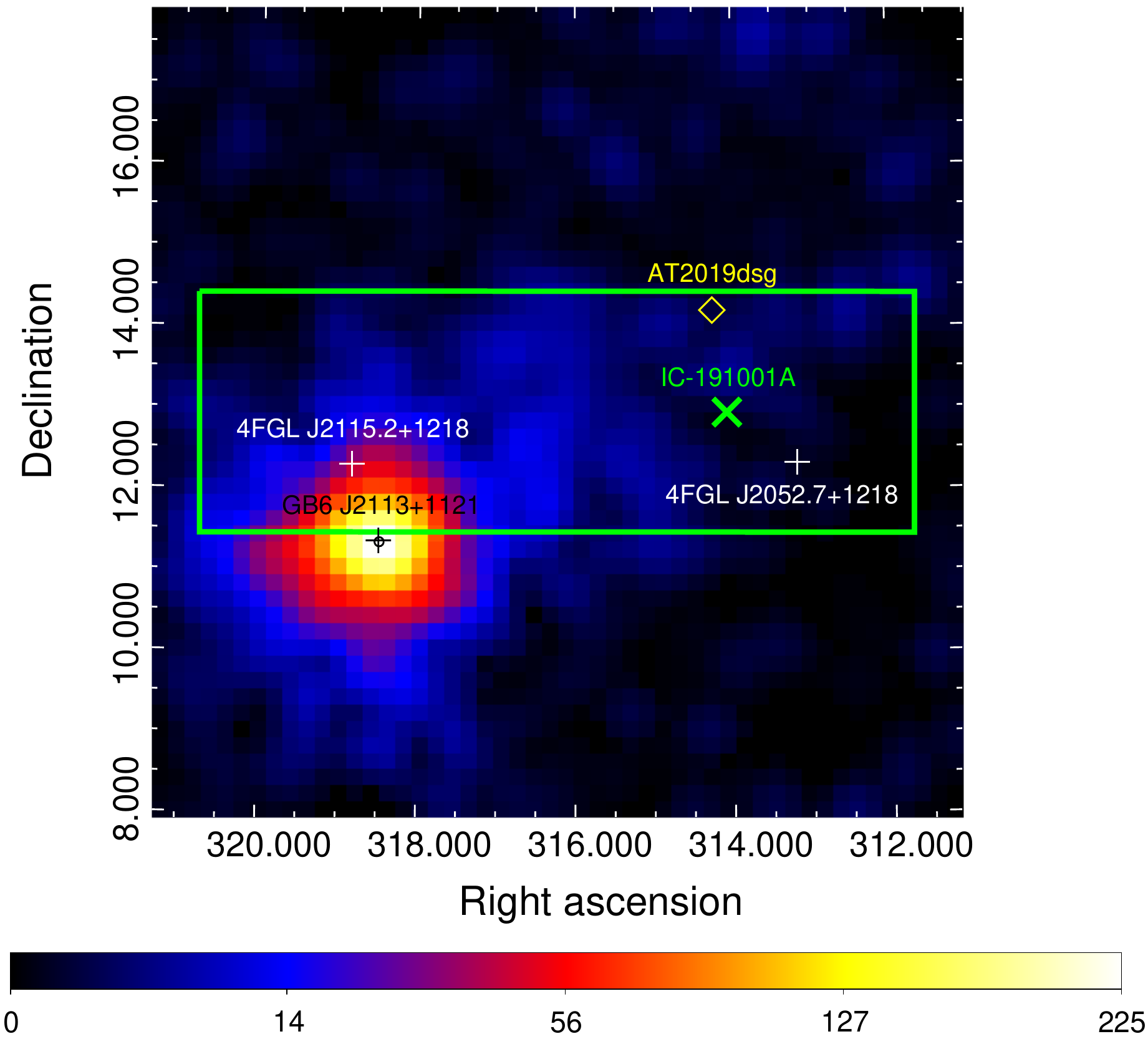}
\includegraphics[scale=0.6]{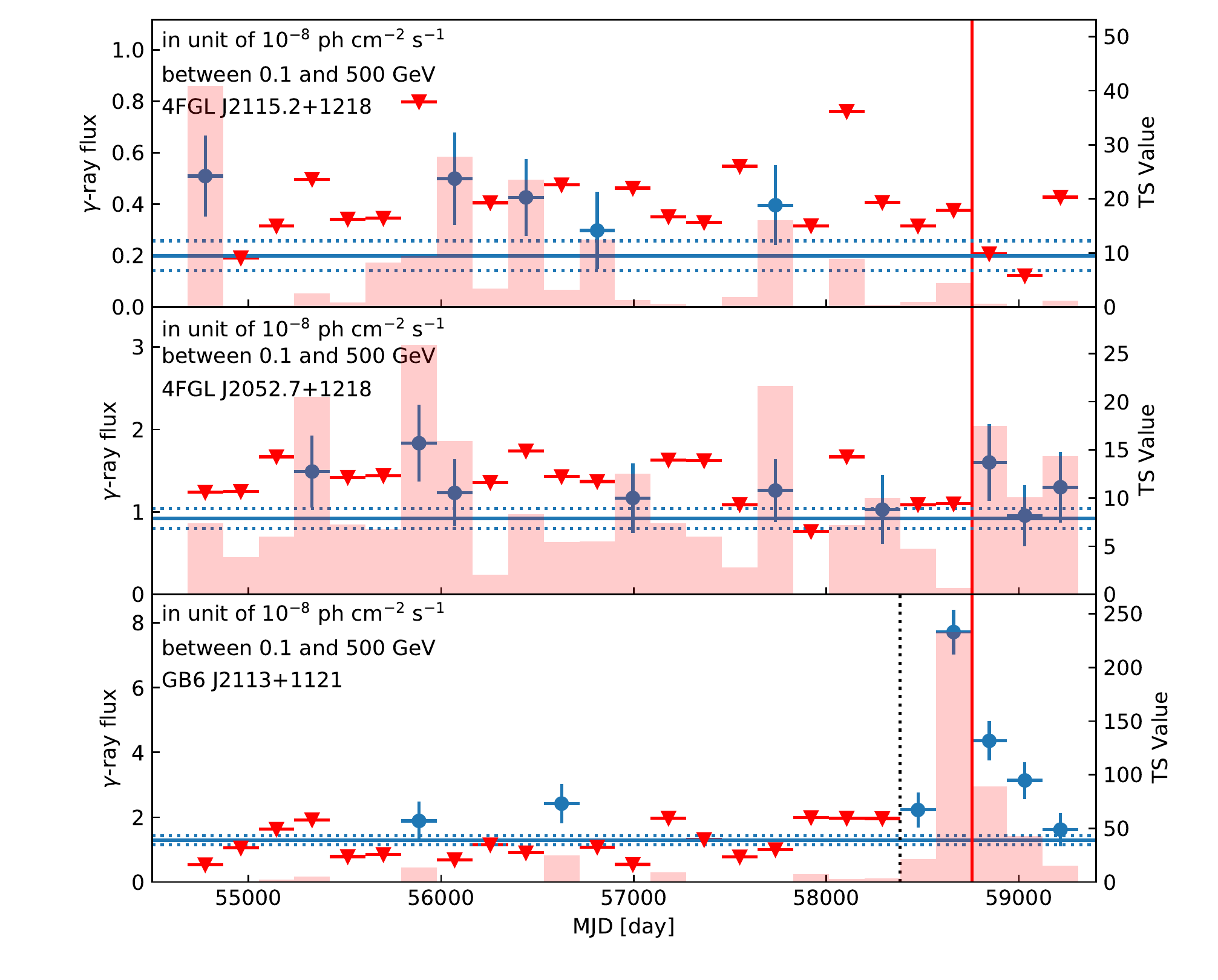}
\caption{{\bf Upper panels:} smoothed $\gamma$-ray TS maps (10$\degr \times 10 \degr$ scale with 0.2$\degr$ per pixel, GB6 J2113+1121 not included in the mode file) of the region that is spatially compatible with the arrival direction of the neutrino IC-191001A. The left one is based on the {\it Fermi}-LAT data with time range between MJD 54683 (i.e. the beginning of its operation) and MJD 58383 (marked as a black vertical dashed line in the bottom panel), while the right one corresponds to the rest of the {\it Fermi}-LAT data. The black circle represents the 95\% C. L. $\gamma$-ray localization error radius of GB6 J2113+1121 and the black cross corresponds to its radio location. {\bf Bottom panel:} Half-year time bin $\gamma$-ray light curves of the three $\gamma$-ray sources spatially in coincidence with the neutrino IC-191001A. Blue circles and red triangles are flux estimations and upper limits, and TS values corresponding to each time bin are also shown. The horizontal lines represent the averaged fluxes of the sources for the entire data set along with the 1$\sigma$ uncertainties. The red vertical line marks the arrival time of the neutrino IC-191001A.}
\label{tsmap}
\end{figure}

A six-month time bin $\gamma$-ray light curve for GB6 J2113+1121 is extracted, see Figure \ref{tsmap}. It is confirmed that the target maintains at a quiescent flux state since the beginning of the \lat~operation for more than 10 years but then a strong $\gamma$-ray flare comes out. Besides the target, the temporal behaviors of the background sources are also examined, no similar behaviors to that of the target are found. Therefore, the variability of the target is suggested to be intrinsic rather than artificial caused by the background sources. Light curves of the two 4FGL sources (i.e. 4FGL J2052.7+1218 and 4FGL 2115.2+1218) that fall into the neutrino localization error box are also extracted, from which no significant ($<$ 3$\sigma$, \citealt{2012ApJS..199...31N}) variability is found. Moreover, it is confirmed that they are not detected by \lat~(i.e. TS $<$ 20) by selecting one year data centered at the arrival time of the neutrino \citep{2021NatAs...5..510S}. Therefore, they are not preferred to be the neutrino counterpart. We divide the time range of the entire data set into two separate parts at MJD 58383 corresponding to different flux states of GB6 J2113+1121. The target is barely detected by \lat~ (TS = 24.7) in the former period, with a flux estimation of (4.7 $\pm$ 1.1)$\rm \times 10^{-9}$ ph $\rm cm^{-2}$ $\rm s^{-1}$. On the other hand, analyses of the last period reveal a significant $\gamma$-ray source (TS = 327.6)  with flux reaching (3.2 $\pm$ 0.3)$\rm \times 10^{-8}$ ph $\rm cm^{-2}$ $\rm s^{-1}$, roughly seven times of the value in the quiescent flux state. Following localization analysis confirms that GB6 J2113+1121 remains to be within the $\gamma$-ray location error radius then, also see Figure \ref{tsmap}. To further investigate the $\gamma$-ray variability properties, monthly light curve focusing on the data in the high flux state is extracted, see Figure 2. The target is detectable lasting for nearly one year. It begins with a rapid flux ascent that the target turns to be bright with a peaking flux of (1.1 $\pm$ 0.2)$\rm \times 10^{-7}$ ph $\rm cm^{-2}$ $\rm s^{-1}$ while the background emission is dominant one month ago. The decline phase is longer than the ascend phase and it is disappeared around MJD 58750. Interestingly, a mild second flare that the flux maintains at $\sim$ 4$\rm \times 10^{-8}$ ph $\rm cm^{-2}$ $\rm s^{-1}$ is followed, then the target is back to the quiescent state a few months later. Spectral analysis of this one-year period suggests a probable spectral hardness (i.e. $\Gamma_{\gamma}$ = 2.37 $\pm$ 0.07) compared with that obtained from the entire data set. Sophisticated spectral template rather than the single power-law function is adopted, no significant spectral curvature is found. Energy of the most energetic $\gamma$-ray photons received from GB6 J2113+1121 then is about 8~GeV. A 10-day time bin light curve focusing on the rapid ascend phase are also extracted and the peaking time of the first flare is around MJD 58615, however, no evidences of intraday variability is found.  

\subsection{{\it Swift} Data}
There are in total 9 visits from the Neil Gehrels {\it Swift} Observatory \citep{2004ApJ...611.1005G} on GB6 J2113+1121. The FTOOLS software version 6.28 is adopted to analyze the XRT photon-counting mode data and the UVOT images. For the XRT data, the initial event cleaning {\tt xrtpipeline} procedure with standard quality cuts is carried out. The {\tt xselect} task is used to extract the source spectra from a circular region with a radius of 12 pixels while the background spectra are from a larger circle (i.e. 50 pixels) in a blank area. Then the ancillary response files are created by the response matrix files taken from the calibration database with {\tt xrtmkarf} to facilitate the sequent spectral analysis. We group the spectra to have at least 1 count per bin using the {\tt cstat} approach. Setting the absorption column density as the Galactic value (i.e. $\rm 5.8\times10^{20}$ $\rm cm^{-2}$), an analysis of the entire dataset (totally 107 net photons) gives an averaged unabsorbed 0.3-10.0 keV flux of $\rm 4.3^{+1.3}_{-0.9}\times10^{-13}$ erg $\rm cm^{-2}$ $\rm s^{-1}$ ($\mathcal{C}$-Statistic/d.o.f, 146/105; \citealt{1979ApJ...228..939C}). Meanwhile, a hard spectrum is suggested (i.e. $\Gamma_{x} \sim 1.3$). X-ray temporal properties are investigated, however, no significant variability can be found due to the limited statistics. For the UVOT images, the magnitudes are extracted by the aperture photometry (i.e. the {\tt uvotsource} task), with a 5$\arcsec$ circular aperture for the target along with a background one in a larger source-free region. Significant variability is revealed in the UV emission densities of the target. For instance, a flux rise of 0.85 mag in $U$-band is detected between observations at MJD 58996.6 and 59004.2.
 
\subsection{Optical Data}
\subsubsection{Optical Spectrum}
The spectrum is taken from Double Spectrograph (the dichroic D55) on the Hale 200-inch telescope at the Palomar Observatory on MJD 59383. The exposure is split into four 450-second periods through a 2\arcsec.0 slit. The data are reduced by {\tt Pypeit}\footnote{\url{https://pypeit.readthedocs.io/en/latest/}} \citep{pypeit:joss_pub}. The resolving power is R$\sim$720 and the median signal to noise is 9.6. After the de-redden the Milky Way extinction \citep{1998ApJ...500..525S}, the best fitting continuum model is $f=8.27\times(\lambda/3000)^{-1.04}$. Several emission lines are distinct, marked in Figure \ref{fig:spec}, from which the redshift estimation of 1.316 $\pm$ 0.002 is given confirming the result in \cite{2013ApJ...767...14P}. We get the $\rm log\lambda L_{\lambda}(3000\,\text{\r{A}}) = (24.16 \pm 0.04) \times 10^{44}\rm erg\ s^{-1}$, and the flux and FWHM of Mg \textsc{ii} are (9.8 $\pm$ 1.68) $\times 10^{-16}\rm erg\,cm^{-2}\,s^{-1}$ and (2714 $\pm$ 597) $\rm km\,s^{-1}$, respectively. Using the empirical virial BH mass relation provided by \cite{2008ApJ...680..169S}, we get the $\rm log(M_{BH}/M_{\sun}) \gtrsim 8.2$, in case of a disk-like geometry of the broad line region (BLR). Adopting the bolometric luminosity correction  $L_{bol}\sim(5.18\pm0.19)L_{3000}$ \citep{2012MNRAS.422..478R} which is likely overestimated due to the jet contribution, we can infer the Eddington ratio $L_{bol}/L_{edd}\lesssim0.6$.

\subsubsection{{\it ZTF} Light-curve Data}
\begin{figure}
\centering
\includegraphics[scale=0.9]{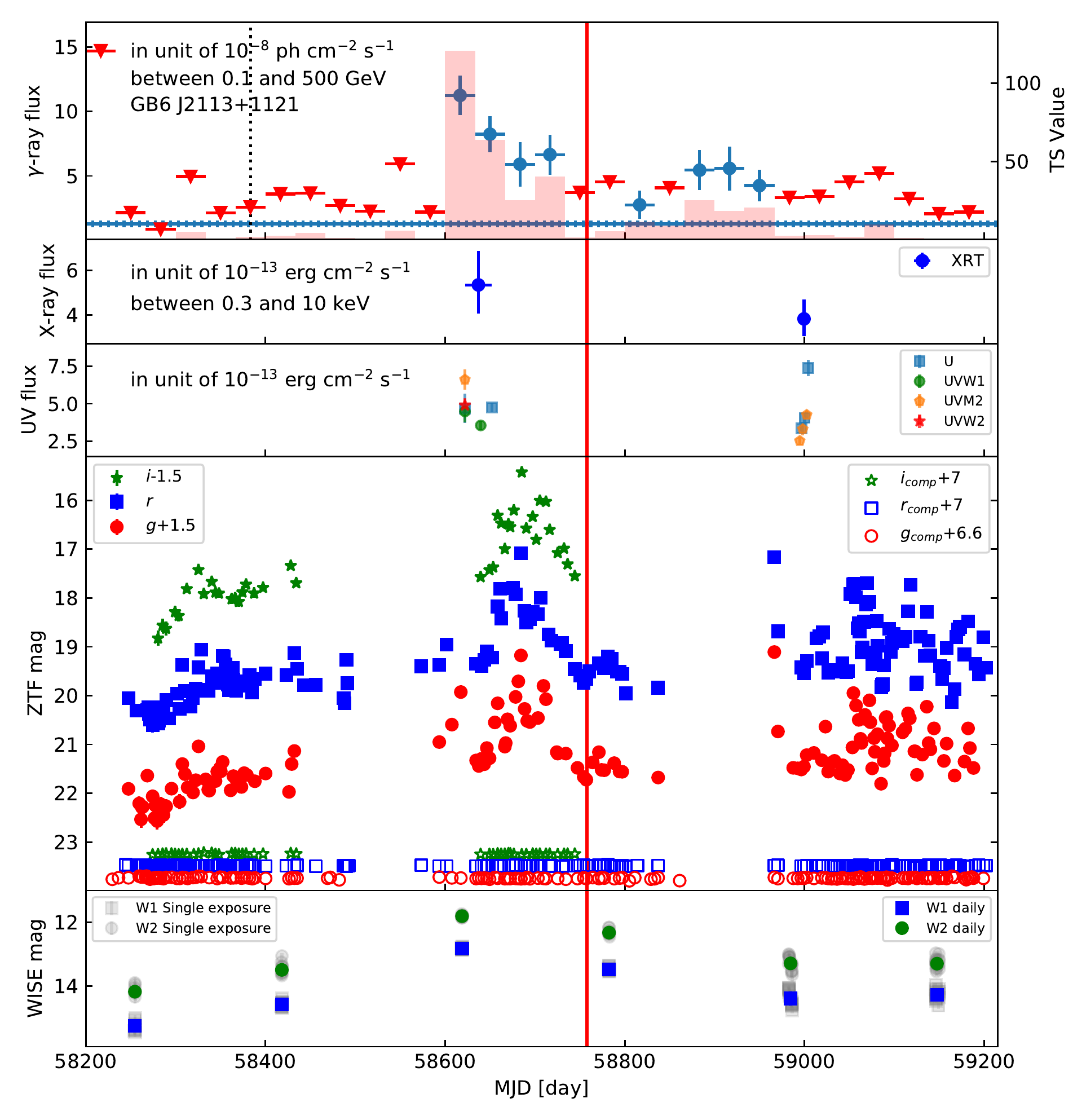}
\caption{Multiwavelength light curves of GB6 J2113+1121. In the monthly $\gamma$-ray light curve panel, the horizontal and black dashed vertical line are the same as those in Figture \ref{tsmap}. For the {\it Swift}-UVOT measurments, corrections of the Galactic absorption have been carried out \citep{1989ApJ...345..245C,2011ApJ...737..103S}. For the ZTF light curves, the solid markers represent the magnitudes of the target, while the hollow ones correspond to averaged values of the comparison stars in the same field. The red vertical line across all the panels marks the arrival time of the neutrino.}
\label{mlc}
\end{figure}

\begin{figure} 
	\label{fig:spec} 
	\centering
	\plotone{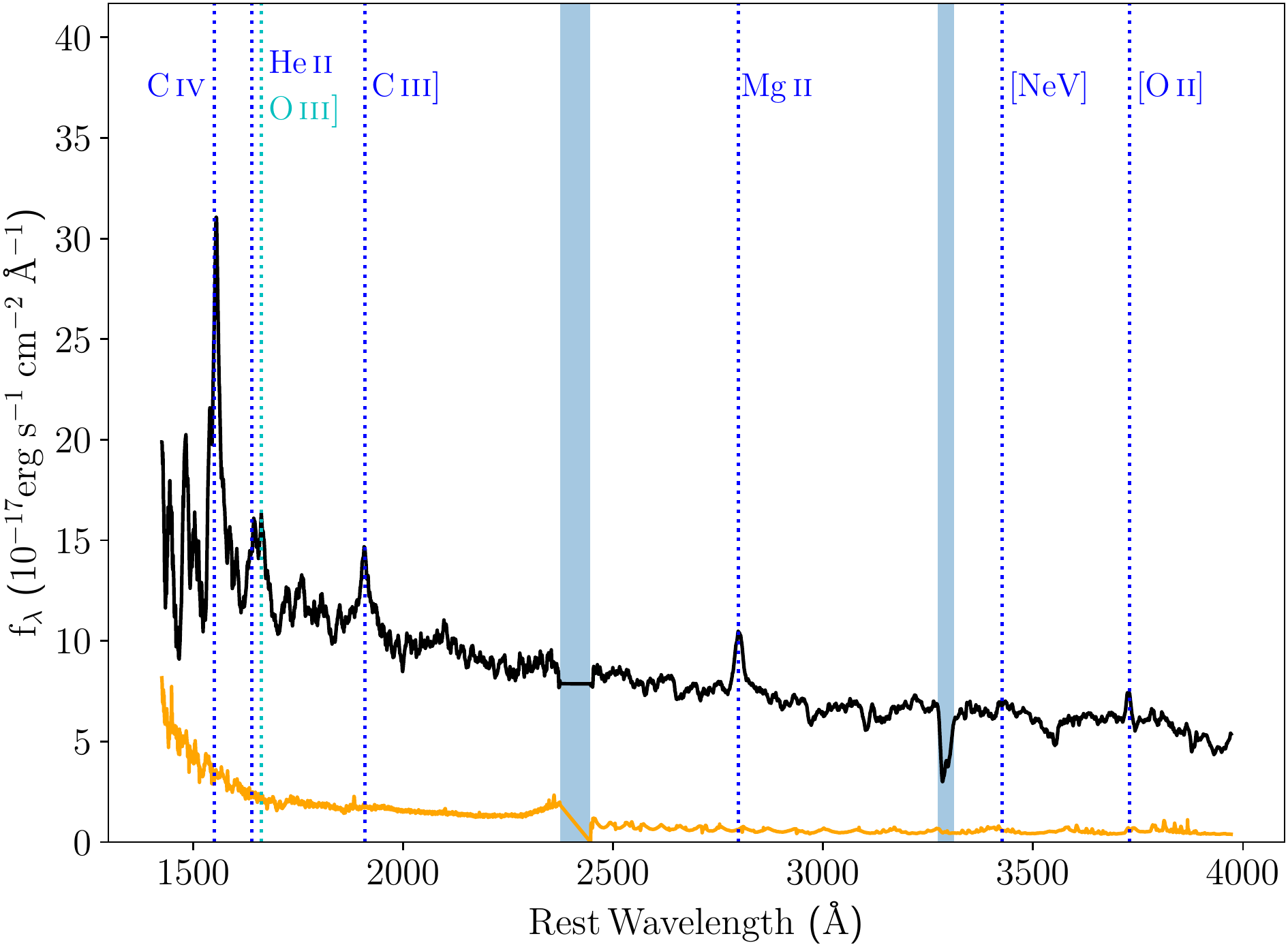} 
	\caption{P200 spectum. The spectrum is plotted in black and smoothed by a ﬂat window with a length of 16\AA, while the uncertainty is in orange. The vertical dotted lines mark the main emission lines. The blue shadow regions represent the position of the dichroic or the uncorrected telluric. } 
\end{figure}

The ZTF \citep{2019PASP..131g8001G,2019PASP..131a8002B,2019PASP..131a8003M,ptf} data from the its Public Data Release 6\footnote{\url{https://www.ztf.caltech.edu/page/dr6}} are collected. Initially, the co-added reference images are examined, from which GB6 J2113+1121 is distinguishable in the $g$, $r$ and $i$ bands. Light curves in these three bands are extracted for objects falling within a 5$\arcsec$ radius from position of the target. Only frames satisfied with {\tt catflags = 0} and {\tt chi < 4} are selected\footnote{\url{http://web.ipac.caltech.edu/staff/fmasci/ztf/extended\_cautionary\_notes.pdf}}. There are totally 149, 201 and 40 exposures in the $g$, $r$ and $i$ bands, respectively, covering time range between MJD 58200 and MJD 59200, see Figure \ref{mlc}. Magnitudes of five comparison stars ($\sim$ 17 mag) among the PTF Photometric Calibrator Catalog \citep{2012PASP..124..854O} lying 10$\arcmin$ from the target are also derived, of which the standard deviations are less than 0.05 mag, also see Figure \ref{mlc}. Significant optical variability of GB6 J2113+1121 is revealed by the ZTF light curves, from which the remarkable signature is a violent flare peaking at MJD 58684 captured by all the three bands. Considering that the target is barely detected in a single exposure at the beginning of the ZTF operation, the brightening for the flare reaches to over 3.0 mag. It is interesting that there is another major flare might peaking around MJD 58966 as violent as the former one, though only its descent phase is caught. Since the first flare is well sampled, it is worthy of attentions. It costs 44 days that flux densities of the target reaches to the peaking values ($g_{peak}$ = 17.68 $\pm$ 0.02, $r_{peak}$ = 17.08 $\pm$ 0.02 and $i_{peak}$ = 16.92 $\pm$0.02 mag) from MJD 58640 ($g$ = 19.80 $\pm$ 0.09, $r$ = 19.39 $\pm$ 0.07 and $i$ = 19.07 $\pm$ 0.07 mag). The flux descent phase begins with a rapid flux density decay (i.e. a faintness of $\simeq$ 1.1 mag within 4 days in $g$ and $r$ bands). Then a slow decay lasting for about 2-month is followed ($g$ = 20.17 $\pm$ 0.11 and $r$ = 19.74 $\pm$ 0.09 mag at MJD 58754). Between the two major flares, no significant activities are detected and the source stays in a low state ($g$ $\sim$ 20.0 and $r$ $\sim$ 19.5 mag) . Based on the limited data, the descent phase of the second flare only lasts about 20 days, from MJD 58966 ($g_{peak}$ = 17.61 $\pm$ 0.02 mag) to MJD 58987 ($g$ = 19.98 $\pm$ 0.10 mag). Interestingly, since the $g$ band magnitude at MJD 58970 is 19.23  $\pm$ 0.06 mag, a rapid flux density decay (i.e. a dimming of 1.6 mag within 4 days) is detected. Since then optical emissions of GB6 J2113+1121 in year 2020 are still active, however, no flares as violent as the major flares have been detected. 

\subsection{{\it WISE} Data}
We retrieve single exposure photometric data in \emph{W1} and \emph{W2} bands (centered at 3.4, 4.6 $\mu$m in the rest frame) from Wide-field Infrared Survey Explorer (\emph{WISE}; \citealt{2010AJ....140.1868W,allwise}) and the Near-Earth Object {\it WISE} Reactivation mission (\emph{NEOWISE-R}; \citealt{2014ApJ...792...30M,neowise}). Following our previous work \citep{2017ApJ...846L...7S,2020ApJ...889...46S}, we filter the bad data points with poor image quality (``qi\_fact"$<$1), a small separation to South Atlantic Anomaly (``SAA"$<$5) and the flagged moon masking (``moon mask"=1). Firstly we bin the data in each epoch (nearly half year) using median value to probe the long-term variability of the target, while the sample standard deviation value is taken as the conservative estimate of the uncertainty. Violent long time-scale variability is clearly exhibited in the {\it WISE} light curves, see Figure \ref{mlc}. For instance, comparing the observations at MJD 58255 and 58619, brightening of 2.4 mag in both \emph{W1} and \emph{W2} bands is detected. In fact, the variability amplitude can be larger considering that the target is dim at the beginning operation of {\it WISE} (i.e. \emph{W1} = 16.2 $\pm$ 0.3 mag, \emph{W2} = 14.9 $\pm$ 0.2 mag at MJD 55332). On the other hand, rapid IR variability with amplitudes of about 0.5 mag within 0.5 day (\emph{W1}) and 0.3 day (\emph{W2}) at MJD 59146 is observed. The variability of the IR color is also checked, no such evidences are found. 

\subsection{Implications of multi-wavelength properties of GB6 J2113+1121}
Here a thorough investigation of multi-wavelength behaviors of the flat spectral radio source GB6 J2113+1121 has been performed. The most distinct feature is the emergence of a giant flare in IR, optical and $\gamma$-ray long-term light curves. The flare brings large variability amplitudes, roughly 25-fold of flux increase in IR and $\gamma$-ray regimes, and 17-fold in optical bands. The violent variability can be naturally explained as an activity induced by the AGN jet \citep[e.g.,][]{2016ARA&A..54..725M,2019ARA&A..57..467B}. The peaking times (around MJD 58600) of the giant flare at different wavelengths are consistent, though the data sampling of the WISE light curves is sparse. Meanwhile, no activity that is as violent as the giant flare are found from the first 10-yr monitoring of {\it Fermi}-LAT as well as observations of {\it WISE} and iPTF then. The similar shape of multi-wavelength light curves provides a decisive evidence supporting the association between the $\gamma$-ray source and the low-energy counterpart and hence GB6 J2113+1121 is a $\gamma$-ray emitting FSRQ. 

In addition to the long-term variability, rapid variations have been found in IR and optical domains, from which the doubling timescale at the source frame can be simply constrained as $\tau_{doub, source} = \Delta t\times ln2/ln(F_{1}/F_{2})/(1+z)$. Therefore, $\tau_{doub, source}$ are inferred as $\sim$ 19-hr for the sharp decline in ZTF $g$ band around MJD 58970 and $\sim$ 5-hr for the dense monitoring of {\it WISE} at MJD 59146. In the perspective of the short variability timescale together with its strong $\gamma$-ray emission, value of the Doppler factor $\delta$ of the emitting jet blob should be high enough to avoid severe attenuation on $\gamma$-rays from soft photons via the $\gamma\gamma$ process. The absorption opacity of can be calculated as \citep{1995MNRAS.273..583D}:
     \begin{equation}
         \tau_{\gamma\gamma}(x^{\prime})=\frac{\sigma_{\rm T}}{5}n^{\prime}(x^{\prime}_{\rm t})x^{\prime}_{\rm t}R^{\prime},
     \end{equation}
     where $x^{\prime}$ = $h\nu^{\prime}/m_{e}c^{2}$ is the dimensionless energy of the $\gamma$ rays and $x^{\prime}_{\rm t}$ for the target soft photon in the co-moving frame, $n^{\prime}(x^{\prime}_{\rm t})$ is the co-moving differential number density of the target photon per energy, $\sigma_{\rm T}$ is the scattering Thomson cross section, and $R^{\prime}$ is the absorption length. The absorbing soft photons could be from the jet itself and hence in this case the absorption length is equal to radius of the emitting jet blob, $R_{\gamma}^{\prime} = c\tau^{\prime}_{var}\lesssim\delta c\tau_{doub,source}$. Since the most energetic $\gamma$-ray photon is at $\simeq$~8~GeV, adopting the soft radiation ($\sim 5\times10^{45}$ erg $\rm s^{-1}$ at several keVs) from the {\it Swift}-XRT observation and $t_{IR,source}$ of $\sim$ 5-hr, a constraint of $\delta \gtrsim$ 5 can be obtained. On the other hand, since significant spectral features are revealed by the P200 spectrum, the soft photons could be external to the jet (e.g. from accretion disk or broad emission lines), depending on the location of the jet dissipation region. Assuming a conical jet geometry (i.e. $\Gamma\theta$=1, where $\Gamma$ is the jet bulk Lorentz factor and $\theta$ is the jet opening angle), the location can be determined as $r_{\gamma} = \delta R_{\gamma}^{\prime} \sim$ 0.05 pc, from which a typical Doppler factor value of 15 \citep[e.g.,][]{2017MNRAS.466.4625L} is used. Comparing with the typical radius of the BLR  (i.e. $\sim$ 0.1 pc, \citealt{2010MNRAS.405L..94T}), the emitting jet blob is likely embedded in the radiation of the broad emission lines. However, due to insufficient information of the the broad emission lines in UV band \citep{2006ApJ...653.1089L,2010ApJ...717L.118P}, constraints from the external absorption to the $\gamma$ rays can not be achieved.

\begin{figure}
\centering
\includegraphics[scale=0.8]{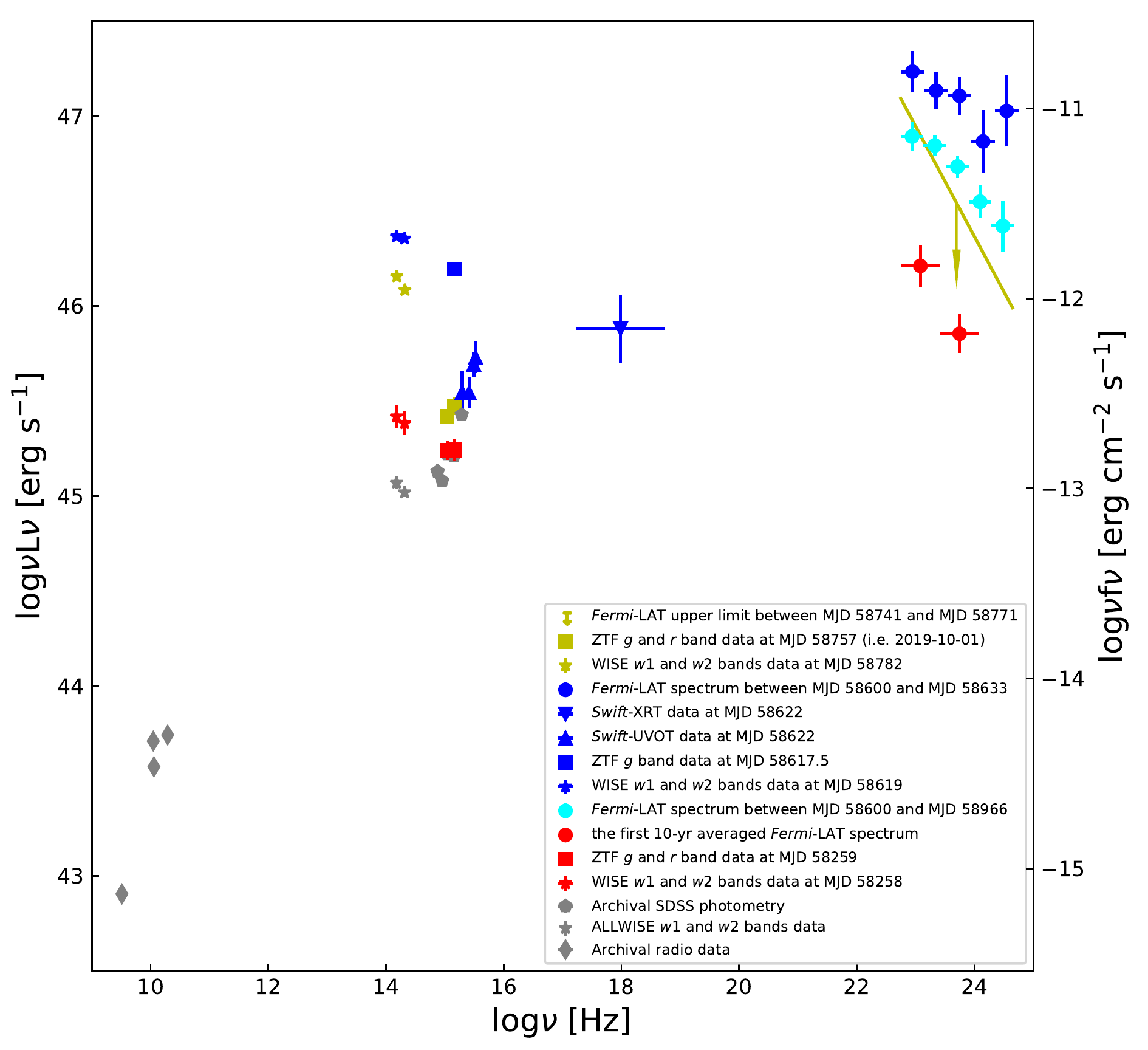}
\caption{The broadband SED of GB6 J2113+1121. Simultaneous data are colored. The blue ones represent the high flux state and red one correspond to the low flux state. The one year averaged $\gamma$-ray spectrum corresponding to the long term high flux state is plotted in cyan symbols. Multi-wavelength data of GB6 J2113+1121 that are strictly simultaneous with the neutrino arrival are also plotted (in yellow). Un-Simultaneous data are in grey.}
\label{sed}
\end{figure}

The multiwavelength data allow us to draw broadband SEDs of GB6 J2113+1121. As shown in Figure \ref{sed}, typical two-bump shape SEDs are pictured. Though the ascent parts of the SED bumps are not well sampled, the ratio of the peak luminosities between the high energy bump and the low energy one is likely $\gtrsim$ 1, which is common for FSRQs \citep[e.g.,][]{2013ApJ...763..134F}. Meanwhile, the peak frequency of synchrotron bump is constrains as $\lesssim$~$10^{14}$~Hz, and hence it is classified as a low-synchrotron-peaked source (LSP, \citealt{1995ApJ...444..567P,2010ApJ...716...30A}). Comparing SEDs between different flux states, the remarkable feature is violent variations in IR, optical and $\gamma$-ray domains. The IR domain is close to the peak frequency of the synchrotron bump and significant dilution of the big blue bump is avoided there. In the optical band, emission of the accretion disk is dominant at low flux state, however, the jet emission is overwhelming at high flux state. Interestingly, a drop-out of the jet emission follows and the {\it Swift}-UVOT data in the high flux state SED is well consistent with the extrapolation of archival SDSS data and ZTF measurements at quiescent state. For the high energy bump peaking in the $\gamma$-ray domain, the variability amplitude is comparable with the low energy one. SED peaks of blazars usually move to shorter wavelengths when it becomes brighter \citep[e.g.,][]{2013MNRAS.432L..66G}, future simultaneous sub-mm and hard X-ray observations are crucial for investigations on evolution of the SED of GB6 J2113+1121 during different flux states.

\section{DISCUSSION and SUMMARY} \label{sec:diss}
TDEs act as sights that activities of the massive black holes (MBHs) lurking in the galaxy centers are triggered when an orbiting star is disrupted and captured \citep{1988Natur.333..523R}. Given the evolution timescale of TDEs is as short as months, the formation of an accretion disk and launching of a possibly accompanying jet or outflow can be witnessed in real time. Benefited from the advanced multi-wavelength (messenger) observations, especially the time-domain surveys, the number of detected TDEs is significantly increased \citep{2021arXiv210414580G}. Among those, TDEs with radio detections are of special interest since the radio emissions are tightly connected to the material ejections from the MBHs \citep{2020SSRv..216...81A}. The unique case is Sw J1644+57, from which luminous X-ray (i.e. $\sim 10^{48}$ erg $\rm s^{-1}$) and radio emissions (i.e. $\sim 10^{44}$ erg $\rm s^{-1}$), a hard X-ray spectrum, and more importantly the rapid X-ray variability have been detected, suggesting emergence of a well-aligned relativistic jet that is alike of blazars \citep{2011Sci...333..203B,2011Natur.476..421B,2011Sci...333..199L,2011Natur.476..425Z}. In principle, off-axis jetted TDE could be also detected, for instance, IGR J12580+0134 \citep{2016ApJ...816...20L} and Arp 299-B AT1 \citep{2018Sci...361..482M}, from which hard X-ray flare and compact core-jet resolved by VLBI are found, respectively. Another well studied radio-emitting TDE is ASASSN-14li, from which thermal X-ray emissions together with weak radio emissions are observed, and the latter is proposed to be from the wind \citep{2016ApJ...819L..25A,2018MNRAS.474.3593K}. It is interesting to probe the location where AT2019dsg lies by comparing it with other TDEs. The X-ray (i.e. $\sim 10^{43}$ erg $\rm s^{-1}$) and radio (i.e. $\sim 10^{39}$ erg $\rm s^{-1}$) emissions of AT2019dsg are mild compared with Sw J1644+57 \citep{2021NatAs...5..510S}. Its X-ray spectrum is well described by a blackbody radiation, and no signs of rapid variability in radio and X-ray band are reported \citep{2021NatAs...5..510S}. Therefore, the radio emission of AT2019dsg is proposed to be from the sub-relativistic wind, which is likely akin to ASASSN-14li \citep{2020SSRv..216...81A,2021NatAs...5..510S}.

Jetted AGNs are widely accepted as one of the most energetic accelerators in the universe \citep[e.g.,][]{2020NatAs...4..124B}, and it is not surprising that blazars are the dominant population of extragalactic $\gamma$-ray sky \citep{2020ApJS..247...33A,2020ApJ...892..105A}. On the other hand, powerful sub-relativistic winds are commonly seen in the luminous AGNs \citep{2015ARA&A..53..115K}, however, they are not notable for generating strong $\gamma$-ray emissions (i.e. in GeV and TeV energies). Nevertheless, neutrinos are proposed be from the accretion flares in the super-Eddington systems, from which the lack of detecting $\gamma$-ray emissions are due to the severe $\gamma$$\gamma$ absorption \citep{2021arXiv211109391V}. Actually, evidences that the incoming neutrino events are associated with flaring blazars are cumulating. Prior to the famous TXS 0506+056 case \citep{2018Sci...361.1378I}, an arriving neutrino (i.e. HESE-35) has been detected in direction of a FSRQ PKS B1424-418 from which ejection of a new jet blob is exhibited at the same time \citep{2016NatPh..12..807K}. GB6 J1040+0617, a LSP BL Lac, has been reported to be a plausible candidate of the neutrino event IC-141209A \citep{2019ApJ...880..103G}. Interestingly, a $\gamma$-ray high-synchrotron-peaked (HSP) BL Lac, BZB J0955+3551, is found to be spatially and temporally (i.e. a X-ray flare) coincident with the IceCube neutrino event IC-200107A \citep{2020A&A...640L...4G,2020ApJ...902...29P}. Considering the proximity between the energies of $\gamma$-ray photons and the neutrinos, searching for the neutrino\footnote{IC-191001A is excluded there due to the relatively large angular uncertainty.} counterparts from the GeV blazars detected by {\it Fermi}-LAT has been performed. MG3 J225517+2409 together with 1H 0323+342 stand out because of detections of simultaneous $\gamma$-ray and optical flares when the neutrinos arriving, though for the latter much stronger flares without detection of neutrino have been seen before \citep{2020ApJ...893..162F}.  Apparently, only PKS B1424-418 and 1H 0323+342 appears to exhibit strong emission lines among these candidates. In fact, the featureless optical spectra are not only due to the inefficient accretion but also the severe dilution of the non-thermal jet radiation. For instance, TXS 0506+056 is suggested as an ``intrinsic" FSRQ because of detection the faint [OII] line \citep{2019MNRAS.484L.104P}. Due to the dense radiation fields external to the jets  (e.g. emissions from BLR), absorption of the $\gamma$-ray photons of LSPs, especially the LSP FSRQs, could be significant and hence the link between observed $\gamma$ rays here and the neutrinos are complicated \citep{2021ApJ...911L..18K}. Albeit the role that the external photons act as obstacles for detecting $\gamma$ rays, they can also be additional target photons for p$\gamma$ interactions and possibly enhance the neutrino generation \citep[e.g.,][]{2017MNRAS.467L..16P}.

Let us focus on the neutrino IC-191001A. Firstly, GB6 J2113+1121 is the only $\gamma$-ray source spatially coincident with the neutrino event. The by chance probability of such a coincidence is calculated. Since the sensitivity of IceCube improves with  increasing sky declination \citep{2020PhRvL.124e1103A}, only 4FGL-DR3\footnote{\url{https://fermi.gsfc.nasa.gov/ssc/data/access/lat/12yr_catalog/}} blazars (candidates) locate with $DEC. >$ $-5^{\circ}$ are chosen. Meanwhile, by excluding the Galactic plane (i.e. $|b| < 10^{\circ}$), we estimate the surface density of blazars to be $\sim 0.09$ per square degree. Therefore, the expectation number of $\gamma$-ray blazars that spatially coincide with a neutrino event like IC 191001A is roughly 2.4, obviously the spatial coincidence alone is not sufficient. Then a Monte-Carlo simulation is carried out to take in account of the $\gamma$-ray flare. The central location of IC 191001A is randomized in the sky region as mentioned, whereas the size of the localization error box remains. On the other hand, additional temporal constraints are applied when selecting corresponding blazars (candidates). Like GB6 J2113+1121, they are significantly $\gamma$-ray variable, and more importantly, exhibit a clear flux enhancement at the last bin of the public 12-year long light curve sampled in yearly bin. After $\rm 10^{4}$ times simulations, we derive the probability of the chance-coincidence as $\sim$ 0.03, supporting that GB6 J2113+1121 is a likely neutrino emitter.

Since relativistic particles accelerated in relativistic jet emit both neutrinos and $\gamma$-ray photons, these two outputs could be proportional. Therefore, broadband temporal behaviors of GB6 J2113+1121 is compared with the incoming of the neutrino. Initially, an epoch of one month  corresponding to the $\gamma$-ray high flux state based on the 10-day bin light curve is identified by the Bayesian block approach \citep{2013ApJ...764..167S}, which is actually the bin with the largest TS value in the monthly light curve. The flux then is one order of magnitude of that during the entire IceCube operation time (i.e. since MJD 55470). Meanwhile, $\gamma$-ray flux of GB6 J2113+1121 in the time range between the discovery of the TDE and arrival time of the neutrino is also extracted, (6.3 $\pm$ 0.8)$\rm \times 10^{-8}$ ph $\rm cm^{-2}$ $\rm s^{-1}$, roughly half of the flux in the flaring epoch. Note that compared with the discovery time of the TDE (i.e. MJD 58582), the onset time of the blazar $\gamma$-ray flaring epoch (i.e. MJD 58600) is about 18 day closer to the arrival time of IC 191001A (i.e. MJD 58758). Interestingly, the recorded onset time of the giant optical flare is even closer (i.e. MJD 58634). For the WISE light curves, despite the sparse sampling the infrared emissions maintain in a high state at MJD 58782 that is within one month after the arrival time of the neutrino, although gamma-ray and optical emissions had already descended from its outbursts. Note that for TXS 0506+056, there is also a time lag of several months between the $\gamma$-ray flux peak and the detection IC-170922A \citep{2019ApJ...880..103G}. Such a time mismatch could be due to the low event rate of the neutrino detection. Nevertheless, GB6 J2113+1121 is among the a few {\it Fermi} blazars that are not only spatially coincident with a detected IceCube neutrino but also with detections of qusi-simultaneous multi-wavelength flares at the arrival time the neutrino. Based on these facts, we propose that GB6 J2113+1121 is a possible counterpart of the neutrino IC-191001A. Behaviors of GB6 J2113+1121 is encouraging for the argument that flaring blazars can act as counterparts of IceCube neutrinos \citep[e.g.,][]{2016ApJ...831...12H}. 

\begin{figure}
\centering
\includegraphics[scale=0.3]{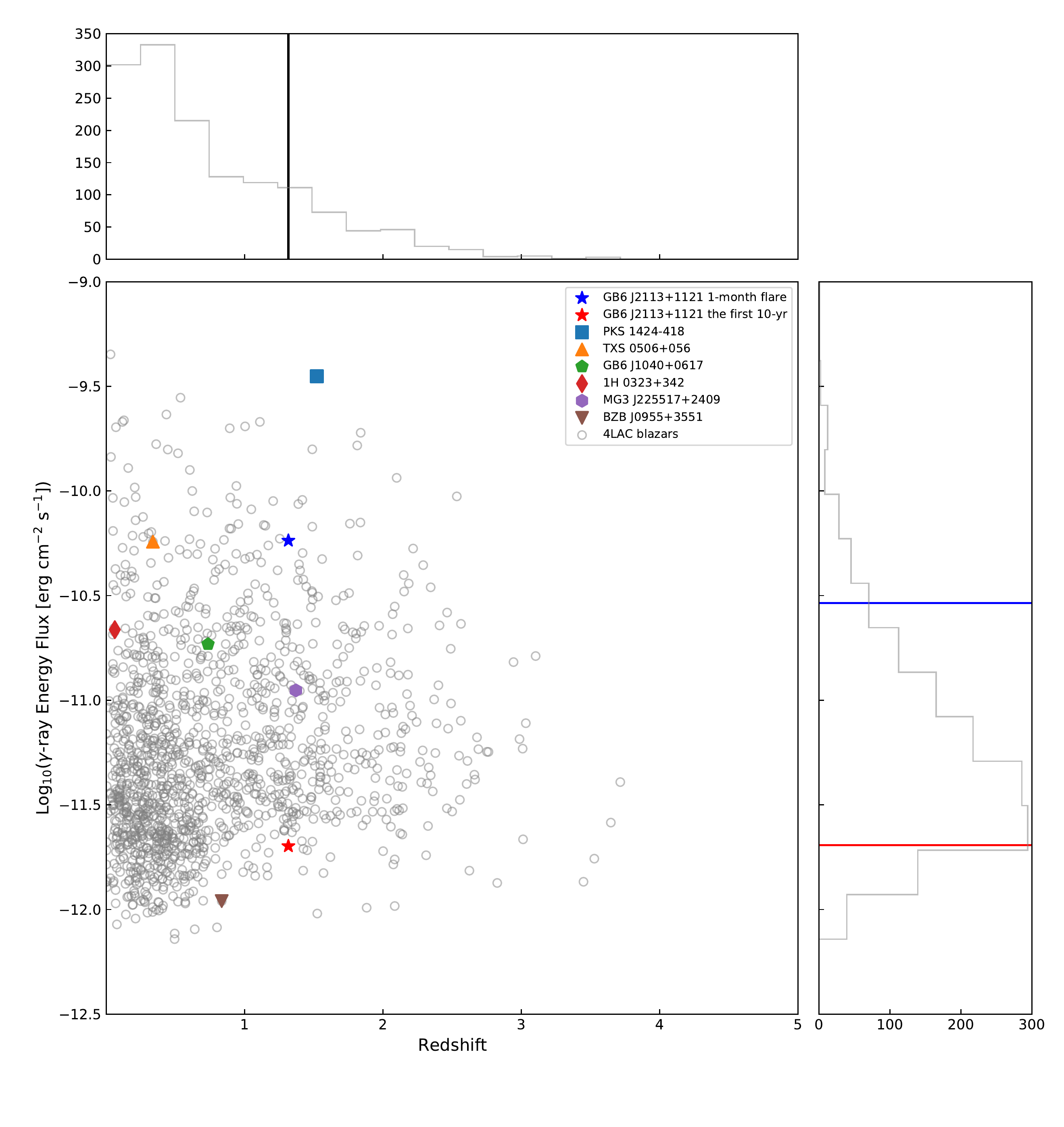}
\includegraphics[scale=0.3]{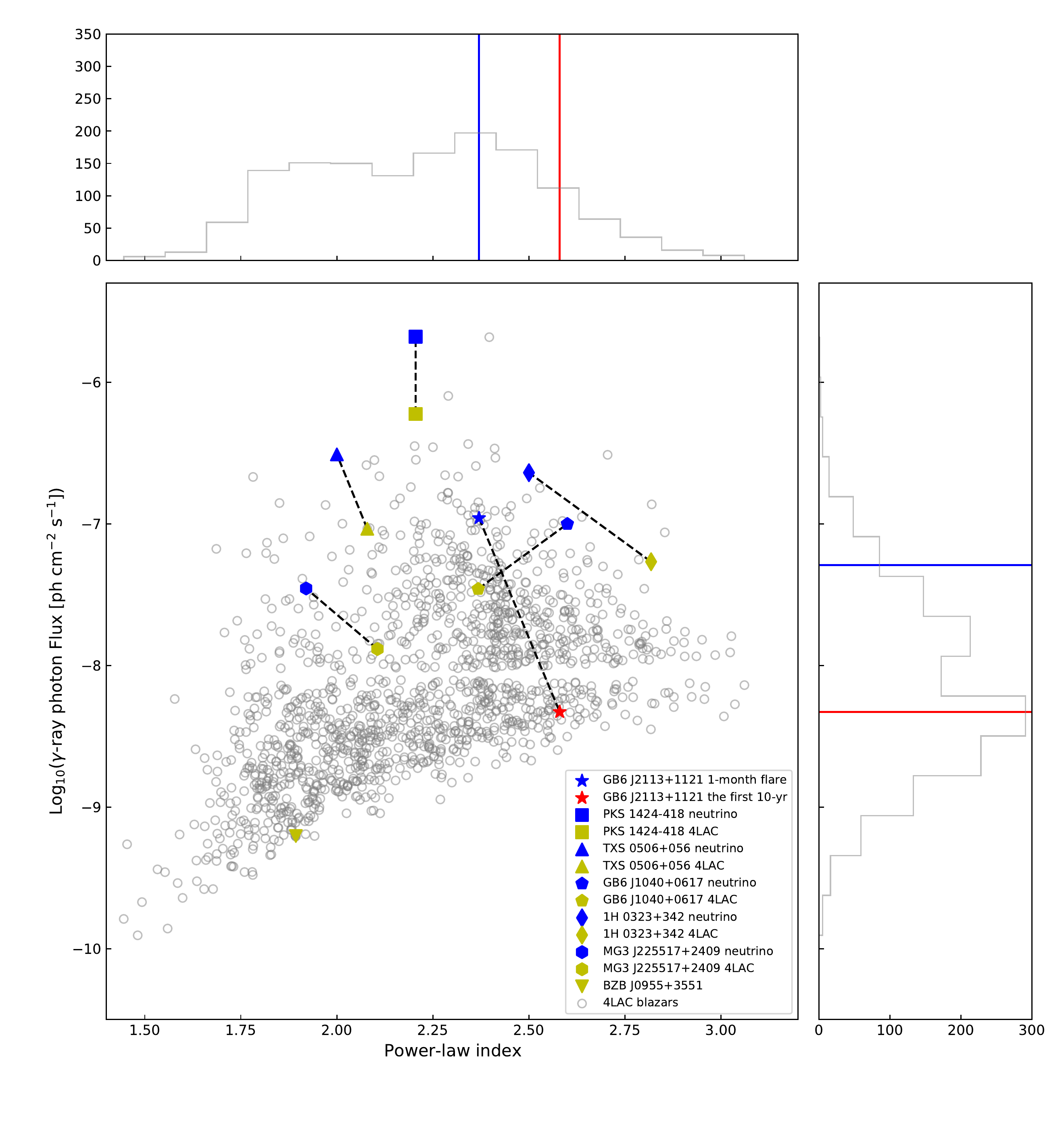}
\caption{Comparisons between GB6 J2113+1121 and 4LAC blazars \citep{2020ApJ...892..105A}, in which the known neutrino-emitting blazar candidates are highlighted. The energy ranges correspond to the energy flux and the photon flux are between 100~MeV and 100~GeV. In the right panel, 4LAC values of the candidates are colored in yellow, meanwhile, values corresponding to the flares temporally coincident with the arrival of neutrinos \citep{2016NatPh..12..807K,2018Sci...361.1378I,2019ApJ...880..103G,2020ApJ...893..162F} are plotted in blue. For PKS B1424-418, no available spectral information in the high flux state is found.}
\label{comp}
\end{figure}

It is interesting to compare GB6 J2113+1121 with the 4LAC blazars \citep{2020ApJ...892..105A}, especially the neutrino-emitting blazar candidates, see Figure \ref{comp}. GB6 J2113+1121 bares a $\gamma$-ray spectral slope typically for 4LAC FSRQs (i.e. 2.44 $\pm$ 0.20). The most luminous one among the candidates is PKS 1424-418 while the faintest one is BZB J0955+3551 with a very hard spectrum. TXS 0506+056 is also a very luminous candidate. 1H 0323+342 stands out because of its rather soft spectrum, meanwhile, it is the only radio-loud narrow line Seyfert I galaxy. The redshift distribution of the candidates is from 0.03 (i.e. 1H 0323+342) to 1.522 (i.e PKS 1424-418). Though the flux level of GB6 J2113+1121 in the quiescent state is low, benefited from the large variability amplitude, its flux level in high flux state is comparable with those of other candidates. Detailed theoretical calculation on neutrino production by GB6 J2113+1121 is beyond the scope of this study, in perspective of the potential attenuation of the observed $\gamma$ rays as well as the tangling between the leptonic and hadronic contributions \citep[e.g.,][]{2017MNRAS.467L..16P,2018ApJ...863L..10A}. Future observational evidences, such as orphan $\gamma$-ray flare or X-ray flare that is possible induced by the secondary particles from the hadronic cascade, are helpful to further investigate the neutrino generation processes.

In summary, multi-wavelength properties of GB6 J2113+1121 and its potential as a possible neutrino emitter corresponding to IC-191001A are investigated. It is confirmed as the only spatial coincident $\gamma$-ray source of the neutrino when it arrives. In May 2019, a strong $\gamma$-ray flare with flux increase reaching to 20-fold and possible spectral hardness then emerges. Violent optical flares in ZTF $g$, $r$ and $i$ bands together with IR ones in {\it WISE} {\it W1} and {\it W2} bands have been also detected. Compared with the detection time of the TDE AT2019dsg, the onset times of the strong $\gamma$-ray and optical flares are closer to the arrival time of the neutrino. Meanwhile, the infrared emissions of GB6 J2113+1121 remain in the high flux state at the time that is within one month from the arrival time of the neutrino. Benefited from the temporal information, the by chance of the coincidence between GB6 J2113+1121 and IC-191001A is estimated as $\sim$ 0.03. Therefore, in addition to the radio-emitting TDE AT2019dsg, the possibility GB6 J2113+1121 being the counterpart of IC-191001A needs to be considered. Since the blazar is a persistent source unlike the TDE, in perspective of the “orphan” neutrino flare of TXS 0506+056 \citep{2018Sci...361..147I}, it would be interesting to perform a specific analysis of archival IceCube data to check whether there is weak neutrino excess in this direction before the TDE. Meanwhile, future detections of more neutrino events towards this direction would be supportive for GB6 J2113+1121 as a neutrino emitter.

\begin{acknowledgments}
This research uses data obtained through the Telescope Access Program (TAP). Observations obtained with the Hale Telescope at Palomar Observatory were obtained as part of an agreement between the National Astronomical Observatories, Chinese Academy of Sciences, and the California Institute of Technology. Tapio Pursimo is appreciated for sharing their optical spectrum of GB6 J2113+1121. This research has made use of data obtained from the High Energy Astrophysics Science Archive Research Center (HEASARC), provided by $\rm NASA^{\prime}$s Goddard Space Flight Center. This research makes use of data products from the Wide-field Infrared Survey Explorer, which is a joint project of the University of California, Los Angeles, and the Jet Propulsion Laboratory/California Institute of Technology, funded by the National Aeronautics and Space Administration. This research also makes use of data products from NEOWISE-R, which is a project of the Jet Propulsion Laboratory/California Institute of Technology, funded by the Planetary Science Division of the National Aeronautics and Space Administration. This study use data based on observations obtained with the Samuel Oschin Telescope 48-inch and the 60-inch Telescope at the Palomar Observatory as part of the Zwicky Transient Facility project. ZTF is supported by the National Science Foundation under Grant No. AST-2034437 and a collaboration including Caltech, IPAC, the Weizmann Institute for Science, the Oskar Klein Center at Stockholm University, the University of Maryland, Deutsches Elektronen-Synchrotron and Humboldt University, the TANGO Consortium of Taiwan, the University of Wisconsin at Milwaukee, Trinity College Dublin, Lawrence Livermore National Laboratories, and IN2P3, France. Operations are conducted by COO, IPAC, and UW.

This work was supported in part by the NSFC under grants 11703093, U2031120, 11833007, 12103048 and 12073025, as well as the Fundamental Research Funds for the Central Universities: WK2030000023. This work was also supported in part by the Special Natural Science Fund of Guizhou University (grant No. 201911A) and the First-class Physics Promotion Programme (2019) of Guizhou University.
\end{acknowledgments}

\vspace{5mm}
\facilities{{\it Fermi} (LAT); {\it Swift}}; ZTF; Hale (DBSP); {\it WISE}

\software{Astropy \citep{2018AJ....156..123A}}
\clearpage
\bibliographystyle{aasjournal}
\bibliography{refs}

\begin{thebibliography}{}
\expandafter\ifx\csname natexlab\endcsname\relax\def\natexlab#1{#1}\fi
\providecommand{\url}[1]{\href{#1}{#1}}
\providecommand{\dodoi}[1]{doi:~\href{http://doi.org/#1}{\nolinkurl{#1}}}
\providecommand{\doeprint}[1]{\href{http://ascl.net/#1}{\nolinkurl{http://ascl.net/#1}}}
\providecommand{\doarXiv}[1]{\href{https://arxiv.org/abs/#1}{\nolinkurl{https://arxiv.org/abs/#1}}}

\bibitem[{{Aartsen} {et~al.}(2015){Aartsen}, {Ackermann}, {Adams}, {Aguilar},
  {Ahlers}, {Ahrens}, {Altmann}, {Anderson}, {Archinger}, {Arguelles}, {Arlen},
  {Auffenberg}, {Bai}, {Barwick}, {Baum}, {Bay}, {Baker}, {Beatty}, {Becker
  Tjus}, {Becker}, {BenZvi}, {Berghaus}, {Berley}, {Bernardini}, {Bernhard},
  {Besson}, {Binder}, {Bindig}, {Bissok}, {Blaufuss}, {Blumenthal}, {Boersma},
  {Bohm}, {Bos}, {Bose}, {B{\"o}ser}, {Botner}, {Brayeur}, {Bretz}, {Brown},
  {Buzinsky}, {Casey}, {Casier}, {Cheung}, {Chirkin}, {Christov}, {Christy},
  {Clark}, {Classen}, {Clevermann}, {Coenders}, {Cowen}, {Cruz Silva},
  {Daughhetee}, {Davis}, {Day}, {de Andr{\'e}}, {De Clercq}, {Dembinski}, {De
  Ridder}, {Desiati}, {de Vries}, {de Wasseige}, {de With}, {DeYoung},
  {D{\'\i}az-V{\'e}lez}, {Dumm}, {Dunkman}, {Eagan}, {Eberhardt}, {Ehrhardt},
  {Eichmann}, {Eisch}, {Euler}, {Evenson}, {Fadiran}, {Fazely}, {Fedynitch},
  {Feintzeig}, {Felde}, {Filimonov}, {Finley}, {Fischer-Wasels}, {Flis},
  {Frantzen}, {Fuchs}, {Gaisser}, {Gaior}, {Gallagher}, {Gerhardt}, {Gier},
  {Gladstone}, {Gl{\"u}senkamp}, {Goldschmidt}, {Golup}, {Gonzalez}, {Goodman},
  {G{\'o}ra}, {Grant}, {Gretskov}, {Groh}, {Gro{\ss}}, {Ha}, {Haack}, {Haj
  Ismail}, {Hallen}, {Hallgren}, {Halzen}, {Hanson}, {Hebecker}, {Heereman},
  {Heinen}, {Helbing}, {Hellauer}, {Hellwig}, {Hickford}, {Hignight}, {Hill},
  {Hoffman}, {Hoffmann}, {Homeier}, {Hoshina}, {Huang}, {Huelsnitz}, {Hulth},
  {Hultqvist}, {In}, {Ishihara}, {Jacobi}, {Jacobsen}, {Japaridze}, {Jero},
  {Jurkovic}, {Kaminsky}, {Kappes}, {Karg}, {Karle}, {Kauer}, {Keivani},
  {Kelley}, {Kheirandish}, {Kiryluk}, {Kl{\"a}s}, {Klein}, {K{\"o}hne},
  {Kohnen}, {Kolanoski}, {Koob}, {K{\"o}pke}, {Kopper}, {Kopper}, {Koskinen},
  {Kowalski}, {Krings}, {Kroll}, {Kroll}, {Kunnen}, {Kurahashi}, {Kuwabara},
  {Labare}, {Lanfranchi}, {Larsen}, {Larson}, {Lesiak-Bzdak}, {Leuermann},
  {L{\"u}nemann}, {Madsen}, {Maggi}, {Mahn}, {Maruyama}, {Mase}, {Matis},
  {Maunu}, {McNally}, {Meagher}, {Medici}, {Meli}, {Meures}, {Miarecki},
  {Middell}, {Middlemas}, {Milke}, {Miller}, {Mohrmann}, {Montaruli}, {Morse},
  {Nahnhauer}, {Naumann}, {Niederhausen}, {Nowicki}, {Nygren}, {Obertacke},
  {Olivas}, {Omairat}, {O'Murchadha}, {Palczewski}, {Paul}, {Pepper},
  {P{\'e}rez de los Heros}, {Pfendner}, {Pieloth}, {Pinat}, {Posselt}, {Price},
  {Przybylski}, {P{\"u}tz}, {Quinnan}, {R{\"a}del}, {Rameez}, {Rawlins},
  {Redl}, {Rees}, {Reimann}, {Relich}, {Resconi}, {Rhode}, {Richman}, {Riedel},
  {Robertson}, {Rodrigues}, {Rongen}, {Rott}, {Ruhe}, {Ruzybayev}, {Ryckbosch},
  {Saba}, {Sander}, {Sandroos}, {Santander}, {Sarkar}, {Schatto}, {Scheriau},
  {Schmidt}, {Schmitz}, {Schoenen}, {Sch{\"o}neberg}, {Sch{\"o}nwald},
  {Schukraft}, {Schulte}, {Schulz}, {Seckel}, {Sestayo}, {Seunarine},
  {Shanidze}, {Smith}, {Soldin}, {Spiczak}, {Spiering}, {Stamatikos}, {Stanev},
  {Stanisha}, {Stasik}, {Stezelberger}, {Stokstad}, {St{\"o}{\ss}l},
  {Strahler}, {Str{\"o}m}, {Strotjohann}, {Sullivan}, {Sutherland}, {Taavola},
  {Taboada}, {Tamburro}, {Ter-Antonyan}, {Terliuk}, {Te{\v{s}}i{\'c}}, {Tilav},
  {Toale}, {Tobin}, {Tosi}, {Tselengidou}, {Unger}, {Usner}, {Vallecorsa}, {van
  Eijndhoven}, {Vandenbroucke}, {van Santen}, {Vanheule}, {Vehring}, {Voge},
  {Vraeghe}, {Walck}, {Wallraff}, {Weaver}, {Wellons}, {Wendt}, {Westerhoff},
  {Whelan}, {Whitehorn}, {Wichary}, {Wiebe}, {Wiebusch}, {Williams}, {Wissing},
  {Wolf}, {Wood}, {Woschnagg}, {Xu}, {Xu}, {Xu}, {Yanez}, {Yodh}, {Yoshida},
  {Zarzhitsky}, {Ziemann}, {Zoll}, \& {IceCube
  Collaboration}}]{2015ApJ...807...46A}
{Aartsen}, M.~G., {Ackermann}, M., {Adams}, J., {et~al.} 2015, \apj, 807, 46,
  \dodoi{10.1088/0004-637X/807/1/46}

\bibitem[{{Aartsen} {et~al.}(2017{\natexlab{a}}){Aartsen}, {Ackermann},
  {Adams}, {Aguilar}, {Ahlers}, {Ahrens}, {Altmann}, {Andeen}, {Anderson},
  {Ansseau}, {Anton}, {Archinger}, {Arg{\"u}elles}, {Auer}, {Auffenberg},
  {Axani}, {Baccus}, {Bai}, {Barnet}, {Barwick}, {Baum}, {Bay}, {Beattie},
  {Beatty}, {Becker Tjus}, {Becker}, {Bendfelt}, {BenZvi}, {Berley},
  {Bernardini}, {Bernhard}, {Besson}, {Binder}, {Bindig}, {Bissok}, {Blaufuss},
  {Blot}, {Boersma}, {Bohm}, {B{\"o}rner}, {Bos}, {Bose}, {B{\"o}ser},
  {Botner}, {Bouchta}, {Braun}, {Brayeur}, {Bretz}, {Bron}, {Burgman},
  {Burreson}, {Carver}, {Casier}, {Cheung}, {Chirkin}, {Christov}, {Clark},
  {Classen}, {Coenders}, {Collin}, {Conrad}, {Cowen}, {Cross}, {Day}, {Day},
  {de Andr{\'e}}, {De Clercq}, {del Pino Rosendo}, {Dembinski}, {De Ridder},
  {Descamps}, {Desiati}, {de Vries}, {de Wasseige}, {de With}, {DeYoung},
  {D{\'\i}az-V{\'e}lez}, {di Lorenzo}, {Dujmovic}, {Dumm}, {Dunkman},
  {Eberhardt}, {Edwards}, {Ehrhardt}, {Eichmann}, {Eller}, {Euler}, {Evenson},
  {Fahey}, {Fazely}, {Feintzeig}, {Felde}, {Filimonov}, {Finley}, {Flis},
  {F{\"o}sig}, {Franckowiak}, {Fr{\`e}re}, {Friedman}, {Fuchs}, {Gaisser},
  {Gallagher}, {Gerhardt}, {Ghorbani}, {Giang}, {Gladstone}, {Glauch},
  {Glowacki}, {Gl{\"u}senkamp}, {Goldschmidt}, {Gonzalez}, {Grant}, {Griffith},
  {Gustafsson}, {Haack}, {Hallgren}, {Halzen}, {Hansen}, {Hansmann}, {Hanson},
  {Haugen}, {Hebecker}, {Heereman}, {Helbing}, {Hellauer}, {Heller},
  {Hickford}, {Hignight}, {Hill}, {Hoffman}, {Hoffmann}, {Hoshina}, {Huang},
  {Huber}, {Hulth}, {Hultqvist}, {In}, {Inaba}, {Ishihara}, {Jacobi},
  {Jacobsen}, {Japaridze}, {Jeong}, {Jero}, {Jones}, {Jones}, {Joseph}, {Kang},
  {Kappes}, {Karg}, {Karle}, {Katz}, {Kauer}, {Keivani}, {Kelley}, {Kemp},
  {Kheirandish}, {Kim}, {Kim}, {Kintscher}, {Kiryluk}, {Kitamura}, {Kittler},
  {Klein}, {Kleinfelder}, {Kleist}, {Kohnen}, {Koirala}, {Kolanoski},
  {Konietz}, {K{\"o}pke}, {Kopper}, {Kopper}, {Koskinen}, {Kowalski},
  {Krasberg}, {Krings}, {Kroll}, {Kr{\"u}ckl}, {Kr{\"u}ger}, {Kunnen},
  {Kunwar}, {Kurahashi}, {Kuwabara}, {Labare}, {Laihem}, {Landsman},
  {Lanfranchi}, {Larson}, {Lauber}, {Laundrie}, {Lennarz}, {Leich},
  {Lesiak-Bzdak}, {Leuermann}, {Lu}, {Ludwig}, {L{\"u}nemann}, {Mackenzie},
  {Madsen}, {Maggi}, {Mahn}, {Mancina}, {Mandelartz}, {Maruyama}, {Mase},
  {Matis}, {Maunu}, {McNally}, {McParland}, {Meade}, {Meagher}, {Medici},
  {Meier}, {Meli}, {Menne}, {Merino}, {Meures}, {Miarecki}, {Minor},
  {Montaruli}, {Moulai}, {Murray}, {Nahnhauer}, {Naumann}, {Neer}, {Newcomb},
  {Niederhausen}, {Nowicki}, {Nygren}, {Obertacke Pollmann}, {Olivas},
  {O'Murchadha}, {Palczewski}, {Pandya}, {Pankova}, {Patton}, {Peiffer},
  {Penek}, {Pepper}, {P{\'e}rez de los Heros}, {Pettersen}, {Pieloth}, {Pinat},
  {Price}, {Przybylski}, {Quinnan}, {Raab}, {R{\"a}del}, {Rameez}, {Rawlins},
  {Reimann}, {Relethford}, {Relich}, {Resconi}, {Rhode}, {Richman}, {Riedel},
  {Robertson}, {Rongen}, {Roucelle}, {Rott}, {Ruhe}, {Ryckbosch}, {Rysewyk},
  {Sabbatini}, {Sanchez Herrera}, {Sandrock}, {Sandroos}, {Sandstrom},
  {Sarkar}, {Satalecka}, {Schlunder}, {Schmidt}, {Schoenen}, {Sch{\"o}neberg},
  {Schukraft}, {Schumacher}, {Seckel}, {Seunarine}, {Solarz}, {Soldin}, {Song},
  {Spiczak}, {Spiering}, {Stanev}, {Stasik}, {Stettner}, {Steuer},
  {Stezelberger}, {Stokstad}, {St{\"o}{\ss}l}, {Str{\"o}m}, {Strotjohann},
  {Sulanke}, {Sullivan}, {Sutherland}, {Taavola}, {Taboada}, {Tatar},
  {Tenholt}, {Ter-Antonyan}, {Terliuk}, {Te{\v{s}}i{\'c}}, {Thollander},
  {Tilav}, {Toale}, {Tobin}, {Toscano}, {Tosi}, {Tselengidou}, {Turcati},
  {Unger}, {Usner}, {Vandenbroucke}, {van Eijndhoven}, {Vanheule}, {van
  Rossem}, {van Santen}, {Vehring}, {Voge}, {Vogel}, {Vraeghe}, {Wahl},
  {Walck}, {Wallace}, {Wallraff}, {Wandkowsky}, {Weaver}, {Weiss}, {Wendt},
  {Westerhoff}, {Wharton}, {Whelan}, {Wickmann}, {Wiebe}, {Wiebusch}, {Wille},
  {Williams}, {Wills}, {Wisniewski}, {Wolf}, {Wood}, {Woolsey}, {Woschnagg},
  {Xu}, {Xu}, {Xu}, {Yanez}, {Yodh}, {Yoshida}, \&
  {Zoll}}]{2017JInst..12P3012A}
---. 2017{\natexlab{a}}, Journal of Instrumentation, 12, P03012,
  \dodoi{10.1088/1748-0221/12/03/P03012}

\bibitem[{{Aartsen} {et~al.}(2017{\natexlab{b}}){Aartsen}, {Abraham},
  {Ackermann}, {Adams}, {Aguilar}, {Ahlers}, {Ahrens}, {Altmann}, {Andeen},
  {Anderson}, {Ansseau}, {Anton}, {Archinger}, {Arg{\"u}elles}, {Auffenberg},
  {Axani}, {Bai}, {Barwick}, {Baum}, {Bay}, {Beatty}, {Becker Tjus}, {Becker},
  {BenZvi}, {Berley}, {Bernardini}, {Bernhard}, {Besson}, {Binder}, {Bindig},
  {Bissok}, {Blaufuss}, {Blot}, {Bohm}, {B{\"o}rner}, {Bos}, {Bose},
  {B{\"o}ser}, {Botner}, {Braun}, {Brayeur}, {Bretz}, {Bron}, {Burgman},
  {Carver}, {Casier}, {Cheung}, {Chirkin}, {Christov}, {Clark}, {Classen},
  {Coenders}, {Collin}, {Conrad}, {Cowen}, {Cross}, {Day}, {de Andr{\'e}}, {De
  Clercq}, {del Pino Rosendo}, {Dembinski}, {De Ridder}, {Desiati}, {de Vries},
  {de Wasseige}, {de With}, {DeYoung}, {D{\'\i}az-V{\'e}lez}, {di Lorenzo},
  {Dujmovic}, {Dumm}, {Dunkman}, {Eberhardt}, {Ehrhardt}, {Eichmann}, {Eller},
  {Euler}, {Evenson}, {Fahey}, {Fazely}, {Feintzeig}, {Felde}, {Filimonov},
  {Finley}, {Flis}, {F{\"o}sig}, {Franckowiak}, {Friedman}, {Fuchs}, {Gaisser},
  {Gallagher}, {Gerhardt}, {Ghorbani}, {Giang}, {Gladstone}, {Glauch},
  {Gl{\"u}senkamp}, {Goldschmidt}, {Golup}, {Gonzalez}, {Grant}, {Griffith},
  {Haack}, {Haj Ismail}, {Hallgren}, {Halzen}, {Hansen}, {Hansmann}, {Hanson},
  {Hebecker}, {Heereman}, {Helbing}, {Hellauer}, {Hickford}, {Hignight},
  {Hill}, {Hoffman}, {Hoffmann}, {Holzapfel}, {Hoshina}, {Huang}, {Huber},
  {Hultqvist}, {In}, {Ishihara}, {Jacobi}, {Japaridze}, {Jeong}, {Jero},
  {Jones}, {Jurkovic}, {Kappes}, {Karg}, {Karle}, {Katz}, {Kauer}, {Keivani},
  {Kelley}, {Kheirandish}, {Kim}, {Kintscher}, {Kiryluk}, {Kittler}, {Klein},
  {Kohnen}, {Koirala}, {Kolanoski}, {Konietz}, {K{\"o}pke}, {Kopper}, {Kopper},
  {Koskinen}, {Kowalski}, {Krings}, {Kroll}, {Kr{\"u}ckl}, {Kr{\"u}ger},
  {Kunnen}, {Kunwar}, {Kurahashi}, {Kuwabara}, {Labare}, {Lanfranchi},
  {Larson}, {Lauber}, {Lennarz}, {Lesiak-Bzdak}, {Leuermann}, {Lu},
  {L{\"u}nemann}, {Madsen}, {Maggi}, {Mahn}, {Mancina}, {Mandelartz},
  {Maruyama}, {Mase}, {Maunu}, {McNally}, {Meagher}, {Medici}, {Meier}, {Meli},
  {Menne}, {Merino}, {Meures}, {Miarecki}, {Mohrmann}, {Montaruli}, {Moulai},
  {Nahnhauer}, {Naumann}, {Neer}, {Niederhausen}, {Nowicki}, {Nygren},
  {Obertacke Pollmann}, {Olivas}, {O'Murchadha}, {Palczewski}, {Pandya},
  {Pankova}, {Peiffer}, {Penek}, {Pepper}, {P{\'e}rez de los Heros}, {Pieloth},
  {Pinat}, {Price}, {Przybylski}, {Quinnan}, {Raab}, {R{\"a}del}, {Rameez},
  {Rawlins}, {Reimann}, {Relethford}, {Relich}, {Resconi}, {Rhode}, {Richman},
  {Riedel}, {Robertson}, {Rongen}, {Rott}, {Ruhe}, {Ryckbosch}, {Rysewyk},
  {Sabbatini}, {Sanchez Herrera}, {Sandrock}, {Sandroos}, {Sarkar},
  {Satalecka}, {Schlunder}, {Schmidt}, {Schoenen}, {Sch{\"o}neberg},
  {Schumacher}, {Seckel}, {Seunarine}, {Soldin}, {Song}, {Spiczak}, {Spiering},
  {Stanev}, {Stasik}, {Stettner}, {Steuer}, {Stezelberger}, {Stokstad},
  {St{\"o}ssl}, {Str{\"o}m}, {Strotjohann}, {Sullivan}, {Sutherland},
  {Taavola}, {Taboada}, {Tatar}, {Tenholt}, {Ter-Antonyan}, {Terliuk},
  {Te{\v{s}}i{\'c}}, {Tilav}, {Toale}, {Tobin}, {Toscano}, {Tosi},
  {Tselengidou}, {Turcati}, {Unger}, {Usner}, {Vandenbroucke}, {van
  Eijndhoven}, {Vanheule}, {van Rossem}, {van Santen}, {Veenkamp}, {Vehring},
  {Voge}, {Vogel}, {Vraeghe}, {Walck}, {Wallace}, {Wallraff}, {Wandkowsky},
  {Weaver}, {Weiss}, {Wendt}, {Westerhoff}, {Whelan}, {Wickmann}, {Wiebe},
  {Wiebusch}, {Wille}, {Williams}, {Wills}, {Wolf}, {Wood}, {Woolsey},
  {Woschnagg}, {Xu}, {Xu}, {Xu}, {Yanez}, {Yodh}, {Yoshida}, {Zoll}, \&
  {IceCube Collaboration}}]{2017ApJ...835..151A}
{Aartsen}, M.~G., {Abraham}, K., {Ackermann}, M., {et~al.} 2017{\natexlab{b}},
  \apj, 835, 151, \dodoi{10.3847/1538-4357/835/2/151}

\bibitem[{{Aartsen} {et~al.}(2018){Aartsen}, {Ackermann}, {Adams}, {Aguilar},
  {Ahlers}, {Ahrens}, {Altmann}, {Andeen}, {Anderson}, {Ansseau}, {Anton},
  {Archinger}, {Arg{\"u}elles}, {Auffenberg}, {Axani}, {Bai}, {Barwick},
  {Baum}, {Bay}, {Beatty}, {Becker Tjus}, {Becker}, {BenZvi}, {Berley},
  {Bernardini}, {Bernhard}, {Besson}, {Binder}, {Bindig}, {Bissok}, {Blaufuss},
  {Blot}, {Bohm}, {B{\"o}rner}, {Bos}, {Bose}, {B{\"o}ser}, {Botner}, {Braun},
  {Brayeur}, {Bretz}, {Bron}, {Burgman}, {Carver}, {Casier}, {Cheung},
  {Chirkin}, {Christov}, {Clark}, {Classen}, {Coenders}, {Collin}, {Conrad},
  {Cowen}, {Cross}, {Day}, {de Andr{\'e}}, {De Clercq}, {del Pino Rosendo},
  {Dembinski}, {De Ridder}, {Desiati}, {de Vries}, {de Wasseige}, {de With},
  {DeYoung}, {D{\'\i}az-V{\'e}lez}, {di Lorenzo}, {Dujmovic}, {Dumm},
  {Dunkman}, {Eberhardt}, {Ehrhardt}, {Eichmann}, {Eller}, {Euler}, {Evenson},
  {Fahey}, {Fazely}, {Feintzeig}, {Felde}, {Filimonov}, {Finley}, {Flis},
  {F{\"o}sig}, {Franckowiak}, {Friedman}, {Fuchs}, {Gaisser}, {Gallagher},
  {Gerhardt}, {Ghorbani}, {Giang}, {Gladstone}, {Glauch}, {Gl{\"u}senkamp},
  {Goldschmidt}, {Gonzalez}, {Grant}, {Griffith}, {Haack}, {Hallgren},
  {Halzen}, {Hansen}, {Hansmann}, {Hanson}, {Hebecker}, {Heereman}, {Helbing},
  {Hellauer}, {Hickford}, {Hignight}, {Hill}, {Hoffman}, {Hoffmann},
  {Holzapfel}, {Hoshina}, {Huang}, {Huber}, {Hultqvist}, {In}, {Ishihara},
  {Jacobi}, {Japaridze}, {Jeong}, {Jero}, {Jones}, {Jurkovic}, {Kang},
  {Kappes}, {Karg}, {Karle}, {Katz}, {Kauer}, {Keivani}, {Kelley},
  {Kheirandish}, {Kim}, {Kim}, {Kintscher}, {Kiryluk}, {Kittler}, {Klein},
  {Kohnen}, {Koirala}, {Kolanoski}, {Konietz}, {K{\"o}pke}, {Kopper}, {Kopper},
  {Koskinen}, {Kowalski}, {Krings}, {Kroll}, {Kr{\"u}ckl}, {Kr{\"u}ger},
  {Kunnen}, {Kunwar}, {Kurahashi}, {Kuwabara}, {Labare}, {Lanfranchi},
  {Larson}, {Lauber}, {Lennarz}, {Lesiak-Bzdak}, {Leuermann}, {Lu},
  {L{\"u}nemann}, {Madsen}, {Maggi}, {Mahn}, {Mancina}, {Mandelartz},
  {Maruyama}, {Mase}, {Maunu}, {McNally}, {Meagher}, {Medici}, {Meier}, {Meli},
  {Menne}, {Merino}, {Meures}, {Miarecki}, {Montaruli}, {Moulai}, {Nahnhauer},
  {Naumann}, {Neer}, {Niederhausen}, {Nowicki}, {Nygren}, {Obertacke Pollmann},
  {Olivas}, {O'Murchadha}, {Palczewski}, {Pandya}, {Pankova}, {Peiffer},
  {Penek}, {Pepper}, {P{\'e}rez de los Heros}, {Pieloth}, {Pinat}, {Price},
  {Przybylski}, {Quinnan}, {Raab}, {R{\"a}del}, {Rameez}, {Rawlins}, {Reimann},
  {Relethford}, {Relich}, {Resconi}, {Rhode}, {Richman}, {Riedel}, {Robertson},
  {Rongen}, {Rott}, {Ruhe}, {Ryckbosch}, {Rysewyk}, {Sabbatini}, {Sanchez
  Herrera}, {Sandrock}, {Sandroos}, {Sarkar}, {Satalecka}, {Schlunder},
  {Schmidt}, {Schoenen}, {Sch{\"o}neberg}, {Schumacher}, {Seckel}, {Seunarine},
  {Soldin}, {Song}, {Spiczak}, {Spiering}, {Stanev}, {Stasik}, {Stettner},
  {Steuer}, {Stezelberger}, {Stokstad}, {St{\"o}{\ss}l}, {Str{\"o}m},
  {Strotjohann}, {Sullivan}, {Sutherland}, {Taavola}, {Taboada}, {Tatar},
  {Tenholt}, {Ter-Antonyan}, {Terliuk}, {Te{\v{s}}i{\'c}}, {Tilav}, {Toale},
  {Tobin}, {Toscano}, {Tosi}, {Tselengidou}, {Turcati}, {Unger}, {Usner},
  {Vandenbroucke}, {van Eijndhoven}, {Vanheule}, {van Rossem}, {van Santen},
  {Veenkamp}, {Vehring}, {Voge}, {Vogel}, {Vraeghe}, {Walck}, {Wallace},
  {Wallraff}, {Wandkowsky}, {Weaver}, {Weiss}, {Wendt}, {Westerhoff}, {Whelan},
  {Wickmann}, {Wiebe}, {Wiebusch}, {Wille}, {Williams}, {Wills}, {Wolf},
  {Wood}, {Woolsey}, {Woschnagg}, {Xu}, {Xu}, {Xu}, {Yanez}, {Yodh}, {Yoshida},
  \& {Zoll}}]{2018AdSpR..62.2902A}
{Aartsen}, M.~G., {Ackermann}, M., {Adams}, J., {et~al.} 2018, Advances in
  Space Research, 62, 2902, \dodoi{10.1016/j.asr.2017.05.030}

\bibitem[{{Aartsen} {et~al.}(2020){Aartsen}, {Ackermann}, {Adams}, {Aguilar},
  {Ahlers}, {Ahrens}, {Alispach}, {Andeen}, {Anderson}, {Ansseau}, {Anton},
  {Arg{\"u}elles}, {Auffenberg}, {Axani}, {Backes}, {Bagherpour}, {Bai},
  {Balagopal}, {Barbano}, {Barwick}, {Bastian}, {Baum}, {Baur}, {Bay},
  {Beatty}, {Becker}, {Becker Tjus}, {BenZvi}, {Berley}, {Bernardini},
  {Besson}, {Binder}, {Bindig}, {Blaufuss}, {Blot}, {Bohm}, {B{\"o}rner},
  {B{\"o}ser}, {Botner}, {B{\"o}ttcher}, {Bourbeau}, {Bourbeau}, {Bradascio},
  {Braun}, {Bron}, {Brostean-Kaiser}, {Burgman}, {Buscher}, {Busse}, {Carver},
  {Chen}, {Cheung}, {Chirkin}, {Choi}, {Clark}, {Classen}, {Coleman}, {Collin},
  {Conrad}, {Coppin}, {Correa}, {Cowen}, {Cross}, {Dave}, {De Clercq},
  {DeLaunay}, {Dembinski}, {Deoskar}, {De Ridder}, {Desiati}, {de Vries}, {de
  Wasseige}, {de With}, {DeYoung}, {Diaz}, {D{\'\i}az-V{\'e}lez}, {Dujmovic},
  {Dunkman}, {Dvorak}, {Eberhardt}, {Ehrhardt}, {Eller}, {Engel}, {Evenson},
  {Fahey}, {Fazely}, {Felde}, {Filimonov}, {Finley}, {Fox}, {Franckowiak},
  {Friedman}, {Fritz}, {Gaisser}, {Gallagher}, {Ganster}, {Garrappa},
  {Gerhardt}, {Ghorbani}, {Glauch}, {Gl{\"u}senkamp}, {Goldschmidt},
  {Gonzalez}, {Grant}, {Griffith}, {Griswold}, {G{\"u}nder}, {G{\"u}nd{\"u}z},
  {Haack}, {Hallgren}, {Halliday}, {Halve}, {Halzen}, {Hanson}, {Haungs},
  {Hebecker}, {Heereman}, {Heix}, {Helbing}, {Hellauer}, {Henningsen},
  {Hickford}, {Hignight}, {Hill}, {Hoffman}, {Hoffmann}, {Hoinka},
  {Hokanson-Fasig}, {Hoshina}, {Huang}, {Huber}, {Huber}, {Hultqvist},
  {H{\"u}nnefeld}, {Hussain}, {In}, {Iovine}, {Ishihara}, {Japaridze}, {Jeong},
  {Jero}, {Jones}, {Jonske}, {Joppe}, {Kang}, {Kang}, {Kappes}, {Kappesser},
  {Karg}, {Karl}, {Karle}, {Katz}, {Kauer}, {Kelley}, {Kheirandish}, {Kim},
  {Kintscher}, {Kiryluk}, {Kittler}, {Klein}, {Koirala}, {Kolanoski},
  {K{\"o}pke}, {Kopper}, {Kopper}, {Koskinen}, {Kowalski}, {Krings},
  {Kr{\"u}ckl}, {Kulacz}, {Kurahashi}, {Kyriacou}, {Labare}, {Lanfranchi},
  {Larson}, {Lauber}, {Lazar}, {Leonard}, {Leszczy{\'n}ska}, {Leuermann},
  {Liu}, {Lohfink}, {Lozano Mariscal}, {Lu}, {Lucarelli}, {L{\"u}nemann},
  {Luszczak}, {Lyu}, {Ma}, {Madsen}, {Maggi}, {Mahn}, {Makino}, {Mallik},
  {Mallot}, {Mancina}, {Mari{\c{s}}}, {Maruyama}, {Mase}, {Matis}, {Maunu},
  {McNally}, {Meagher}, {Medici}, {Medina}, {Meier}, {Meighen-Berger}, {Menne},
  {Merino}, {Meures}, {Micallef}, {Mockler}, {Moment{\'e}}, {Montaruli},
  {Moore}, {Morse}, {Moulai}, {Muth}, {Nagai}, {Naumann}, {Neer},
  {Niederhausen}, {Nisa}, {Nowicki}, {Nygren}, {Obertacke Pollmann}, {Oehler},
  {Olivas}, {O'Murchadha}, {O'Sullivan}, {Palczewski}, {Pandya}, {Pankova},
  {Park}, {Peiffer}, {P{\'e}rez de los Heros}, {Philippen}, {Pieloth}, {Pinat},
  {Pizzuto}, {Plum}, {Porcelli}, {Price}, {Przybylski}, {Raab}, {Raissi},
  {Rameez}, {Rauch}, {Rawlins}, {Rea}, {Reimann}, {Relethford}, {Renschler},
  {Renzi}, {Resconi}, {Rhode}, {Richman}, {Robertson}, {Rongen}, {Rott},
  {Ruhe}, {Ryckbosch}, {Rysewyk}, {Safa}, {Sanchez Herrera}, {Sandrock},
  {Sandroos}, {Santander}, {Sarkar}, {Sarkar}, {Satalecka}, {Schaufel},
  {Schieler}, {Schlunder}, {Schmidt}, {Schneider}, {Schneider}, {Schr{\"o}der},
  {Schumacher}, {Sclafani}, {Seckel}, {Seunarine}, {Shefali}, {Silva},
  {Snihur}, {Soedingrekso}, {Soldin}, {Song}, {Spiczak}, {Spiering},
  {Stachurska}, {Stamatikos}, {Stanev}, {Stein}, {Steinm{\"u}ller}, {Stettner},
  {Steuer}, {Stezelberger}, {Stokstad}, {St{\"o}{\ss}l}, {Strotjohann},
  {St{\"u}rwald}, {Stuttard}, {Sullivan}, {Taboada}, {Tenholt}, {Ter-Antonyan},
  {Terliuk}, {Tilav}, {Tollefson}, {Tomankova}, {T{\"o}nnis}, {Toscano},
  {Tosi}, {Trettin}, {Tselengidou}, {Tung}, {Turcati}, {Turcotte}, {Turley},
  {Ty}, {Unger}, {Unland Elorrieta}, {Usner}, {Vandenbroucke}, {Van Driessche},
  {van Eijk}, {van Eijndhoven}, {Vanheule}, {van Santen}, {Vraeghe}, {Walck},
  {Wallace}, {Wallraff}, {Wandkowsky}, {Watson}, {Weaver}, {Weindl}, {Weiss},
  {Weldert}, {Wendt}, {Werthebach}, {Whelan}, {Whitehorn}, {Wiebe}, {Wiebusch},
  {Wille}, {Williams}, {Wills}, {Wolf}, {Wood}, {Wood}, {Woschnagg}, {Wrede},
  {Xu}, {Xu}, {Xu}, {Yanez}, {Yodh}, {Yoshida}, {Yuan}, \&
  {Z{\"o}cklein}}]{2020PhRvL.124e1103A}
---. 2020, \prl, 124, 051103, \dodoi{10.1103/PhysRevLett.124.051103}

\bibitem[{{Abdo} {et~al.}(2010){Abdo}, {Ackermann}, {Agudo}, {Ajello}, {Aller},
  {Aller}, {Angelakis}, {Arkharov}, {Axelsson}, {Bach}, {Baldini}, {Ballet},
  {Barbiellini}, {Bastieri}, {Baughman}, {Bechtol}, {Bellazzini}, {Benitez},
  {Berdyugin}, {Berenji}, {Blandford}, {Bloom}, {Boettcher}, {Bonamente},
  {Borgland}, {Bregeon}, {Brez}, {Brigida}, {Bruel}, {Burnett}, {Burrows},
  {Buson}, {Caliandro}, {Calzoletti}, {Cameron}, {Capalbi}, {Caraveo},
  {Carosati}, {Casandjian}, {Cavazzuti}, {Cecchi}, {{\c{C}}elik}, {Charles},
  {Chaty}, {Chekhtman}, {Chen}, {Chiang}, {Chincarini}, {Ciprini}, {Claus},
  {Cohen-Tanugi}, {Colafrancesco}, {Cominsky}, {Conrad}, {Costamante},
  {Cutini}, {D'ammando}, {Deitrick}, {D'Elia}, {Dermer}, {de Angelis}, {de
  Palma}, {Digel}, {Donnarumma}, {Silva}, {Drell}, {Dubois}, {Dultzin},
  {Dumora}, {Falcone}, {Farnier}, {Favuzzi}, {Fegan}, {Focke}, {Forn{\'e}},
  {Fortin}, {Frailis}, {Fuhrmann}, {Fukazawa}, {Funk}, {Fusco}, {G{\'o}mez},
  {Gargano}, {Gasparrini}, {Gehrels}, {Germani}, {Giebels}, {Giglietto},
  {Giommi}, {Giordano}, {Giuliani}, {Glanzman}, {Godfrey}, {Grenier},
  {Gronwall}, {Grove}, {Guillemot}, {Guiriec}, {Gurwell}, {Hadasch},
  {Hanabata}, {Harding}, {Hayashida}, {Hays}, {Healey}, {Heidt}, {Hiriart},
  {Horan}, {Hoversten}, {Hughes}, {Itoh}, {Jackson}, {J{\'o}hannesson},
  {Johnson}, {Johnson}, {Jorstad}, {Kadler}, {Kamae}, {Katagiri}, {Kataoka},
  {Kawai}, {Kennea}, {Kerr}, {Kimeridze}, {Kn{\"o}dlseder}, {Kocian},
  {Kopatskaya}, {Koptelova}, {Konstantinova}, {Kovalev}, {Kovalev},
  {Kurtanidze}, {Kuss}, {Lande}, {Larionov}, {Latronico}, {Leto}, {Lindfors},
  {Longo}, {Loparco}, {Lott}, {Lovellette}, {Lubrano}, {Madejski}, {Makeev},
  {Marchegiani}, {Marscher}, {Marshall}, {Max-Moerbeck}, {Mazziotta},
  {McConville}, {McEnery}, {Meurer}, {Michelson}, {Mitthumsiri}, {Mizuno},
  {Moiseev}, {Monte}, {Monzani}, {Morselli}, {Moskalenko}, {Murgia},
  {Nestoras}, {Nilsson}, {Nizhelsky}, {Nolan}, {Norris}, {Nuss}, {Ohsugi},
  {Ojha}, {Omodei}, {Orlando}, {Ormes}, {Osborne}, {Ozaki}, {Pacciani},
  {Padovani}, {Pagani}, {Page}, {Paneque}, {Panetta}, {Parent}, {Pasanen},
  {Pavlidou}, {Pelassa}, {Pepe}, {Perri}, {Pesce-Rollins}, {Piranomonte},
  {Piron}, {Pittori}, {Porter}, {Puccetti}, {Rahoui}, {Rain{\`o}}, {Raiteri},
  {Rando}, {Razzano}, {Reimer}, {Reimer}, {Reposeur}, {Richards}, {Ritz},
  {Rochester}, {Rodriguez}, {Romani}, {Ros}, {Roth}, {Roustazadeh}, {Ryde},
  {Sadrozinski}, {Sadun}, {Sanchez}, {Sander}, {Saz Parkinson}, {Scargle},
  {Sellerholm}, {Sgr{\`o}}, {Shaw}, {Sigua}, {Siskind}, {Smith}, {Smith},
  {Spandre}, {Spinelli}, {Starck}, {Stevenson}, {Stratta}, {Strickman},
  {Suson}, {Tajima}, {Takahashi}, {Takahashi}, {Takalo}, {Tanaka}, {Thayer},
  {Thayer}, {Thompson}, {Tibaldo}, {Torres}, {Tosti}, {Tramacere}, {Uchiyama},
  {Usher}, {Vasileiou}, {Verrecchia}, {Vilchez}, {Villata}, {Vitale}, {Waite},
  {Wang}, {Winer}, {Wood}, {Ylinen}, {Zensus}, {Zhekanis}, \&
  {Ziegler}}]{2010ApJ...716...30A}
{Abdo}, A.~A., {Ackermann}, M., {Agudo}, I., {et~al.} 2010, \apj, 716, 30,
  \dodoi{10.1088/0004-637X/716/1/30}

\bibitem[{{Abdollahi} {et~al.}(2020){Abdollahi}, {Acero}, {Ackermann},
  {Ajello}, {Atwood}, {Axelsson}, {Baldini}, {Ballet}, {Barbiellini},
  {Bastieri}, {Becerra Gonzalez}, {Bellazzini}, {Berretta}, {Bissaldi}, {Bland
  ford}, {Bloom}, {Bonino}, {Bottacini}, {Brandt}, {Bregeon}, {Bruel},
  {Buehler}, {Burnett}, {Buson}, {Cameron}, {Caputo}, {Caraveo}, {Casandjian},
  {Castro}, {Cavazzuti}, {Charles}, {Chaty}, {Chen}, {Cheung}, {Chiaro},
  {Ciprini}, {Cohen-Tanugi}, {Cominsky}, {Coronado-Bl{\'a}zquez}, {Costantin},
  {Cuoco}, {Cutini}, {D'Ammando}, {DeKlotz}, {Torre Luque}, {de Palma},
  {Desai}, {Digel}, {Lalla}, {Mauro}, {Venere}, {Dom{\'\i}nguez}, {Dumora},
  {Dirirsa}, {Fegan}, {Ferrara}, {Franckowiak}, {Fukazawa}, {Funk}, {Fusco},
  {Gargano}, {Gasparrini}, {Giglietto}, {Giommi}, {Giordano}, {Giroletti},
  {Glanzman}, {Green}, {Grenier}, {Griffin}, {Grondin}, {Grove}, {Guiriec},
  {Harding}, {Hayashi}, {Hays}, {Hewitt}, {Horan}, {J{\'o}hannesson},
  {Johnson}, {Kamae}, {Kerr}, {Kocevski}, {Kovac'evic'}, {Kuss}, {Landriu},
  {Larsson}, {Latronico}, {Lemoine-Goumard}, {Li}, {Liodakis}, {Longo},
  {Loparco}, {Lott}, {Lovellette}, {Lubrano}, {Madejski}, {Maldera},
  {Malyshev}, {Manfreda}, {Marchesini}, {Marcotulli}, {Mart{\'\i}-Devesa},
  {Martin}, {Massaro}, {Mazziotta}, {McEnery}, {Mereu}, {Meyer}, {Michelson},
  {Mirabal}, {Mizuno}, {Monzani}, {Morselli}, {Moskalenko}, {Negro}, {Nuss},
  {Ojha}, {Omodei}, {Orienti}, {Orlando}, {Ormes}, {Palatiello}, {Paliya},
  {Paneque}, {Pei}, {Pe{\~n}a-Herazo}, {Perkins}, {Persic}, {Pesce-Rollins},
  {Petrosian}, {Petrov}, {Piron}, {Poon}, {Porter}, {Principe}, {Rain{\`o}},
  {Rando}, {Razzano}, {Razzaque}, {Reimer}, {Reimer}, {Remy}, {Reposeur},
  {Romani}, {Parkinson}, {Schinzel}, {Serini}, {Sgr{\`o}}, {Siskind}, {Smith},
  {Spandre}, {Spinelli}, {Strong}, {Suson}, {Tajima}, {Takahashi}, {Tak},
  {Thayer}, {Thompson}, {Tibaldo}, {Torres}, {Torresi}, {Valverde}, {Klaveren},
  {Zyl}, {Wood}, {Yassine}, \& {Zaharijas}}]{2020ApJS..247...33A}
{Abdollahi}, S., {Acero}, F., {Ackermann}, M., {et~al.} 2020, \apjs, 247, 33,
  \dodoi{10.3847/1538-4365/ab6bcb}

\bibitem[{{Adelman-McCarthy} {et~al.}(2007){Adelman-McCarthy}, {Ag{\"u}eros},
  {Allam}, {Anderson}, {Anderson}, {Annis}, {Bahcall}, {Bailer-Jones},
  {Baldry}, {Barentine}, {Beers}, {Belokurov}, {Berlind}, {Bernardi},
  {Blanton}, {Bochanski}, {Boroski}, {Bramich}, {Brewington}, {Brinchmann},
  {Brinkmann}, {Brunner}, {Budav{\'a}ri}, {Carey}, {Carliles}, {Carr},
  {Castander}, {Connolly}, {Cool}, {Cunha}, {Csabai}, {Dalcanton}, {Doi},
  {Eisenstein}, {Evans}, {Evans}, {Fan}, {Finkbeiner}, {Friedman}, {Frieman},
  {Fukugita}, {Gillespie}, {Gilmore}, {Glazebrook}, {Gray}, {Grebel}, {Gunn},
  {de Haas}, {Hall}, {Harvanek}, {Hawley}, {Hayes}, {Heckman}, {Hendry},
  {Hennessy}, {Hindsley}, {Hirata}, {Hogan}, {Hogg}, {Holtzman}, {Ichikawa},
  {Ichikawa}, {Ivezi{\'c}}, {Jester}, {Johnston}, {Jorgensen}, {Juri{\'c}},
  {Kauffmann}, {Kent}, {Kleinman}, {Knapp}, {Kniazev}, {Kron}, {Krzesinski},
  {Kuropatkin}, {Lamb}, {Lampeitl}, {Lee}, {Leger}, {Lima}, {Lin}, {Long},
  {Loveday}, {Lupton}, {Mandelbaum}, {Margon}, {Mart{\'\i}nez-Delgado},
  {Matsubara}, {McGehee}, {McKay}, {Meiksin}, {Munn}, {Nakajima}, {Nash},
  {Neilsen}, {Newberg}, {Nichol}, {Nieto-Santisteban}, {Nitta}, {Oyaizu},
  {Okamura}, {Ostriker}, {Padmanabhan}, {Park}, {Peoples}, {Pier}, {Pope},
  {Pourbaix}, {Quinn}, {Raddick}, {Re Fiorentin}, {Richards}, {Richmond},
  {Rix}, {Rockosi}, {Schlegel}, {Schneider}, {Scranton}, {Seljak}, {Sheldon},
  {Shimasaku}, {Silvestri}, {Smith}, {Smol{\v{c}}i{\'c}}, {Snedden},
  {Stebbins}, {Stoughton}, {Strauss}, {SubbaRao}, {Suto}, {Szalay}, {Szapudi},
  {Szkody}, {Tegmark}, {Thakar}, {Tremonti}, {Tucker}, {Uomoto}, {Vanden Berk},
  {Vandenberg}, {Vidrih}, {Vogeley}, {Voges}, {Vogt}, {Weinberg}, {West},
  {White}, {Wilhite}, {Yanny}, {Yocum}, {York}, {Zehavi}, {Zibetti}, \&
  {Zucker}}]{2007ApJS..172..634A}
{Adelman-McCarthy}, J.~K., {Ag{\"u}eros}, M.~A., {Allam}, S.~S., {et~al.} 2007,
  \apjs, 172, 634, \dodoi{10.1086/518864}

\bibitem[{{Ahlers} \& {Halzen}(2015)}]{2015RPPh...78l6901A}
{Ahlers}, M., \& {Halzen}, F. 2015, Reports on Progress in Physics, 78, 126901,
  \dodoi{10.1088/0034-4885/78/12/126901}

\bibitem[{{Ajello} {et~al.}(2020){Ajello}, {Angioni}, {Axelsson}, {Ballet},
  {Barbiellini}, {Bastieri}, {Becerra Gonzalez}, {Bellazzini}, {Bissaldi},
  {Bloom}, {Bonino}, {Bottacini}, {Bruel}, {Buson}, {Cafardo}, {Cameron},
  {Cavazzuti}, {Chen}, {Cheung}, {Ciprini}, {Costantin}, {Cutini}, {D'Ammando},
  {de la Torre Luque}, {de Menezes}, {de Palma}, {Desai}, {Di Lalla}, {Di
  Venere}, {Dom{\'\i}nguez}, {Dirirsa}, {Ferrara}, {Finke}, {Franckowiak},
  {Fukazawa}, {Funk}, {Fusco}, {Gargano}, {Garrappa}, {Gasparrini},
  {Giglietto}, {Giordano}, {Giroletti}, {Green}, {Grenier}, {Guiriec},
  {Harita}, {Hays}, {Horan}, {Itoh}, {J{\'o}hannesson}, {Kovac'evic'},
  {Krauss}, {Kreter}, {Kuss}, {Larsson}, {Leto}, {Li}, {Liodakis}, {Longo},
  {Loparco}, {Lott}, {Lovellette}, {Lubrano}, {Madejski}, {Maldera},
  {Manfreda}, {Mart{\'\i}-Devesa}, {Massaro}, {Mazziotta}, {Mereu}, {Meyer},
  {Migliori}, {Mirabal}, {Mizuno}, {Monzani}, {Morselli}, {Moskalenko},
  {Negro}, {Nemmen}, {Nuss}, {Ojha}, {Ojha}, {Omodei}, {Orienti}, {Orlando},
  {Ormes}, {Paliya}, {Pei}, {Pe{\~n}a-Herazo}, {Persic}, {Pesce-Rollins},
  {Petrov}, {Piron}, {Poon}, {Principe}, {Rain{\`o}}, {Rando}, {Rani},
  {Razzano}, {Razzaque}, {Reimer}, {Reimer}, {Schinzel}, {Serini}, {Sgr{\`o}},
  {Siskind}, {Spandre}, {Spinelli}, {Suson}, {Tachibana}, {Thompson}, {Torres},
  {Torresi}, {Troja}, {Valverde}, {van Zyl}, \&
  {Yassine}}]{2020ApJ...892..105A}
{Ajello}, M., {Angioni}, R., {Axelsson}, M., {et~al.} 2020, \apj, 892, 105,
  \dodoi{10.3847/1538-4357/ab791e}

\bibitem[{{Alexander} {et~al.}(2016){Alexander}, {Berger}, {Guillochon},
  {Zauderer}, \& {Williams}}]{2016ApJ...819L..25A}
{Alexander}, K.~D., {Berger}, E., {Guillochon}, J., {Zauderer}, B.~A., \&
  {Williams}, P.~K.~G. 2016, \apjl, 819, L25,
  \dodoi{10.3847/2041-8205/819/2/L25}

\bibitem[{{Alexander} {et~al.}(2020){Alexander}, {van Velzen}, {Horesh}, \&
  {Zauderer}}]{2020SSRv..216...81A}
{Alexander}, K.~D., {van Velzen}, S., {Horesh}, A., \& {Zauderer}, B.~A. 2020,
  \ssr, 216, 81, \dodoi{10.1007/s11214-020-00702-w}

\bibitem[{{Ansoldi} {et~al.}(2018){Ansoldi}, {Antonelli}, {Arcaro}, {Baack},
  {Babi{\'c}}, {Banerjee}, {Bangale}, {Barres de Almeida}, {Barrio}, {Becerra
  Gonz{\'a}lez}, {Bednarek}, {Bernardini}, {Berse}, {Berti}, {Besenrieder},
  {Bhattacharyya}, {Bigongiari}, {Biland}, {Blanch}, {Bonnoli}, {Carosi},
  {Ceribella}, {Chatterjee}, {Colak}, {Colin}, {Colombo}, {Contreras},
  {Cortina}, {Covino}, {Cumani}, {D'Elia}, {Da Vela}, {Dazzi}, {De Angelis},
  {De Lotto}, {Delfino}, {Delgado}, {Di Pierro}, {Dom{\'\i}nguez}, {Dominis
  Prester}, {Dorner}, {Doro}, {Einecke}, {Elsaesser}, {Fallah Ramazani},
  {Fattorini}, {Fern{\'a}ndez-Barral}, {Ferrara}, {Fidalgo}, {Foffano},
  {Fonseca}, {Font}, {Fruck}, {Gallozzi}, {Garc{\'\i}a L{\'o}pez},
  {Garczarczyk}, {Gaug}, {Giammaria}, {Godinovi{\'c}}, {Guberman}, {Hadasch},
  {Hahn}, {Hassan}, {Hayashida}, {Herrera}, {Hoang}, {Hrupec}, {Inoue},
  {Ishio}, {Iwamura}, {Konno}, {Kubo}, {Kushida}, {Lamastra}, {Lelas}, {Leone},
  {Lindfors}, {Lombardi}, {Longo}, {L{\'o}pez}, {Maggio}, {Majumdar},
  {Makariev}, {Maneva}, {Manganaro}, {Mannheim}, {Maraschi}, {Mariotti},
  {Mart{\'\i}nez}, {Masuda}, {Mazin}, {Mielke}, {Minev}, {Miranda}, {Mirzoyan},
  {Moralejo}, {Moreno}, {Moretti}, {Neustroev}, {Niedzwiecki}, {Nievas
  Rosillo}, {Nigro}, {Nilsson}, {Ninci}, {Nishijima}, {Noda}, {Nogu{\'e}s},
  {Paiano}, {Palacio}, {Paneque}, {Paoletti}, {Paredes}, {Pedaletti},
  {Pe{\~n}il}, {Peresano}, {Persic}, {Pfrang}, {Prada Moroni}, {Prandini},
  {Puljak}, {Garcia}, {Rhode}, {Rib{\'o}}, {Rico}, {Righi}, {Rugliancich},
  {Saha}, {Saito}, {Satalecka}, {Schweizer}, {Sitarek}, {{\v{S}}nidari{\'c}},
  {Sobczynska}, {Stamerra}, {Strzys}, {Suri{\'c}}, {Tavecchio}, {Temnikov},
  {Terzi{\'c}}, {Teshima}, {Torres-Alb{\'a}}, {Tsujimoto}, {Vanzo}, {Vazquez
  Acosta}, {Vovk}, {Ward}, {Will}, {Zari{\'c}}, \&
  {Cerruti}}]{2018ApJ...863L..10A}
{Ansoldi}, S., {Antonelli}, L.~A., {Arcaro}, C., {et~al.} 2018, \apjl, 863,
  L10, \dodoi{10.3847/2041-8213/aad083}

\bibitem[{{Astropy Collaboration} {et~al.}(2018){Astropy Collaboration},
  {Price-Whelan}, {Sip{\H{o}}cz}, {G{\"u}nther}, {Lim}, {Crawford}, {Conseil},
  {Shupe}, {Craig}, {Dencheva}, {Ginsburg}, {VanderPlas}, {Bradley},
  {P{\'e}rez-Su{\'a}rez}, {de Val-Borro}, {Aldcroft}, {Cruz}, {Robitaille},
  {Tollerud}, {Ardelean}, {Babej}, {Bach}, {Bachetti}, {Bakanov}, {Bamford},
  {Barentsen}, {Barmby}, {Baumbach}, {Berry}, {Biscani}, {Boquien}, {Bostroem},
  {Bouma}, {Brammer}, {Bray}, {Breytenbach}, {Buddelmeijer}, {Burke},
  {Calderone}, {Cano Rodr{\'\i}guez}, {Cara}, {Cardoso}, {Cheedella}, {Copin},
  {Corrales}, {Crichton}, {D'Avella}, {Deil}, {Depagne}, {Dietrich}, {Donath},
  {Droettboom}, {Earl}, {Erben}, {Fabbro}, {Ferreira}, {Finethy}, {Fox},
  {Garrison}, {Gibbons}, {Goldstein}, {Gommers}, {Greco}, {Greenfield},
  {Groener}, {Grollier}, {Hagen}, {Hirst}, {Homeier}, {Horton}, {Hosseinzadeh},
  {Hu}, {Hunkeler}, {Ivezi{\'c}}, {Jain}, {Jenness}, {Kanarek}, {Kendrew},
  {Kern}, {Kerzendorf}, {Khvalko}, {King}, {Kirkby}, {Kulkarni}, {Kumar},
  {Lee}, {Lenz}, {Littlefair}, {Ma}, {Macleod}, {Mastropietro}, {McCully},
  {Montagnac}, {Morris}, {Mueller}, {Mumford}, {Muna}, {Murphy}, {Nelson},
  {Nguyen}, {Ninan}, {N{\"o}the}, {Ogaz}, {Oh}, {Parejko}, {Parley}, {Pascual},
  {Patil}, {Patil}, {Plunkett}, {Prochaska}, {Rastogi}, {Reddy Janga},
  {Sabater}, {Sakurikar}, {Seifert}, {Sherbert}, {Sherwood-Taylor}, {Shih},
  {Sick}, {Silbiger}, {Singanamalla}, {Singer}, {Sladen}, {Sooley},
  {Sornarajah}, {Streicher}, {Teuben}, {Thomas}, {Tremblay}, {Turner},
  {Terr{\'o}n}, {van Kerkwijk}, {de la Vega}, {Watkins}, {Weaver}, {Whitmore},
  {Woillez}, {Zabalza}, \& {Astropy Contributors}}]{2018AJ....156..123A}
{Astropy Collaboration}, {Price-Whelan}, A.~M., {Sip{\H{o}}cz}, B.~M., {et~al.}
  2018, \aj, 156, 123, \dodoi{10.3847/1538-3881/aabc4f}

\bibitem[{{Atoyan} \& {Dermer}(2001)}]{2001PhRvL..87v1102A}
{Atoyan}, A., \& {Dermer}, C.~D. 2001, \prl, 87, 221102,
  \dodoi{10.1103/PhysRevLett.87.221102}

\bibitem[{{Becker Tjus} {et~al.}(2014){Becker Tjus}, {Eichmann}, {Halzen},
  {Kheirandish}, \& {Saba}}]{2014PhRvD..89l3005B}
{Becker Tjus}, J., {Eichmann}, B., {Halzen}, F., {Kheirandish}, A., \& {Saba},
  S.~M. 2014, \prd, 89, 123005, \dodoi{10.1103/PhysRevD.89.123005}

\bibitem[{{Bellm} {et~al.}(2019){Bellm}, {Kulkarni}, {Graham}, {Dekany},
  {Smith}, {Riddle}, {Masci}, {Helou}, {Prince}, {Adams}, {Barbarino},
  {Barlow}, {Bauer}, {Beck}, {Belicki}, {Biswas}, {Blagorodnova}, {Bodewits},
  {Bolin}, {Brinnel}, {Brooke}, {Bue}, {Bulla}, {Burruss}, {Cenko}, {Chang},
  {Connolly}, {Coughlin}, {Cromer}, {Cunningham}, {De}, {Delacroix}, {Desai},
  {Duev}, {Eadie}, {Farnham}, {Feeney}, {Feindt}, {Flynn}, {Franckowiak},
  {Frederick}, {Fremling}, {Gal-Yam}, {Gezari}, {Giomi}, {Goldstein},
  {Golkhou}, {Goobar}, {Groom}, {Hacopians}, {Hale}, {Henning}, {Ho}, {Hover},
  {Howell}, {Hung}, {Huppenkothen}, {Imel}, {Ip}, {Ivezi{\'c}}, {Jackson},
  {Jones}, {Juric}, {Kasliwal}, {Kaspi}, {Kaye}, {Kelley}, {Kowalski},
  {Kramer}, {Kupfer}, {Landry}, {Laher}, {Lee}, {Lin}, {Lin}, {Lunnan},
  {Giomi}, {Mahabal}, {Mao}, {Miller}, {Monkewitz}, {Murphy}, {Ngeow},
  {Nordin}, {Nugent}, {Ofek}, {Patterson}, {Penprase}, {Porter}, {Rauch},
  {Rebbapragada}, {Reiley}, {Rigault}, {Rodriguez}, {van Roestel}, {Rusholme},
  {van Santen}, {Schulze}, {Shupe}, {Singer}, {Soumagnac}, {Stein}, {Surace},
  {Sollerman}, {Szkody}, {Taddia}, {Terek}, {Van Sistine}, {van Velzen},
  {Vestrand}, {Walters}, {Ward}, {Ye}, {Yu}, {Yan}, \&
  {Zolkower}}]{2019PASP..131a8002B}
{Bellm}, E.~C., {Kulkarni}, S.~R., {Graham}, M.~J., {et~al.} 2019, \pasp, 131,
  018002, \dodoi{10.1088/1538-3873/aaecbe}

\bibitem[{{Bennett} {et~al.}(1986){Bennett}, {Lawrence}, {Burke}, {Hewitt}, \&
  {Mahoney}}]{1986ApJS...61....1B}
{Bennett}, C.~L., {Lawrence}, C.~R., {Burke}, B.~F., {Hewitt}, J.~N., \&
  {Mahoney}, J. 1986, \apjs, 61, 1, \dodoi{10.1086/191108}

\bibitem[{{Biteau} {et~al.}(2020){Biteau}, {Prandini}, {Costamante}, {Lemoine},
  {Padovani}, {Pueschel}, {Resconi}, {Tavecchio}, {Taylor}, \&
  {Zech}}]{2020NatAs...4..124B}
{Biteau}, J., {Prandini}, E., {Costamante}, L., {et~al.} 2020, Nature
  Astronomy, 4, 124, \dodoi{10.1038/s41550-019-0988-4}

\bibitem[{{Blandford} {et~al.}(2019){Blandford}, {Meier}, \&
  {Readhead}}]{2019ARA&A..57..467B}
{Blandford}, R., {Meier}, D., \& {Readhead}, A. 2019, \araa, 57, 467,
  \dodoi{10.1146/annurev-astro-081817-051948}

\bibitem[{{Blandford} \& {Rees}(1978)}]{1978bllo.conf..328B}
{Blandford}, R.~D., \& {Rees}, M.~J. 1978, in BL Lac Objects, ed. A.~M.
  {Wolfe}, 328--341

\bibitem[{{B{\l}a{\.z}ejowski} {et~al.}(2000){B{\l}a{\.z}ejowski}, {Sikora},
  {Moderski}, \& {Madejski}}]{2000ApJ...545..107B}
{B{\l}a{\.z}ejowski}, M., {Sikora}, M., {Moderski}, R., \& {Madejski}, G.~M.
  2000, \apj, 545, 107, \dodoi{10.1086/317791}

\bibitem[{{Bloom} {et~al.}(2011){Bloom}, {Giannios}, {Metzger}, {Cenko},
  {Perley}, {Butler}, {Tanvir}, {Levan}, {O'Brien}, {Strubbe}, {De Colle},
  {Ramirez-Ruiz}, {Lee}, {Nayakshin}, {Quataert}, {King}, {Cucchiara},
  {Guillochon}, {Bower}, {Fruchter}, {Morgan}, \& {van der
  Horst}}]{2011Sci...333..203B}
{Bloom}, J.~S., {Giannios}, D., {Metzger}, B.~D., {et~al.} 2011, Science, 333,
  203, \dodoi{10.1126/science.1207150}

\bibitem[{{B{\"o}ttcher} {et~al.}(2013){B{\"o}ttcher}, {Reimer}, {Sweeney}, \&
  {Prakash}}]{2013ApJ...768...54B}
{B{\"o}ttcher}, M., {Reimer}, A., {Sweeney}, K., \& {Prakash}, A. 2013, \apj,
  768, 54, \dodoi{10.1088/0004-637X/768/1/54}

\bibitem[{{Burrows} {et~al.}(2011){Burrows}, {Kennea}, {Ghisellini}, {Mangano},
  {Zhang}, {Page}, {Eracleous}, {Romano}, {Sakamoto}, {Falcone}, {Osborne},
  {Campana}, {Beardmore}, {Breeveld}, {Chester}, {Corbet}, {Covino},
  {Cummings}, {D'Avanzo}, {D'Elia}, {Esposito}, {Evans}, {Fugazza}, {Gelbord},
  {Hiroi}, {Holland}, {Huang}, {Im}, {Israel}, {Jeon}, {Jeon}, {Jun}, {Kawai},
  {Kim}, {Krimm}, {Marshall}, {P. M{\'e}sz{\'a}ros}, {Negoro}, {Omodei},
  {Park}, {Perkins}, {Sugizaki}, {Sung}, {Tagliaferri}, {Troja}, {Ueda},
  {Urata}, {Usui}, {Antonelli}, {Barthelmy}, {Cusumano}, {Giommi}, {Melandri},
  {Perri}, {Racusin}, {Sbarufatti}, {Siegel}, \&
  {Gehrels}}]{2011Natur.476..421B}
{Burrows}, D.~N., {Kennea}, J.~A., {Ghisellini}, G., {et~al.} 2011, \nat, 476,
  421, \dodoi{10.1038/nature10374}

\bibitem[{{Cardelli} {et~al.}(1989){Cardelli}, {Clayton}, \&
  {Mathis}}]{1989ApJ...345..245C}
{Cardelli}, J.~A., {Clayton}, G.~C., \& {Mathis}, J.~S. 1989, \apj, 345, 245,
  \dodoi{10.1086/167900}

\bibitem[{{Cash}(1979)}]{1979ApJ...228..939C}
{Cash}, W. 1979, \apj, 228, 939, \dodoi{10.1086/156922}

\bibitem[{{Condon} {et~al.}(1998){Condon}, {Cotton}, {Greisen}, {Yin},
  {Perley}, {Taylor}, \& {Broderick}}]{1998AJ....115.1693C}
{Condon}, J.~J., {Cotton}, W.~D., {Greisen}, E.~W., {et~al.} 1998, \aj, 115,
  1693, \dodoi{10.1086/300337}

\bibitem[{{Dermer} \& {Schlickeiser}(1993)}]{1993ApJ...416..458D}
{Dermer}, C.~D., \& {Schlickeiser}, R. 1993, \apj, 416, 458,
  \dodoi{10.1086/173251}

\bibitem[{{Dondi} \& {Ghisellini}(1995)}]{1995MNRAS.273..583D}
{Dondi}, L., \& {Ghisellini}, G. 1995, \mnras, 273, 583,
  \dodoi{10.1093/mnras/273.3.583}

\bibitem[{{Finke}(2013)}]{2013ApJ...763..134F}
{Finke}, J.~D. 2013, \apj, 763, 134, \dodoi{10.1088/0004-637X/763/2/134}

\bibitem[{{Franckowiak} {et~al.}(2020){Franckowiak}, {Garrappa}, {Paliya},
  {Shappee}, {Stein}, {Strotjohann}, {Kowalski}, {Buson}, {Kiehlmann},
  {Max-Moerbeck}, \& {Angioni}}]{2020ApJ...893..162F}
{Franckowiak}, A., {Garrappa}, S., {Paliya}, V., {et~al.} 2020, \apj, 893, 162,
  \dodoi{10.3847/1538-4357/ab8307}

\bibitem[{{Garrappa} {et~al.}(2019{\natexlab{a}}){Garrappa}, {Buson}, \&
  {Fermi-LAT Collaboration}}]{2019GCN.25932....1G}
{Garrappa}, S., {Buson}, S., \& {Fermi-LAT Collaboration}. 2019{\natexlab{a}},
  GRB Coordinates Network, 25932, 1

\bibitem[{{Garrappa} {et~al.}(2019{\natexlab{b}}){Garrappa}, {Buson},
  {Franckowiak}, {Fermi-LAT Collaboration}, {Shappee}, {Beacom}, {Dong},
  {Holoien}, {Kochanek}, {Prieto}, {Stanek}, {Thompson}, {ASAS-SN
  Collaboration}, {Aartsen}, {Ackermann}, {Adams}, {Aguilar}, {Ahlers},
  {Ahrens}, {Alispach}, {Andeen}, {Anderson}, {Ansseau}, {Anton},
  {Arg{\"u}elles}, {Auffenberg}, {Axani}, {Backes}, {Bagherpour}, {Bai},
  {Barbano}, {Barwick}, {Baum}, {Bay}, {Beatty}, {Becker}, {Becker Tjus},
  {BenZvi}, {Berley}, {Bernardini}, {Besson}, {Binder}, {Bindig}, {Blaufuss},
  {Blot}, {Bohm}, {B{\"o}rner}, {B{\"o}ser}, {Botner}, {Bourbeau}, {Bourbeau},
  {Bradascio}, {Braun}, {Bretz}, {Bron}, {Brostean-Kaiser}, {Burgman}, {Busse},
  {Carver}, {Chen}, {Cheung}, {Chirkin}, {Clark}, {Classen}, {Collin},
  {Conrad}, {Coppin}, {Correa}, {Cowen}, {Cross}, {Dave}, {de Andr{\'e}}, {De
  Clercq}, {DeLaunay}, {Dembinski}, {Deoskar}, {De Ridder}, {Desiati}, {de
  Vries}, {de Wasseige}, {de With}, {DeYoung}, {Diaz}, {D{\'\i}az-V{\'e}lez},
  {Dujmovic}, {Dunkman}, {Dvorak}, {Eberhardt}, {Ehrhardt}, {Eller}, {Evenson},
  {Fahey}, {Fazely}, {Felde}, {Filimonov}, {Finley}, {Franckowiak}, {Friedman},
  {Fritz}, {Gaisser}, {Gallagher}, {Ganster}, {Garrappa}, {Gerhardt},
  {Ghorbani}, {Glauch}, {Gl{\"u}senkamp}, {Goldschmidt}, {Gonzalez}, {Grant},
  {Griffith}, {G{\"u}nder}, {G{\"u}nd{\"u}z}, {Haack}, {Hallgren}, {Halve},
  {Halzen}, {Hanson}, {Hebecker}, {Heereman}, {Helbing}, {Hellauer},
  {Henningsen}, {Hickford}, {Hignight}, {Hill}, {Hoffman}, {Hoffmann},
  {Hoinka}, {Hokanson-Fasig}, {Hoshina}, {Huang}, {Huber}, {Hultqvist},
  {H{\"u}nnefeld}, {Hussain}, {In}, {Iovine}, {Ishihara}, {Jacobi},
  {Japaridze}, {Jeong}, {Jero}, {Jones}, {Kang}, {Kappes}, {Kappesser}, {Karg},
  {Karl}, {Karle}, {Katz}, {Kauer}, {Keivani}, {Kelley}, {Kheirandish}, {Kim},
  {Kintscher}, {Kiryluk}, {Kittler}, {Klein}, {Koirala}, {Kolanoski},
  {K{\"o}pke}, {Kopper}, {Kopper}, {Koskinen}, {Kowalski}, {Krings},
  {Kr{\"u}ckl}, {Kulacz}, {Kunwar}, {Kurahashi}, {Kyriacou}, {Labare},
  {Lanfranchi}, {Larson}, {Lauber}, {Lazar}, {Leonard}, {Leuermann}, {Liu},
  {Lohfink}, {Lozano Mariscal}, {Lu}, {Lucarelli}, {L{\"u}nemann}, {Luszczak},
  {Madsen}, {Maggi}, {Mahn}, {Makino}, {Mallot}, {Mancina}, {Mari{\c{s}}},
  {Maruyama}, {Mase}, {Maunu}, {Meagher}, {Medici}, {Medina}, {Meier},
  {Meighen-Berger}, {Menne}, {Merino}, {Meures}, {Miarecki}, {Micallef},
  {Moment{\'e}}, {Montaruli}, {Moore}, {Moulai}, {Nagai}, {Nahnhauer},
  {Nakarmi}, {Naumann}, {Neer}, {Niederhausen}, {Nowicki}, {Nygren}, {Obertacke
  Pollmann}, {Olivas}, {O'Murchadha}, {O'Sullivan}, {Palczewski}, {Pandya},
  {Pankova}, {Park}, {Peiffer}, {P{\'e}rez de los Heros}, {Pieloth}, {Pinat},
  {Pizzuto}, {Plum}, {Price}, {Przybylski}, {Raab}, {Raissi}, {Rameez},
  {Rauch}, {Rawlins}, {Rea}, {Reimann}, {Relethford}, {Renzi}, {Resconi},
  {Rhode}, {Richman}, {Robertson}, {Rongen}, {Rott}, {Ruhe}, {Ryckbosch},
  {Rysewyk}, {Safa}, {Sanchez Herrera}, {Sandrock}, {Sandroos}, {Santander},
  {Sarkar}, {Sarkar}, {Satalecka}, {Schaufel}, {Schlunder}, {Schmidt},
  {Schneider}, {Schneider}, {Schumacher}, {Sclafani}, {Seckel}, {Seunarine},
  {Silva}, {Snihur}, {Soedingrekso}, {Soldin}, {Song}, {Spiczak}, {Spiering},
  {Stachurska}, {Stamatikos}, {Stanev}, {Stasik}, {Stein}, {Stettner},
  {Steuer}, {Stezelberger}, {Stokstad}, {St{\"o}{\ss}l}, {Strotjohann},
  {Stuttard}, {Sullivan}, {Sutherland}, {Taboada}, {Tenholt}, {Ter-Antonyan},
  {Terliuk}, {Tilav}, {Tomankova}, {T{\"o}nnis}, {Toscano}, {Tosi},
  {Tselengidou}, {Tung}, {Turcati}, {Turcotte}, {Turley}, {Ty}, {Unger},
  {Unland Elorrieta}, {Usner}, {Vandenbroucke}, {Van Driessche}, {van Eijk},
  {van Eijndhoven}, {Vanheule}, {van Santen}, {Vraeghe}, {Walck}, {Wallace},
  {Wallraff}, {Wandkowsky}, {Watson}, {Weaver}, {Weiss}, {Weldert}, {Wendt},
  {Werthebach}, {Westerhoff}, {Whelan}, {Whitehorn}, {Wiebe}, {Wiebusch},
  {Wille}, {Williams}, {Wills}, {Wolf}, {Wood}, {Wood}, {Woschnagg}, {Wrede},
  {Xu}, {Xu}, {Xu}, {Yanez}, {Yodh}, {Yoshida}, {Yuan}, \& {IceCube
  Collaboration}}]{2019ApJ...880..103G}
{Garrappa}, S., {Buson}, S., {Franckowiak}, A., {et~al.} 2019{\natexlab{b}},
  \apj, 880, 103, \dodoi{10.3847/1538-4357/ab2ada}

\bibitem[{{Gehrels} {et~al.}(2004){Gehrels}, {Chincarini}, {Giommi}, {Mason},
  {Nousek}, {Wells}, {White}, {Barthelmy}, {Burrows}, {Cominsky}, {Hurley},
  {Marshall}, {M{\'e}sz{\'a}ros}, {Roming}, {Angelini}, {Barbier}, {Belloni},
  {Campana}, {Caraveo}, {Chester}, {Citterio}, {Cline}, {Cropper}, {Cummings},
  {Dean}, {Feigelson}, {Fenimore}, {Frail}, {Fruchter}, {Garmire}, {Gendreau},
  {Ghisellini}, {Greiner}, {Hill}, {Hunsberger}, {Krimm}, {Kulkarni}, {Kumar},
  {Lebrun}, {Lloyd-Ronning}, {Markwardt}, {Mattson}, {Mushotzky}, {Norris},
  {Osborne}, {Paczynski}, {Palmer}, {Park}, {Parsons}, {Paul}, {Rees},
  {Reynolds}, {Rhoads}, {Sasseen}, {Schaefer}, {Short}, {Smale}, {Smith},
  {Stella}, {Tagliaferri}, {Takahashi}, {Tashiro}, {Townsley}, {Tueller},
  {Turner}, {Vietri}, {Voges}, {Ward}, {Willingale}, {Zerbi}, \&
  {Zhang}}]{2004ApJ...611.1005G}
{Gehrels}, N., {Chincarini}, G., {Giommi}, P., {et~al.} 2004, \apj, 611, 1005,
  \dodoi{10.1086/422091}

\bibitem[{{Gezari}(2021)}]{2021arXiv210414580G}
{Gezari}, S. 2021, arXiv e-prints, arXiv:2104.14580.
\newblock \doarXiv{2104.14580}

\bibitem[{{Ghisellini} {et~al.}(2013){Ghisellini}, {Tavecchio}, {Foschini},
  {Bonnoli}, \& {Tagliaferri}}]{2013MNRAS.432L..66G}
{Ghisellini}, G., {Tavecchio}, F., {Foschini}, L., {Bonnoli}, G., \&
  {Tagliaferri}, G. 2013, \mnras, 432, L66, \dodoi{10.1093/mnrasl/slt041}

\bibitem[{{Giommi} {et~al.}(2020){Giommi}, {Padovani}, {Oikonomou}, {Glauch},
  {Paiano}, \& {Resconi}}]{2020A&A...640L...4G}
{Giommi}, P., {Padovani}, P., {Oikonomou}, F., {et~al.} 2020, \aap, 640, L4,
  \dodoi{10.1051/0004-6361/202038423}

\bibitem[{{Graham} {et~al.}(2019){Graham}, {Kulkarni}, {Bellm}, {Adams},
  {Barbarino}, {Blagorodnova}, {Bodewits}, {Bolin}, {Brady}, {Cenko}, {Chang},
  {Coughlin}, {De}, {Eadie}, {Farnham}, {Feindt}, {Franckowiak}, {Fremling},
  {Gezari}, {Ghosh}, {Goldstein}, {Golkhou}, {Goobar}, {Ho}, {Huppenkothen},
  {Ivezi{\'c}}, {Jones}, {Juric}, {Kaplan}, {Kasliwal}, {Kelley}, {Kupfer},
  {Lee}, {Lin}, {Lunnan}, {Mahabal}, {Miller}, {Ngeow}, {Nugent}, {Ofek},
  {Prince}, {Rauch}, {van Roestel}, {Schulze}, {Singer}, {Sollerman}, {Taddia},
  {Yan}, {Ye}, {Yu}, {Barlow}, {Bauer}, {Beck}, {Belicki}, {Biswas}, {Brinnel},
  {Brooke}, {Bue}, {Bulla}, {Burruss}, {Connolly}, {Cromer}, {Cunningham},
  {Dekany}, {Delacroix}, {Desai}, {Duev}, {Feeney}, {Flynn}, {Frederick},
  {Gal-Yam}, {Giomi}, {Groom}, {Hacopians}, {Hale}, {Helou}, {Henning},
  {Hover}, {Hillenbrand}, {Howell}, {Hung}, {Imel}, {Ip}, {Jackson}, {Kaspi},
  {Kaye}, {Kowalski}, {Kramer}, {Kuhn}, {Landry}, {Laher}, {Mao}, {Masci},
  {Monkewitz}, {Murphy}, {Nordin}, {Patterson}, {Penprase}, {Porter},
  {Rebbapragada}, {Reiley}, {Riddle}, {Rigault}, {Rodriguez}, {Rusholme}, {van
  Santen}, {Shupe}, {Smith}, {Soumagnac}, {Stein}, {Surace}, {Szkody}, {Terek},
  {Van Sistine}, {van Velzen}, {Vestrand}, {Walters}, {Ward}, {Zhang}, \&
  {Zolkower}}]{2019PASP..131g8001G}
{Graham}, M.~J., {Kulkarni}, S.~R., {Bellm}, E.~C., {et~al.} 2019, \pasp, 131,
  078001, \dodoi{10.1088/1538-3873/ab006c}

\bibitem[{{Halzen} \& {Kheirandish}(2016)}]{2016ApJ...831...12H}
{Halzen}, F., \& {Kheirandish}, A. 2016, \apj, 831, 12,
  \dodoi{10.3847/0004-637X/831/1/12}

\bibitem[{{Healey} {et~al.}(2008){Healey}, {Romani}, {Cotter}, {Michelson},
  {Schlafly}, {Readhead}, {Giommi}, {Chaty}, {Grenier}, \&
  {Weintraub}}]{2008ApJS..175...97H}
{Healey}, S.~E., {Romani}, R.~W., {Cotter}, G., {et~al.} 2008, \apjs, 175, 97,
  \dodoi{10.1086/523302}

\bibitem[{{IceCube Collaboration}(2013)}]{2013Sci...342E...1I}
{IceCube Collaboration}. 2013, Science, 342, 1242856,
  \dodoi{10.1126/science.1242856}

\bibitem[{{IceCube Collaboration}(2019)}]{2019GCN.25913....1I}
---. 2019, GRB Coordinates Network, 25913, 1

\bibitem[{{IceCube Collaboration} {et~al.}(2018{\natexlab{a}}){IceCube
  Collaboration}, {Aartsen}, {Ackermann}, {Adams}, {Aguilar}, {Ahlers},
  {Ahrens}, {Al Samarai}, {Altmann}, {Andeen}, {Anderson}, {Ansseau}, {Anton},
  {Arg{\"u}elles}, {Auffenberg}, {Axani}, {Bagherpour}, {Bai}, {Barron},
  {Barwick}, {Baum}, {Bay}, {Beatty}, {Becker Tjus}, {Becker}, {BenZvi},
  {Berley}, {Bernardini}, {Besson}, {Binder}, {Bindig}, {Blaufuss}, {Blot},
  {Bohm}, {B{\"o}rner}, {Bos}, {B{\"o}ser}, {Botner}, {Bourbeau}, {Bourbeau},
  {Bradascio}, {Braun}, {Brenzke}, {Bretz}, {Bron}, {Brostean-Kaiser},
  {Burgman}, {Busse}, {Carver}, {Cheung}, {Chirkin}, {Christov}, {Clark},
  {Classen}, {Coenders}, {Collin}, {Conrad}, {Coppin}, {Correa}, {Cowen},
  {Cross}, {Dave}, {Day}, {de Andr{\'e}}, {De Clercq}, {DeLaunay}, {Dembinski},
  {De Ridder}, {Desiati}, {de Vries}, {de Wasseige}, {de With}, {DeYoung},
  {D{\'\i}az-V{\'e}lez}, {di Lorenzo}, {Dujmovic}, {Dumm}, {Dunkman}, {Dvorak},
  {Eberhardt}, {Ehrhardt}, {Eichmann}, {Eller}, {Evenson}, {Fahey}, {Fazely},
  {Felde}, {Filimonov}, {Finley}, {Flis}, {Franckowiak}, {Friedman}, {Fritz},
  {Gaisser}, {Gallagher}, {Gerhardt}, {Ghorbani}, {Glauch}, {Gl{\"u}senkamp},
  {Goldschmidt}, {Gonzalez}, {Grant}, {Griffith}, {Haack}, {Hallgren},
  {Halzen}, {Hanson}, {Hebecker}, {Heereman}, {Helbing}, {Hellauer},
  {Hickford}, {Hignight}, {Hill}, {Hoffman}, {Hoffmann}, {Hoinka},
  {Hokanson-Fasig}, {Hoshina}, {Huang}, {Huber}, {Hultqvist}, {H{\"u}nnefeld},
  {Hussain}, {In}, {Iovine}, {Ishihara}, {Jacobi}, {Japaridze}, {Jeong},
  {Jero}, {Jones}, {Kalaczynski}, {Kang}, {Kappes}, {Kappesser}, {Karg},
  {Karle}, {Katz}, {Kauer}, {Keivani}, {Kelley}, {Kheirandish}, {Kim}, {Kim},
  {Kintscher}, {Kiryluk}, {Kittler}, {Klein}, {Koirala}, {Kolanoski},
  {K{\"o}pke}, {Kopper}, {Kopper}, {Koschinsky}, {Koskinen}, {Kowalski},
  {Krings}, {Kroll}, {Kr{\"u}ckl}, {Kunwar}, {Kurahashi}, {Kuwabara},
  {Kyriacou}, {Labare}, {Lanfranchi}, {Larson}, {Lauber}, {Leonard},
  {Lesiak-Bzdak}, {Leuermann}, {Liu}, {Lozano Mariscal}, {Lu}, {L{\"u}nemann},
  {Luszczak}, {Madsen}, {Maggi}, {Mahn}, {Mancina}, {Maruyama}, {Mase},
  {Maunu}, {Meagher}, {Medici}, {Meier}, {Menne}, {Merino}, {Meures},
  {Miarecki}, {Micallef}, {Moment{\'e}}, {Montaruli}, {Moore}, {S}, {Morse},
  {Moulai}, {Nahnhauer}, {Nakarmi}, {Naumann}, {Neer}, {Niederhausen},
  {Nowicki}, {Nygren}, {Obertacke Pollmann}, {Olivas}, {O'Murchadha},
  {O'Sullivan}, {Palczewski}, {Pandya}, {Pankova}, {Peiffer}, {Pepper},
  {P{\'e}rez de los Heros}, {Pieloth}, {Pinat}, {Plum}, {Price}, {Przybylski},
  {Raab}, {R{\"a}del}, {Rameez}, {Rauch}, {Rawlins}, {Rea}, {Reimann},
  {Relethford}, {Relich}, {Resconi}, {Rhode}, {Richman}, {Robertson}, {Rongen},
  {Rott}, {Ruhe}, {Ryckbosch}, {Rysewyk}, {Safa}, {S{\"a}lzer}, {Sanchez
  Herrera}, {Sandrock}, {Sandroos}, {Santander}, {Sarkar}, {Sarkar},
  {Satalecka}, {Schlunder}, {Schmidt}, {Schneider}, {Schoenen},
  {Sch{\"o}neberg}, {Schumacher}, {Sclafani}, {Seckel}, {Seunarine},
  {Soedingrekso}, {Soldin}, {Song}, {Spiczak}, {Spiering}, {Stachurska},
  {Stamatikos}, {Stanev}, {Stasik}, {Stein}, {Stettner}, {Steuer},
  {Stezelberger}, {Stokstad}, {St{\"o}{\ss}l}, {Strotjohann}, {Stuttard},
  {Sullivan}, {Sutherland}, {Taboada}, {Tatar}, {Tenholt}, {Ter-Antonyan},
  {Terliuk}, {Tilav}, {Toale}, {Tobin}, {Toennis}, {Toscano}, {Tosi},
  {Tselengidou}, {Tung}, {Turcati}, {Turley}, {Ty}, {Unger}, {Usner},
  {Vandenbroucke}, {Van Driessche}, {van Eijk}, {van Eijndhoven}, {Vanheule},
  {van Santen}, {Vogel}, {Vraeghe}, {Walck}, {Wallace}, {Wallraff}, {Wandler},
  {Wandkowsky}, {Waza}, {Weaver}, {Weiss}, {Wendt}, {Werthebach}, {Westerhoff},
  {Whelan}, {Whitehorn}, {Wiebe}, {Wiebusch}, {Wille}, {Williams}, {Wills},
  {Wolf}, {Wood}, {Wood}, {Woschnagg}, {Xu}, {Xu}, {Xu}, {Yanez}, {Yodh},
  {Yoshida}, {Yuan}, {Fermi-LAT Collaboration}, {Abdollahi}, {Ajello},
  {Angioni}, {Baldini}, {Ballet}, {Barbiellini}, {Bastieri}, {Bechtol},
  {Bellazzini}, {Berenji}, {Bissaldi}, {Blandford}, {Bonino}, {Bottacini},
  {Bregeon}, {Bruel}, {Buehler}, {Burnett}, {Burns}, {Buson}, {Cameron},
  {Caputo}, {Caraveo}, {Cavazzuti}, {Charles}, {Chen}, {Cheung}, {Chiang},
  {Chiaro}, {Ciprini}, {Cohen-Tanugi}, {Conrad}, {Costantin}, {Cutini},
  {D'Ammando}, {de Palma}, {Digel}, {Di Lalla}, {Di Mauro}, {Di Venere},
  {Dom{\'\i}nguez}, {Favuzzi}, {Franckowiak}, {Fukazawa}, {Funk}, {Fusco},
  {Gargano}, {Gasparrini}, {Giglietto}, {Giomi}, {Giommi}, {Giordano},
  {Giroletti}, {Glanzman}, {Green}, {Grenier}, {Grondin}, {Guiriec}, {Harding},
  {Hayashida}, {Hays}, {Hewitt}, {Horan}, {J{\'o}hannesson}, {Kadler},
  {Kensei}, {Kocevski}, {Krauss}, {Kreter}, {Kuss}, {La Mura}, {Larsson},
  {Latronico}, {Lemoine-Goumard}, {Li}, {Longo}, {Loparco}, {Lovellette},
  {Lubrano}, {Magill}, {Maldera}, {Malyshev}, {Manfreda}, {Mazziotta},
  {McEnery}, {Meyer}, {Michelson}, {Mizuno}, {Monzani}, {Morselli},
  {Moskalenko}, {Negro}, {Nuss}, {Ojha}, {Omodei}, {Orienti}, {Orlando},
  {Palatiello}, {Paliya}, {Perkins}, {Persic}, {Pesce-Rollins}, {Piron},
  {Porter}, {Principe}, {Rain{\`o}}, {Rando}, {Rani}, {Razzano}, {Razzaque},
  {Reimer}, {Reimer}, {Renault-Tinacci}, {Ritz}, {Rochester}, {Saz Parkinson},
  {Sgr{\`o}}, {Siskind}, {Spandre}, {Spinelli}, {Suson}, {Tajima}, {Takahashi},
  {Tanaka}, {Thayer}, {Thompson}, {Tibaldo}, {Torres}, {Torresi}, {Tosti},
  {Troja}, {Valverde}, {Vianello}, {Vogel}, {Wood}, {Wood}, {Zaharijas}, {MAGIC
  Collaboration}, {Ahnen}, {Ansoldi}, {Antonelli}, {Arcaro}, {Baack},
  {Babi{\'c}}, {Banerjee}, {Bangale}, {Barres de Almeida}, {Barrio}, {Becerra
  Gonz{\'a}lez}, {Bednarek}, {Bernardini}, {Berti}, {Bhattacharyya}, {Biland},
  {Blanch}, {Bonnoli}, {Carosi}, {Carosi}, {Ceribella}, {Chatterjee}, {Colak},
  {Colin}, {Colombo}, {Contreras}, {Cortina}, {Covino}, {Cumani}, {Da Vela},
  {Dazzi}, {De Angelis}, {De Lotto}, {Delfino}, {Delgado}, {Di Pierro},
  {Dom{\'\i}nguez}, {Dominis Prester}, {Dorner}, {Doro}, {Einecke},
  {Elsaesser}, {Fallah Ramazani}, {Fern{\'a}ndez-Barral}, {Fidalgo}, {Foffano},
  {Pfrang}, {Fonseca}, {Font}, {Franceschini}, {Fruck}, {Galindo}, {Gallozzi},
  {Garc{\'\i}a L{\'o}pez}, {Garczarczyk}, {Gaug}, {Giammaria}, {Godinovi{\'c}},
  {Gora}, {Guberman}, {Hadasch}, {Hahn}, {Hassan}, {Hayashida}, {Herrera},
  {Hose}, {Hrupec}, {Inoue}, {Ishio}, {Konno}, {Kubo}, {Kushida}, {Lelas},
  {Lindfors}, {Lombardi}, {Longo}, {L{\'o}pez}, {Maggio}, {Majumdar},
  {Makariev}, {Maneva}, {Manganaro}, {Mannheim}, {Maraschi}, {Mariotti},
  {Mart{\'\i}nez}, {Masuda}, {Mazin}, {Minev}, {M}, {Mirzoyan}, {Moralejo},
  {Moreno}, {Moretti}, {Nagayoshi}, {Neustroev}, {Niedzwiecki}, {Nievas
  Rosillo}, {Nigro}, {Nilsson}, {Ninci}, {Nishijima}, {Noda}, {Nogu{\'e}s},
  {Paiano}, {Palacio}, {Paneque}, {Paoletti}, {Paredes}, {Pedaletti},
  {Peresano}, {Persic}, {Prada Moroni}, {Prandini}, {Puljak}, {Rodriguez
  Garcia}, {Reichardt}, {Rhode}, {Rib{\'o}}, {Rico}, {Righi}, {Rugliancich},
  {Saito}, {Satalecka}, {Schweizer}, {Sitarek}, {{\v{S}}nidaric ́},
  {Sobczynska}, {Stamerra}, {Strzys}, {Suri{\'c}}, {Takahashi}, {Tavecchio},
  {Temnikov}, {Terzi{\'c}}, {Teshima}, {Torres-Alb{\`a}}, {Treves},
  {Tsujimoto}, {Vanzo}, {Vazquez Acosta}, {Vovk}, {Ward}, {Will}, {S}, {Zaric
  ́}, {AGILE Team}, {Lucarelli}, {Tavani}, {Piano}, {Donnarumma}, {Pittori},
  {Verrecchia}, {Barbiellini}, {Bulgarelli}, {Caraveo}, {Cattaneo},
  {Colafrancesco}, {Costa}, {Di Cocco}, {Ferrari}, {Gianotti}, {Giuliani},
  {Lipari}, {Mereghetti}, {Morselli}, {Pacciani}, {Paoletti}, {Parmiggiani},
  {Pellizzoni}, {Picozza}, {Pilia}, {Rappoldi}, {Trois}, {Vercellone},
  {Vittorini}, {ASAS-SN Team}, {Stanek}, {Kochanek}, {Beacom}, {Thompson},
  {Holoien}, {Dong}, {Prieto}, {Shappee}, {Holmbo}, {HAWC Collaboration},
  {Abeysekara}, {Albert}, {Alfaro}, {Alvarez}, {Arceo},
  {Arteaga-Vel{\'a}zquez}, {Avila Rojas}, {Ayala Solares}, {Becerril},
  {Belmont-Moreno}, {Bernal}, {Caballero-Mora}, {Capistr{\'a}n},
  {Carrami{\~n}ana}, {Casanova}, {Castillo}, {Cotti}, {Cotzomi}, {Couti{\~n}o
  de Le{\'o}n}, {De Le{\'o}n}, {De la Fuente}, {Diaz Hernandez}, {Dichiara},
  {Dingus}, {DuVernois}, {D{\'\i}az-V{\'e}lez}, {Ellsworth}, {Engel},
  {Fiorino}, {Fleischhack}, {Fraija}, {Garc{\'\i}a-Gonz{\'a}lez}, {Garfias},
  {Gonz{\'a}lez Mu{\~n}oz}, {Gonz{\'a}lez}, {Goodman}, {Hampel-Arias},
  {Harding}, {Hernand ez}, {Hona}, {Hueyotl-Zahuantitla}, {Hui},
  {H{\"u}ntemeyer}, {Iriarte}, {Jardin-Blicq}, {Joshi}, {Kaufmann}, {Kunde},
  {Lara}, {Lauer}, {Lee}, {Lennarz}, {Le{\'o}n Vargas}, {Linnemann},
  {Longinotti}, {Luis-Raya}, {Luna-Garc{\'\i}a}, {Malone}, {Marinelli},
  {Martinez}, {Martinez-Castellanos}, {Mart{\'\i}nez-Castro},
  {Mart{\'\i}nez-Huerta}, {Matthews}, {Miranda-Romagnoli}, {Moreno},
  {Mostaf{\'a}}, {Nayerhoda}, {Nellen}, {Newbold}, {Nisa}, {Noriega-Papaqui},
  {Pelayo}, {Pretz}, {P{\'e}rez-P{\'e}rez}, {Ren}, {Rho}, {Rivi{\`e}re},
  {Rosa-Gonz{\'a}lez}, {Rosenberg}, {Ruiz-Velasco}, {Ruiz-Velasco}, {Salesa
  Greus}, {Sandoval}, {Schneider}, {Schoorlemmer}, {Sinnis}, {Smith},
  {Springer}, {Surajbali}, {Tibolla}, {Tollefson}, {Torres}, {Villase{\~n}or},
  {Weisgarber}, {Werner}, {Yapici}, {Gaurang}, {Zepeda}, {Zhou}, {{\'A}lvarez},
  {H.~E.~S.~S. Collaboration}, {Abdalla}, {Ang{\"u}ner}, {Armand}, {Backes},
  {Becherini}, {Berge}, {B{\"o}ttcher}, {Boisson}, {Bolmont}, {Bonnefoy},
  {Bordas}, {Brun}, {B{\"u}chele}, {Bulik}, {Caroff}, {Carosi}, {Casanova},
  {Cerruti}, {Chakraborty}, {Chandra}, {Chen}, {Colafrancesco}, {Davids},
  {Deil}, {Devin}, {Djannati-Ata{\"\i}}, {Egberts}, {Emery}, {Eschbach},
  {Fiasson}, {Fontaine}, {Funk}, {F{\"u}{\ss}ling}, {Gallant}, {Gat{\'e}},
  {Giavitto}, {Glawion}, {Glicenstein}, {Gottschall}, {Grondin}, {Haupt},
  {Henri}, {Hinton}, {Hoischen}, {Holch}, {Huber}, {Jamrozy}, {Jankowsky},
  {Jankowsky}, {Jouvin}, {Jung-Richardt}, {Kerszberg}, {Kh{\'e}lifi}, {King},
  {Klepser}, {Kluz ́niak}, {Komin}, {Kraus}, {Lefaucheur}, {Lemi{\`e}re},
  {Lemoine-Goumard}, {Lenain}, {Leser}, {Lohse}, {L{\'o}pez-Coto}, {Lorentz},
  {Lypova}, {Marandon}, {Guillem Mart{\'\i}-Devesa}, {Maurin}, {Mitchell},
  {Moderski}, {Mohamed}, {Mohrmann}, {Moulin}, {Murach}, {de Naurois},
  {Niederwanger}, {Niemiec}, {Oakes}, {O'Brien}, {Ohm}, {Ostrowski}, {Oya},
  {Panter}, {Parsons}, {Perennes}, {Piel}, {Pita}, {Poireau}, {Priyana Noel},
  {Prokoph}, {P{\"u}hlhofer}, {Quirrenbach}, {Raab}, {Rauth}, {Renaud},
  {Rieger}, {Rinchiuso}, {Romoli}, {Rowell}, {Rudak}, {Sasaki}, {Sanchez},
  {Schlickeiser}, {Sch{\"u}ssler}, {Schulz}, {Schwanke}, {Seglar-Arroyo},
  {Shafi}, {Simoni}, {Sol}, {Stegmann}, {Steppa}, {Tavernier}, {Taylor},
  {Tiziani}, {Trichard}, {Tsirou}, {van Eldik}, {van Rensburg}, {van Soelen},
  {Veh}, {Vincent}, {Voisin}, {Wagner}, {Wagner}, {Wierzcholska}, {Zanin},
  {Zdziarski}, {Zech}, {Ziegler}, {Zorn}, {{\.Z}ywucka}, {INTEGRAL Team},
  {Savchenko}, {Ferrigno}, {Bazzano}, {Diehl}, {Kuulkers}, {Laurent},
  {Mereghetti}, {Natalucci}, {Panessa}, {Rodi}, {Ubertini}, {Kanata}, Teams,
  {Morokuma}, {Ohta}, {Tanaka}, {Mori}, {Yamanaka}, {Kawabata}, {Utsumi},
  {Nakaoka}, {Kawabata}, {Nagashima}, {Yoshida}, {Matsuoka}, {Itoh}, {Kapteyn
  Team}, {Keel}, {Liverpool Telescope Team}, {Copperwheat}, {Steele},
  {Swift/NuSTAR Team}, {Cenko}, {Cowen}, {DeLaunay}, {Evans}, {Fox}, {Keivani},
  {Kennea}, {Marshall}, {Osborne}, {Santander}, {Tohuvavohu}, {Turley},
  {VERITAS Collaboration}, {Abeysekara}, {Archer}, {Benbow}, {Bird}, {Brill},
  {Brose}, {Buchovecky}, {Buckley}, {Bugaev}, {Christiansen}, {Connolly},
  {Cui}, {Daniel}, {Errando}, {Falcone}, {Feng}, {Finley}, {Fortson},
  {Furniss}, {Gueta}, {H{\"u}tten}, {Hervet}, {Hughes}, {Humensky}, {Johnson},
  {Kaaret}, {Kar}, {Kelley-Hoskins}, {Kertzman}, {Kieda}, {Krause},
  {Krennrich}, {Kumar}, {Lang}, {Lin}, {Maier}, {McArthur}, {Moriarty},
  {Mukherjee}, {Nieto}, {O'Brien}, {Ong}, {Otte}, {Park}, {Petrashyk}, {Pohl},
  {Popkow}, {Pueschel}, {Quinn}, {Ragan}, {Reynolds}, {Richards}, {Roache},
  {Rulten}, {Sadeh}, {Santander}, {Scott}, {Sembroski}, {Shahinyan}, {Sushch},
  {Tr{\'e}panier}, {Tyler}, {Vassiliev}, {Wakely}, {Weinstein}, {Wells},
  {Wilcox}, {Wilhelm}, {Williams}, {Zitzer}, {VLA/B Team}, {Tetarenko},
  {Kimball}, {Miller-Jones}, \& {Sivakoff}}]{2018Sci...361.1378I}
{IceCube Collaboration}, {Aartsen}, M.~G., {Ackermann}, M., {et~al.}
  2018{\natexlab{a}}, Science, 361, eaat1378, \dodoi{10.1126/science.aat1378}

\bibitem[{{IceCube Collaboration} {et~al.}(2018{\natexlab{b}}){IceCube
  Collaboration}, {Aartsen}, {Ackermann}, {Adams}, {Aguilar}, {Ahlers},
  {Ahrens}, {Samarai}, {Altmann}, {Andeen}, {Anderson}, {Ansseau}, {Anton},
  {Arg{\"u}elles}, {Arsioli}, {Auffenberg}, {Axani}, {Bagherpour}, {Bai},
  {Barron}, {Barwick}, {Baum}, {Bay}, {Beatty}, {Becker Tjus}, {Becker},
  {BenZvi}, {Berley}, {Bernardini}, {Besson}, {Binder}, {Bindig}, {Blaufuss},
  {Blot}, {Bohm}, {B{\"o}rner}, {Bos}, {B{\"o}ser}, {Botner}, {Bourbeau},
  {Bourbeau}, {Bradascio}, {Braun}, {Brenzke}, {Bretz}, {Bron},
  {Brostean-Kaiser}, {Burgman}, {Busse}, {Carver}, {Cheung}, {Chirkin},
  {Christov}, {Clark}, {Classen}, {Coenders}, {Collin}, {Conrad}, {Coppin},
  {Correa}, {Cowen}, {Cross}, {Dave}, {Day}, {de Andr{\'e}}, {De Clercq},
  {DeLaunay}, {Dembinski}, {DeRidder}, {Desiati}, {de Vries}, {de Wasseige},
  {de With}, {DeYoung}, {D{\'\i}az-V{\'e}lez}, {di Lorenzo}, {Dujmovic},
  {Dumm}, {Dunkman}, {Dvorak}, {Eberhardt}, {Ehrhardt}, {Eichmann}, {Eller},
  {Evenson}, {Fahey}, {Fazely}, {Felde}, {Filimonov}, {Finley}, {Flis},
  {Franckowiak}, {Friedman}, {Fritz}, {Gaisser}, {Gallagher}, {Gerhardt},
  {Ghorbani}, {Giommi}, {Glauch}, {Gl{\"u}senkamp}, {Goldschmidt}, {Gonzalez},
  {Grant}, {Griffith}, {Haack}, {Hallgren}, {Halzen}, {Hanson}, {Hebecker},
  {Heereman}, {Helbing}, {Hellauer}, {Hickford}, {Hignight}, {Hill}, {Hoffman},
  {Hoffmann}, {Hoinka}, {Hokanson-Fasig}, {Hoshina}, {Huang}, {Huber},
  {Hultqvist}, {H{\"u}nnefeld}, {Hussain}, {In}, {Iovine}, {Ishihara},
  {Jacobi}, {Japaridze}, {Jeong}, {Jero}, {Jones}, {Kalaczynski}, {Kang},
  {Kappes}, {Kappesser}, {Karg}, {Karle}, {Katz}, {Kauer}, {Keivani}, {Kelley},
  {Kheirandish}, {Kim}, {Kim}, {Kintscher}, {Kiryluk}, {Kittler}, {Klein},
  {Koirala}, {Kolanoski}, {K{\"o}pke}, {Kopper}, {Kopper}, {Koschinsky},
  {Koskinen}, {Kowalski}, {Krammer}, {Krings}, {Kroll}, {Kr{\"u}ckl}, {Kunwar},
  {Kurahashi}, {Kuwabara}, {Kyriacou}, {Labare}, {Lanfranchi}, {Larson},
  {Lauber}, {Leonard}, {Lesiak-Bzdak}, {Leuermann}, {Liu}, {Lozano Mariscal},
  {Lu}, {L{\"u}nemann}, {Luszczak}, {Madsen}, {Maggi}, {Mahn}, {Mancina},
  {Maruyama}, {Mase}, {Maunu}, {Meagher}, {Medici}, {Meier}, {Menne}, {Merino},
  {Meures}, {Miarecki}, {Micallef}, {Moment{\'e}}, {Montaruli}, {Moore},
  {Morse}, {Moulai}, {Nahnhauer}, {Nakarmi}, {Naumann}, {Neer}, {Niederhausen},
  {Nowicki}, {Nygren}, {Obertacke Pollmann}, {Olivas}, {O'Murchadha},
  {O'Sullivan}, {Padovani}, {Palczewski}, {Pandya}, {Pankova}, {Peiffer},
  {Pepper}, {P{\'e}rez de los Heros}, {Pieloth}, {Pinat}, {Plum}, {Price},
  {Przybylski}, {Raab}, {R{\"a}del}, {Rameez}, {Rawlins}, {Rea}, {Reimann},
  {Relethford}, {Relich}, {Resconi}, {Rhode}, {Richman}, {Robertson}, {Rongen},
  {Rott}, {Ruhe}, {Ryckbosch}, {Rysewyk}, {Safa}, {Sahakyan}, {S{\"a}lzer},
  {Sanchez Herrera}, {Sandrock}, {Sandroos}, {Santander}, {Sarkar}, {Sarkar},
  {Satalecka}, {Schlunder}, {Schmidt}, {Schneider}, {Schoenen},
  {Sch{\"o}neberg}, {Schumacher}, {Sclafani}, {Seckel}, {Seunarine},
  {Soedingrekso}, {Soldin}, {Song}, {Spiczak}, {Spiering}, {Stachurska},
  {Stamatikos}, {Stanev}, {Stasik}, {Stettner}, {Steuer}, {Stezelberger},
  {Stokstad}, {St{\"o}{\ss}l}, {Strotjohann}, {Stuttard}, {Sullivan},
  {Sutherland}, {Taboada}, {Tatar}, {Tenholt}, {Ter-Antonyan}, {Terliuk},
  {Tilav}, {Toale}, {Tobin}, {Toennis}, {Toscano}, {Tosi}, {Tselengidou},
  {Tung}, {Turcati}, {Turley}, {Ty}, {Unger}, {Usner}, {Vandenbroucke}, {Van
  Driessche}, {van Eijk}, {van Eijndhoven}, {Vanheule}, {van Santen}, {Vogel},
  {Vraeghe}, {Walck}, {Wallace}, {Wallraff}, {Wandler}, {Wandkowsky}, {Waza},
  {Weaver}, {Weiss}, {Wendt}, {Werthebach}, {Westerhoff}, {Whelan},
  {Whitehorn}, {Wiebe}, {Wiebusch}, {Wille}, {Williams}, {Wills}, {Wolf},
  {Wood}, {Wood}, {Woschnagg}, {Xu}, {Xu}, {Xu}, {Yanez}, {Yodh}, {Yoshida}, \&
  {Yuan}}]{2018Sci...361..147I}
---. 2018{\natexlab{b}}, Science, 361, 147, \dodoi{10.1126/science.aat2890}

\bibitem[{{IceCube Collaboration} {et~al.}(2021){IceCube Collaboration},
  {Abbasi}, {Ackermann}, {Adams}, {Aguilar}, {Ahlers}, {Ahrens}, {Alispach},
  {Amin}, {Andeen}, {Anderson}, {Ansseau}, {Anton}, {Arg{\"u}elles}, {Axani},
  {Bai}, {Balagopal V.}, {Barbano}, {Barwick}, {Bastian}, {Basu}, {Baum},
  {Baur}, {Bay}, {Beatty}, {Becker}, {Becker Tjus}, {Bellenghi}, {BenZvi},
  {Berley}, {Bernardini}, {Besson}, {Binder}, {Bindig}, {Blaufuss}, {Blot},
  {Bohm}, {B{\"o}ser}, {Botner}, {B{\"o}ttcher}, {Bourbeau}, {Bourbeau},
  {Bradascio}, {Braun}, {Bron}, {Brostean-Kaiser}, {Burgman}, {Buscher},
  {Busse}, {Campana}, {Carver}, {Chen}, {Cheung}, {Chirkin}, {Choi}, {Clark},
  {Clark}, {Classen}, {Coleman}, {Collin}, {Conrad}, {Coppin}, {Correa},
  {Cowen}, {Cross}, {Dave}, {De Clercq}, {DeLaunay}, {Dembinski}, {Deoskar},
  {De Ridder}, {Desai}, {Desiati}, {de Vries}, {de Wasseige}, {de With},
  {DeYoung}, {Dharani}, {Diaz}, {D{\'\i}az-V{\'e}lez}, {Dujmovic}, {Dunkman},
  {DuVernois}, {Dvorak}, {Ehrhardt}, {Eller}, {Engel}, {Evenson}, {Fahey},
  {Fazely}, {Felde}, {Fienberg}, {Filimonov}, {Finley}, {Fischer}, {Fox},
  {Franckowiak}, {Friedman}, {Fritz}, {Gaisser}, {Gallagher}, {Ganster},
  {Garrappa}, {Gerhardt}, {Ghadimi}, {Glauch}, {Gl{\"u}senkamp}, {Goldschmidt},
  {Gonzalez}, {Goswami}, {Grant}, {Gr{\'e}goire}, {Griffith}, {Griswold},
  {G{\"u}nd{\"u}z}, {Haack}, {Hallgren}, {Halliday}, {Halve}, {Halzen}, {Minh},
  {Hanson}, {Hardin}, {Haungs}, {Hauser}, {Hebecker}, {Heix}, {Helbing},
  {Hellauer}, {Henningsen}, {Hickford}, {Hignight}, {Hill}, {Hill}, {Hoffman},
  {Hoffmann}, {Hoinka}, {Hokanson-Fasig}, {Hoshina}, {Huang}, {Huber}, {Huber},
  {Hultqvist}, {H{\"u}nnefeld}, {Hussain}, {In}, {Iovine}, {Ishihara},
  {Jansson}, {Japaridze}, {Jeong}, {Jones}, {Jonske}, {Joppe}, {Kang}, {Kang},
  {Kang}, {Kappes}, {Kappesser}, {Karg}, {Karl}, {Karle}, {Katz}, {Kauer},
  {Kellermann}, {Kelley}, {Kheirandish}, {Kim}, {Kin}, {Kintscher}, {Kiryluk},
  {Kittler}, {Klein}, {Koirala}, {Kolanoski}, {K{\"o}pke}, {Kopper}, {Kopper},
  {Koskinen}, {Koundal}, {Kovacevich}, {Kowalski}, {Krings}, {Kr{\"u}ckl},
  {Kulacz}, {Kurahashi}, {Kyriacou}, {Lagunas Gualda}, {Lanfranchi}, {Larson},
  {Lauber}, {Lazar}, {Leonard}, {Leszczy{\'n}ska}, {Li}, {Liu}, {Lohfink},
  {Lozano Mariscal}, {Lu}, {Lucarelli}, {Ludwig}, {L{\"u}nemann}, {Luszczak},
  {Lyu}, {Ma}, {Madsen}, {Maggi}, {Mahn}, {Makino}, {Mallik}, {Mancina},
  {Mari{\c{s}}}, {Maruyama}, {Mase}, {Maunu}, {McNally}, {Meagher}, {Medina},
  {Meier}, {Meighen-Berger}, {Merz}, {Micallef}, {Mockler}, {Moment{\'e}},
  {Montaruli}, {Moore}, {Morse}, {Moulai}, {Muth}, {Naab}, {Nagai}, {Naumann},
  {Necker}, {Neer}, {Nguy{\^e}n}, {Niederhausen}, {Nisa}, {Nowicki}, {Nygren},
  {Obertacke Pollmann}, {Oehler}, {Olivas}, {O'Sullivan}, {Pandya}, {Pankova},
  {Park}, {Parker}, {Paudel}, {Peiffer}, {P{\'e}rez de los Heros}, {Philippen},
  {Pieloth}, {Pieper}, {Pizzuto}, {Plum}, {Popovych}, {Porcelli}, {Prado
  Rodriguez}, {Price}, {Przybylski}, {Raab}, {Raissi}, {Rameez}, {Rawlins},
  {Rea}, {Rehman}, {Reimann}, {Renschler}, {Renzi}, {Resconi}, {Reusch},
  {Rhode}, {Richman}, {Riedel}, {Robertson}, {Roellinghoff}, {Rongen}, {Rott},
  {Ruhe}, {Ryckbosch}, {Rysewyk Cantu}, {Safa}, {Sanchez Herrera}, {Sandrock},
  {Sandroos}, {Santander}, {Sarkar}, {Sarkar}, {Satalecka}, {Scharf},
  {Schaufel}, {Schieler}, {Schlunder}, {Schmidt}, {Schneider}, {Schneider},
  {Schr{\"o}der}, {Schumacher}, {Sclafani}, {Seckel}, {Seunarine}, {Shefali},
  {Silva}, {Smithers}, {Snihur}, {Soedingrekso}, {Soldin}, {Song}, {Spiczak},
  {Spiering}, {Stachurska}, {Stamatikos}, {Stanev}, {Stein}, {Stettner},
  {Steuer}, {Stezelberger}, {Stokstad}, {Strotjohann}, {St{\"u}rwald},
  {Stuttard}, {Sullivan}, {Taboada}, {Tenholt}, {Ter-Antonyan}, {Tilav},
  {Tollefson}, {Tomankova}, {T{\"o}nnis}, {Toscano}, {Tosi}, {Trettin},
  {Tselengidou}, {Tung}, {Turcati}, {Turcotte}, {Turley}, {Twagirayezu}, {Ty},
  {Unger}, {Unland Elorrieta}, {Vandenbroucke}, {van Eijk}, {van Eijndhoven},
  {Vannerom}, {van Santen}, {Verpoest}, {Vraeghe}, {Walck}, {Wallace},
  {Watson}, {Weaver}, {Weindl}, {Weiss}, {Weldert}, {Wendt}, {Werthebach},
  {Whelan}, {Whitehorn}, {Wiebe}, {Wiebusch}, {Williams}, {Wolf}, {Wood},
  {Woschnagg}, {Wrede}, {Wulff}, {Xu}, {Xu}, {Yanez}, {Yoshida}, {Yuan},
  {Zhang}, \& {Z{\"o}cklein}}]{2021arXiv210109836I}
{IceCube Collaboration}, {Abbasi}, R., {Ackermann}, M., {et~al.} 2021, arXiv
  e-prints, arXiv:2101.09836.
\newblock \doarXiv{2101.09836}

\bibitem[{{Kadler} {et~al.}(2016){Kadler}, {Krau{\ss}}, {Mannheim}, {Ojha},
  {M{\"u}ller}, {Schulz}, {Anton}, {Baumgartner}, {Beuchert}, {Buson},
  {Carpenter}, {Eberl}, {Edwards}, {Eisenacher Glawion}, {Els{\"a}sser},
  {Gehrels}, {Gr{\"a}fe}, {Gulyaev}, {Hase}, {Horiuchi}, {James}, {Kappes},
  {Kappes}, {Katz}, {Kreikenbohm}, {Kreter}, {Kreykenbohm}, {Langejahn},
  {Leiter}, {Litzinger}, {Longo}, {Lovell}, {McEnery}, {Natusch}, {Phillips},
  {Pl{\"o}tz}, {Quick}, {Ros}, {Stecker}, {Steinbring}, {Stevens}, {Thompson},
  {Tr{\"u}stedt}, {Tzioumis}, {Weston}, {Wilms}, \&
  {Zensus}}]{2016NatPh..12..807K}
{Kadler}, M., {Krau{\ss}}, F., {Mannheim}, K., {et~al.} 2016, Nature Physics,
  12, 807, \dodoi{10.1038/nphys3715}

\bibitem[{{Kara} {et~al.}(2018){Kara}, {Dai}, {Reynolds}, \&
  {Kallman}}]{2018MNRAS.474.3593K}
{Kara}, E., {Dai}, L., {Reynolds}, C.~S., \& {Kallman}, T. 2018, \mnras, 474,
  3593, \dodoi{10.1093/mnras/stx3004}

\bibitem[{{King} \& {Pounds}(2015)}]{2015ARA&A..53..115K}
{King}, A., \& {Pounds}, K. 2015, \araa, 53, 115,
  \dodoi{10.1146/annurev-astro-082214-122316}

\bibitem[{{Koay} {et~al.}(2011){Koay}, {Macquart}, {Rickett}, {Bignall},
  {Lovell}, {Reynolds}, {Jauncey}, {Pursimo}, {Kedziora-Chudczer}, \&
  {Ojha}}]{2011AJ....142..108K}
{Koay}, J.~Y., {Macquart}, J.~P., {Rickett}, B.~J., {et~al.} 2011, \aj, 142,
  108, \dodoi{10.1088/0004-6256/142/4/108}

\bibitem[{{Kun} {et~al.}(2021){Kun}, {Bartos}, {Tjus}, {Biermann}, {Halzen}, \&
  {Mez{\H{o}}}}]{2021ApJ...911L..18K}
{Kun}, E., {Bartos}, I., {Tjus}, J.~B., {et~al.} 2021, \apjl, 911, L18,
  \dodoi{10.3847/2041-8213/abf1ec}

\bibitem[{{Lei} {et~al.}(2016){Lei}, {Yuan}, {Zhang}, \&
  {Wang}}]{2016ApJ...816...20L}
{Lei}, W.-H., {Yuan}, Q., {Zhang}, B., \& {Wang}, D. 2016, \apj, 816, 20,
  \dodoi{10.3847/0004-637X/816/1/20}

\bibitem[{{Levan} {et~al.}(2011){Levan}, {Tanvir}, {Cenko}, {Perley},
  {Wiersema}, {Bloom}, {Fruchter}, {de Ugarte Postigo}, {O'Brien}, {Butler},
  {van der Horst}, {Leloudas}, {Morgan}, {Misra}, {Bower}, {Farihi},
  {Tunnicliffe}, {Modjaz}, {Silverman}, {Hjorth}, {Th{\"o}ne}, {Cucchiara},
  {Cer{\'o}n}, {Castro-Tirado}, {Arnold}, {Bremer}, {Brodie}, {Carroll},
  {Cooper}, {Curran}, {Cutri}, {Ehle}, {Forbes}, {Fynbo}, {Gorosabel},
  {Graham}, {Hoffman}, {Guziy}, {Jakobsson}, {Kamble}, {Kerr}, {Kasliwal},
  {Kouveliotou}, {Kocevski}, {Law}, {Nugent}, {Ofek}, {Poznanski}, {Quimby},
  {Rol}, {Romanowsky}, {S{\'a}nchez-Ram{\'\i}rez}, {Schulze}, {Singh}, {van
  Spaandonk}, {Starling}, {Strom}, {Tello}, {Vaduvescu}, {Wheatley}, {Wijers},
  {Winters}, \& {Xu}}]{2011Sci...333..199L}
{Levan}, A.~J., {Tanvir}, N.~R., {Cenko}, S.~B., {et~al.} 2011, Science, 333,
  199, \dodoi{10.1126/science.1207143}

\bibitem[{{Liodakis} {et~al.}(2017){Liodakis}, {Marchili}, {Angelakis},
  {Fuhrmann}, {Nestoras}, {Myserlis}, {Karamanavis}, {Krichbaum}, {Sievers},
  {Ungerechts}, \& {Zensus}}]{2017MNRAS.466.4625L}
{Liodakis}, I., {Marchili}, N., {Angelakis}, E., {et~al.} 2017, \mnras, 466,
  4625, \dodoi{10.1093/mnras/stx002}

\bibitem[{{Liu} \& {Bai}(2006)}]{2006ApJ...653.1089L}
{Liu}, H.~T., \& {Bai}, J.~M. 2006, \apj, 653, 1089, \dodoi{10.1086/509097}

\bibitem[{{Loeb} \& {Waxman}(2006)}]{2006JCAP...05..003L}
{Loeb}, A., \& {Waxman}, E. 2006, \jcap, 2006, 003,
  \dodoi{10.1088/1475-7516/2006/05/003}

\bibitem[{{Madejski} \& {Sikora}(2016)}]{2016ARA&A..54..725M}
{Madejski}, G.~G., \& {Sikora}, M. 2016, \araa, 54, 725,
  \dodoi{10.1146/annurev-astro-081913-040044}

\bibitem[{{Mainzer} {et~al.}(2014){Mainzer}, {Bauer}, {Cutri}, {Grav},
  {Masiero}, {Beck}, {Clarkson}, {Conrow}, {Dailey}, {Eisenhardt}, {Fabinsky},
  {Fajardo-Acosta}, {Fowler}, {Gelino}, {Grillmair}, {Heinrichsen}, {Kendall},
  {Kirkpatrick}, {Liu}, {Masci}, {McCallon}, {Nugent}, {Papin}, {Rice},
  {Royer}, {Ryan}, {Sevilla}, {Sonnett}, {Stevenson}, {Thompson}, {Wheelock},
  {Wiemer}, {Wittman}, {Wright}, \& {Yan}}]{2014ApJ...792...30M}
{Mainzer}, A., {Bauer}, J., {Cutri}, R.~M., {et~al.} 2014, \apj, 792, 30,
  \dodoi{10.1088/0004-637X/792/1/30}

\bibitem[{{Mannheim}(1993)}]{1993A&A...269...67M}
{Mannheim}, K. 1993, \aap, 269, 67.
\newblock \doarXiv{astro-ph/9302006}

\bibitem[{{Maraschi} {et~al.}(1992){Maraschi}, {Ghisellini}, \&
  {Celotti}}]{1992ApJ...397L...5M}
{Maraschi}, L., {Ghisellini}, G., \& {Celotti}, A. 1992, \apjl, 397, L5,
  \dodoi{10.1086/186531}

\bibitem[{{Masci} {et~al.}(2019){Masci}, {Laher}, {Rusholme}, {Shupe}, {Groom},
  {Surace}, {Jackson}, {Monkewitz}, {Beck}, {Flynn}, {Terek}, {Landry},
  {Hacopians}, {Desai}, {Howell}, {Brooke}, {Imel}, {Wachter}, {Ye}, {Lin},
  {Cenko}, {Cunningham}, {Rebbapragada}, {Bue}, {Miller}, {Mahabal}, {Bellm},
  {Patterson}, {Juri{\'c}}, {Golkhou}, {Ofek}, {Walters}, {Graham}, {Kasliwal},
  {Dekany}, {Kupfer}, {Burdge}, {Cannella}, {Barlow}, {Van Sistine}, {Giomi},
  {Fremling}, {Blagorodnova}, {Levitan}, {Riddle}, {Smith}, {Helou}, {Prince},
  \& {Kulkarni}}]{2019PASP..131a8003M}
{Masci}, F.~J., {Laher}, R.~R., {Rusholme}, B., {et~al.} 2019, \pasp, 131,
  018003, \dodoi{10.1088/1538-3873/aae8ac}

\bibitem[{{Mattila} {et~al.}(2018){Mattila}, {P{\'e}rez-Torres}, {Efstathiou},
  {Mimica}, {Fraser}, {Kankare}, {Alberdi}, {Aloy}, {Heikkil{\"a}}, {Jonker},
  {Lundqvist}, {Mart{\'\i}-Vidal}, {Meikle}, {Romero-Ca{\~n}izales}, {Smartt},
  {Tsygankov}, {Varenius}, {Alonso-Herrero}, {Bondi}, {Fransson},
  {Herrero-Illana}, {Kangas}, {Kotak}, {Ram{\'\i}rez-Olivencia},
  {V{\"a}is{\"a}nen}, {Beswick}, {Clements}, {Greimel}, {Harmanen},
  {Kotilainen}, {Nandra}, {Reynolds}, {Ryder}, {Walton}, {Wiik}, \&
  {{\"O}stlin}}]{2018Sci...361..482M}
{Mattila}, S., {P{\'e}rez-Torres}, M., {Efstathiou}, A., {et~al.} 2018,
  Science, 361, 482, \dodoi{10.1126/science.aao4669}

\bibitem[{{Mattox} {et~al.}(1996){Mattox}, {Bertsch}, {Chiang}, {Dingus},
  {Digel}, {Esposito}, {Fierro}, {Hartman}, {Hunter}, {Kanbach}, {Kniffen},
  {Lin}, {Macomb}, {Mayer-Hasselwander}, {Michelson}, {von Montigny},
  {Mukherjee}, {Nolan}, {Ramanamurthy}, {Schneid}, {Sreekumar}, {Thompson}, \&
  {Willis}}]{1996ApJ...461..396M}
{Mattox}, J.~R., {Bertsch}, D.~L., {Chiang}, J., {et~al.} 1996, \apj, 461, 396,
  \dodoi{10.1086/177068}

\bibitem[{{Murase} {et~al.}(2008){Murase}, {Inoue}, \&
  {Nagataki}}]{2008ApJ...689L.105M}
{Murase}, K., {Inoue}, S., \& {Nagataki}, S. 2008, \apjl, 689, L105,
  \dodoi{10.1086/595882}

\bibitem[{{Murase} {et~al.}(2018){Murase}, {Oikonomou}, \&
  {Petropoulou}}]{2018ApJ...865..124M}
{Murase}, K., {Oikonomou}, F., \& {Petropoulou}, M. 2018, \apj, 865, 124,
  \dodoi{10.3847/1538-4357/aada00}

\bibitem[{{Murase} \& {Waxman}(2016)}]{2016PhRvD..94j3006M}
{Murase}, K., \& {Waxman}, E. 2016, \prd, 94, 103006,
  \dodoi{10.1103/PhysRevD.94.103006}

\bibitem[{{Myers} {et~al.}(2003){Myers}, {Jackson}, {Browne}, {de Bruyn},
  {Pearson}, {Readhead}, {Wilkinson}, {Biggs}, {Blandford}, {Fassnacht},
  {Koopmans}, {Marlow}, {McKean}, {Norbury}, {Phillips}, {Rusin}, {Shepherd},
  \& {Sykes}}]{2003MNRAS.341....1M}
{Myers}, S.~T., {Jackson}, N.~J., {Browne}, I.~W.~A., {et~al.} 2003, \mnras,
  341, 1, \dodoi{10.1046/j.1365-8711.2003.06256.x}

\bibitem[{{Nolan} {et~al.}(2012){Nolan}, {Abdo}, {Ackermann}, {Ajello},
  {Allafort}, {Antolini}, {Atwood}, {Axelsson}, {Baldini}, {Ballet},
  {Barbiellini}, {Bastieri}, {Bechtol}, {Belfiore}, {Bellazzini}, {Berenji},
  {Bignami}, {Blandford}, {Bloom}, {Bonamente}, {Bonnell}, {Borgland},
  {Bottacini}, {Bouvier}, {Brandt}, {Bregeon}, {Brigida}, {Bruel}, {Buehler},
  {Burnett}, {Buson}, {Caliandro}, {Cameron}, {Campana}, {Ca{\~n}adas},
  {Cannon}, {Caraveo}, {Casandjian}, {Cavazzuti}, {Ceccanti}, {Cecchi},
  {{\c{C}}elik}, {Charles}, {Chekhtman}, {Cheung}, {Chiang}, {Chipaux},
  {Ciprini}, {Claus}, {Cohen-Tanugi}, {Cominsky}, {Conrad}, {Corbet}, {Cutini},
  {D'Ammando}, {Davis}, {de Angelis}, {DeCesar}, {DeKlotz}, {De Luca}, {den
  Hartog}, {de Palma}, {Dermer}, {Digel}, {Silva}, {Drell}, {Drlica-Wagner},
  {Dubois}, {Dumora}, {Enoto}, {Escande}, {Fabiani}, {Falletti}, {Favuzzi},
  {Fegan}, {Ferrara}, {Focke}, {Fortin}, {Frailis}, {Fukazawa}, {Funk},
  {Fusco}, {Gargano}, {Gasparrini}, {Gehrels}, {Germani}, {Giebels},
  {Giglietto}, {Giommi}, {Giordano}, {Giroletti}, {Glanzman}, {Godfrey},
  {Grenier}, {Grondin}, {Grove}, {Guillemot}, {Guiriec}, {Gustafsson},
  {Hadasch}, {Hanabata}, {Harding}, {Hayashida}, {Hays}, {Hill}, {Horan},
  {Hou}, {Hughes}, {Iafrate}, {Itoh}, {J{\'o}hannesson}, {Johnson}, {Johnson},
  {Johnson}, {Johnson}, {Kamae}, {Katagiri}, {Kataoka}, {Katsuta}, {Kawai},
  {Kerr}, {Kn{\"o}dlseder}, {Kocevski}, {Kuss}, {Lande}, {Landriu},
  {Latronico}, {Lemoine-Goumard}, {Lionetto}, {Llena Garde}, {Longo},
  {Loparco}, {Lott}, {Lovellette}, {Lubrano}, {Madejski}, {Marelli}, {Massaro},
  {Mazziotta}, {McConville}, {McEnery}, {Mehault}, {Michelson}, {Minuti},
  {Mitthumsiri}, {Mizuno}, {Moiseev}, {Mongelli}, {Monte}, {Monzani},
  {Morselli}, {Moskalenko}, {Murgia}, {Nakamori}, {Naumann-Godo}, {Norris},
  {Nuss}, {Nymark}, {Ohno}, {Ohsugi}, {Okumura}, {Omodei}, {Orlando}, {Ormes},
  {Ozaki}, {Paneque}, {Panetta}, {Parent}, {Perkins}, {Pesce-Rollins},
  {Pierbattista}, {Pinchera}, {Piron}, {Pivato}, {Porter}, {Racusin},
  {Rain{\`o}}, {Rand o}, {Razzano}, {Razzaque}, {Reimer}, {Reimer}, {Reposeur},
  {Ritz}, {Rochester}, {Romani}, {Roth}, {Rousseau}, {Ryde}, {Sadrozinski},
  {Salvetti}, {Sanchez}, {Saz Parkinson}, {Sbarra}, {Scargle}, {Schalk},
  {Sgr{\`o}}, {Shaw}, {Shrader}, {Siskind}, {Smith}, {Spandre}, {Spinelli},
  {Stephens}, {Strickman}, {Suson}, {Tajima}, {Takahashi}, {Takahashi},
  {Tanaka}, {Thayer}, {Thayer}, {Thompson}, {Tibaldo}, {Tibolla}, {Tinebra},
  {Tinivella}, {Torres}, {Tosti}, {Troja}, {Uchiyama}, {Vandenbroucke}, {Van
  Etten}, {Van Klaveren}, {Vasileiou}, {Vianello}, {Vitale}, {Waite},
  {Wallace}, {Wang}, {Werner}, {Winer}, {Wood}, {Wood}, {Wood}, {Yang}, \&
  {Zimmer}}]{2012ApJS..199...31N}
{Nolan}, P.~L., {Abdo}, A.~A., {Ackermann}, M., {et~al.} 2012, \apjs, 199, 31,
  \dodoi{10.1088/0067-0049/199/2/31}

\bibitem[{{Ofek} {et~al.}(2012){Ofek}, {Laher}, {Surace}, {Levitan}, {Sesar},
  {Horesh}, {Law}, {van Eyken}, {Kulkarni}, {Prince}, {Nugent}, {Sullivan},
  {Yaron}, {Pickles}, {Ag{\"u}eros}, {Arcavi}, {Bildsten}, {Bloom}, {Cenko},
  {Gal-Yam}, {Grillmair}, {Helou}, {Kasliwal}, {Poznanski}, \&
  {Quimby}}]{2012PASP..124..854O}
{Ofek}, E.~O., {Laher}, R., {Surace}, J., {et~al.} 2012, \pasp, 124, 854,
  \dodoi{10.1086/666978}

\bibitem[{{Padovani} \& {Giommi}(1995)}]{1995ApJ...444..567P}
{Padovani}, P., \& {Giommi}, P. 1995, \apj, 444, 567, \dodoi{10.1086/175631}

\bibitem[{{Padovani} {et~al.}(2019){Padovani}, {Oikonomou}, {Petropoulou},
  {Giommi}, \& {Resconi}}]{2019MNRAS.484L.104P}
{Padovani}, P., {Oikonomou}, F., {Petropoulou}, M., {Giommi}, P., \& {Resconi},
  E. 2019, \mnras, 484, L104, \dodoi{10.1093/mnrasl/slz011}

\bibitem[{{Paliya} {et~al.}(2020){Paliya}, {B{\"o}ttcher}, {Olmo-Garc{\'\i}a},
  {Dom{\'\i}nguez}, {Gil de Paz}, {Franckowiak}, {Garrappa}, \&
  {Stein}}]{2020ApJ...902...29P}
{Paliya}, V.~S., {B{\"o}ttcher}, M., {Olmo-Garc{\'\i}a}, A., {et~al.} 2020,
  \apj, 902, 29, \dodoi{10.3847/1538-4357/abb46e}

\bibitem[{{Petropoulou} {et~al.}(2017){Petropoulou}, {Nalewajko}, {Hayashida},
  \& {Mastichiadis}}]{2017MNRAS.467L..16P}
{Petropoulou}, M., {Nalewajko}, K., {Hayashida}, M., \& {Mastichiadis}, A.
  2017, \mnras, 467, L16, \dodoi{10.1093/mnrasl/slw252}

\bibitem[{{Planck Collaboration} {et~al.}(2014){Planck Collaboration}, {Ade},
  {Aghanim}, {Armitage-Caplan}, {Arnaud}, {Ashdown}, {Atrio-Barand ela},
  {Aumont}, {Baccigalupi}, {Banday}, {Barreiro}, {Bartlett}, {Battaner},
  {Benabed}, {Beno{\^\i}t}, {Benoit-L{\'e}vy}, {Bernard}, {Bersanelli},
  {Bielewicz}, {Bobin}, {Bock}, {Bonaldi}, {Bond}, {Borrill}, {Bouchet},
  {Bridges}, {Bucher}, {Burigana}, {Butler}, {Calabrese}, {Cappellini},
  {Cardoso}, {Catalano}, {Challinor}, {Chamballu}, {Chary}, {Chen}, {Chiang},
  {Chiang}, {Christensen}, {Church}, {Clements}, {Colombi}, {Colombo},
  {Couchot}, {Coulais}, {Crill}, {Curto}, {Cuttaia}, {Danese}, {Davies},
  {Davis}, {de Bernardis}, {de Rosa}, {de Zotti}, {Delabrouille}, {Delouis},
  {D{\'e}sert}, {Dickinson}, {Diego}, {Dolag}, {Dole}, {Donzelli}, {Dor{\'e}},
  {Douspis}, {Dunkley}, {Dupac}, {Efstathiou}, {Elsner}, {En{\ss}lin},
  {Eriksen}, {Finelli}, {Forni}, {Frailis}, {Fraisse}, {Franceschi}, {Gaier},
  {Galeotta}, {Galli}, {Ganga}, {Giard}, {Giardino}, {Giraud-H{\'e}raud},
  {Gjerl{\o}w}, {Gonz{\'a}lez-Nuevo}, {G{\'o}rski}, {Gratton}, {Gregorio},
  {Gruppuso}, {Gudmundsson}, {Haissinski}, {Hamann}, {Hansen}, {Hanson},
  {Harrison}, {Henrot-Versill{\'e}}, {Hern{\'a}ndez-Monteagudo}, {Herranz},
  {Hildebrand t}, {Hivon}, {Hobson}, {Holmes}, {Hornstrup}, {Hou}, {Hovest},
  {Huffenberger}, {Jaffe}, {Jaffe}, {Jewell}, {Jones}, {Juvela},
  {Keih{\"a}nen}, {Keskitalo}, {Kisner}, {Kneissl}, {Knoche}, {Knox}, {Kunz},
  {Kurki-Suonio}, {Lagache}, {L{\"a}hteenm{\"a}ki}, {Lamarre}, {Lasenby},
  {Lattanzi}, {Laureijs}, {Lawrence}, {Leach}, {Leahy}, {Leonardi},
  {Le{\'o}n-Tavares}, {Lesgourgues}, {Lewis}, {Liguori}, {Lilje},
  {Linden-V{\o}rnle}, {L{\'o}pez-Caniego}, {Lubin}, {Mac{\'\i}as-P{\'e}rez},
  {Maffei}, {Maino}, {Mand olesi}, {Maris}, {Marshall}, {Martin},
  {Mart{\'\i}nez-Gonz{\'a}lez}, {Masi}, {Massardi}, {Matarrese}, {Matthai},
  {Mazzotta}, {Meinhold}, {Melchiorri}, {Melin}, {Mendes}, {Menegoni},
  {Mennella}, {Migliaccio}, {Millea}, {Mitra}, {Miville-Desch{\^e}nes},
  {Moneti}, {Montier}, {Morgante}, {Mortlock}, {Moss}, {Munshi}, {Murphy},
  {Naselsky}, {Nati}, {Natoli}, {Netterfield}, {N{\o}rgaard-Nielsen},
  {Noviello}, {Novikov}, {Novikov}, {O'Dwyer}, {Osborne}, {Oxborrow}, {Paci},
  {Pagano}, {Pajot}, {Paladini}, {Paoletti}, {Partridge}, {Pasian},
  {Patanchon}, {Pearson}, {Pearson}, {Peiris}, {Perdereau}, {Perotto},
  {Perrotta}, {Pettorino}, {Piacentini}, {Piat}, {Pierpaoli}, {Pietrobon},
  {Plaszczynski}, {Platania}, {Pointecouteau}, {Polenta}, {Ponthieu}, {Popa},
  {Poutanen}, {Pratt}, {Pr{\'e}zeau}, {Prunet}, {Puget}, {Rachen}, {Reach},
  {Rebolo}, {Reinecke}, {Remazeilles}, {Renault}, {Ricciardi}, {Riller},
  {Ristorcelli}, {Rocha}, {Rosset}, {Roudier}, {Rowan-Robinson},
  {Rubi{\~n}o-Mart{\'\i}n}, {Rusholme}, {Sandri}, {Santos}, {Savelainen},
  {Savini}, {Scott}, {Seiffert}, {Shellard}, {Spencer}, {Starck}, {Stolyarov},
  {Stompor}, {Sudiwala}, {Sunyaev}, {Sureau}, {Sutton}, {Suur-Uski}, {Sygnet},
  {Tauber}, {Tavagnacco}, {Terenzi}, {Toffolatti}, {Tomasi}, {Tristram},
  {Tucci}, {Tuovinen}, {T{\"u}rler}, {Umana}, {Valenziano}, {Valiviita}, {Van
  Tent}, {Vielva}, {Villa}, {Vittorio}, {Wade}, {Wandelt}, {Wehus}, {White},
  {White}, {Wilkinson}, {Yvon}, {Zacchei}, \& {Zonca}}]{2014A&A...571A..16P}
{Planck Collaboration}, {Ade}, P.~A.~R., {Aghanim}, N., {et~al.} 2014, \aap,
  571, A16, \dodoi{10.1051/0004-6361/201321591}

\bibitem[{{Poutanen} \& {Stern}(2010)}]{2010ApJ...717L.118P}
{Poutanen}, J., \& {Stern}, B. 2010, \apjl, 717, L118,
  \dodoi{10.1088/2041-8205/717/2/L118}

\bibitem[{Prochaska {et~al.}(2020)Prochaska, Hennawi, Westfall, Cooke, Wang,
  Hsyu, Davies, Farina, \& Pelliccia}]{pypeit:joss_pub}
Prochaska, J.~X., Hennawi, J.~F., Westfall, K.~B., {et~al.} 2020, Journal of
  Open Source Software, 5, 2308, \dodoi{10.21105/joss.02308}

\bibitem[{{PTF Team}(2020)}]{ptf}
{PTF Team}. 2020, Palomar Transient Factory Level 1,  IPAC,
  \dodoi{10.26131/IRSA155}

\bibitem[{{Pursimo} {et~al.}(2013){Pursimo}, {Ojha}, {Jauncey}, {Rickett},
  {Dutka}, {Koay}, {Lovell}, {Bignall}, {Kedziora-Chudczer}, \&
  {Macquart}}]{2013ApJ...767...14P}
{Pursimo}, T., {Ojha}, R., {Jauncey}, D.~L., {et~al.} 2013, \apj, 767, 14,
  \dodoi{10.1088/0004-637X/767/1/14}

\bibitem[{{Rees}(1988)}]{1988Natur.333..523R}
{Rees}, M.~J. 1988, \nat, 333, 523, \dodoi{10.1038/333523a0}

\bibitem[{{Runnoe} {et~al.}(2012){Runnoe}, {Brotherton}, \&
  {Shang}}]{2012MNRAS.422..478R}
{Runnoe}, J.~C., {Brotherton}, M.~S., \& {Shang}, Z. 2012, \mnras, 422, 478,
  \dodoi{10.1111/j.1365-2966.2012.20620.x}

\bibitem[{{Scargle} {et~al.}(2013){Scargle}, {Norris}, {Jackson}, \&
  {Chiang}}]{2013ApJ...764..167S}
{Scargle}, J.~D., {Norris}, J.~P., {Jackson}, B., \& {Chiang}, J. 2013, \apj,
  764, 167, \dodoi{10.1088/0004-637X/764/2/167}

\bibitem[{{Schlafly} \& {Finkbeiner}(2011)}]{2011ApJ...737..103S}
{Schlafly}, E.~F., \& {Finkbeiner}, D.~P. 2011, \apj, 737, 103,
  \dodoi{10.1088/0004-637X/737/2/103}

\bibitem[{{Schlegel} {et~al.}(1998){Schlegel}, {Finkbeiner}, \&
  {Davis}}]{1998ApJ...500..525S}
{Schlegel}, D.~J., {Finkbeiner}, D.~P., \& {Davis}, M. 1998, \apj, 500, 525,
  \dodoi{10.1086/305772}

\bibitem[{{Shen} {et~al.}(2008){Shen}, {Greene}, {Strauss}, {Richards}, \&
  {Schneider}}]{2008ApJ...680..169S}
{Shen}, Y., {Greene}, J.~E., {Strauss}, M.~A., {Richards}, G.~T., \&
  {Schneider}, D.~P. 2008, \apj, 680, 169, \dodoi{10.1086/587475}

\bibitem[{{Sheng} {et~al.}(2017){Sheng}, {Wang}, {Jiang}, {Yang}, {Yan}, {Dou},
  \& {Peng}}]{2017ApJ...846L...7S}
{Sheng}, Z., {Wang}, T., {Jiang}, N., {et~al.} 2017, \apjl, 846, L7,
  \dodoi{10.3847/2041-8213/aa85de}

\bibitem[{{Sheng} {et~al.}(2020){Sheng}, {Wang}, {Jiang}, {Ding}, {Cai}, {Guo},
  {Sun}, {Dou}, \& {Yang}}]{2020ApJ...889...46S}
---. 2020, \apj, 889, 46, \dodoi{10.3847/1538-4357/ab5af9}

\bibitem[{{Sikora} {et~al.}(1994){Sikora}, {Begelman}, \&
  {Rees}}]{1994ApJ...421..153S}
{Sikora}, M., {Begelman}, M.~C., \& {Rees}, M.~J. 1994, \apj, 421, 153,
  \dodoi{10.1086/173633}

\bibitem[{{Stecker} {et~al.}(1991){Stecker}, {Done}, {Salamon}, \&
  {Sommers}}]{1991PhRvL..66.2697S}
{Stecker}, F.~W., {Done}, C., {Salamon}, M.~H., \& {Sommers}, P. 1991, \prl,
  66, 2697, \dodoi{10.1103/PhysRevLett.66.2697}

\bibitem[{{Stein} {et~al.}(2019){Stein}, {Franckowiak}, {Necker}, {Gezari},
  {van Velzen}, {Ztf Collaboration}, \& {Growth
  Collaboration}}]{2019GCN.25929....1S}
{Stein}, R., {Franckowiak}, A., {Necker}, J., {et~al.} 2019, GRB Coordinates
  Network, 25929, 1

\bibitem[{{Stein} {et~al.}(2021){Stein}, {Velzen}, {Kowalski}, {Franckowiak},
  {Gezari}, {Miller-Jones}, {Frederick}, {Sfaradi}, {Bietenholz}, {Horesh},
  {Fender}, {Garrappa}, {Ahumada}, {Andreoni}, {Belicki}, {Bellm},
  {B{\"o}ttcher}, {Brinnel}, {Burruss}, {Cenko}, {Coughlin}, {Cunningham},
  {Drake}, {Farrar}, {Feeney}, {Foley}, {Gal-Yam}, {Golkhou}, {Goobar},
  {Graham}, {Hammerstein}, {Helou}, {Hung}, {Kasliwal}, {Kilpatrick}, {Kong},
  {Kupfer}, {Laher}, {Mahabal}, {Masci}, {Necker}, {Nordin}, {Perley},
  {Rigault}, {Reusch}, {Rodriguez}, {Rojas-Bravo}, {Rusholme}, {Shupe},
  {Singer}, {Sollerman}, {Soumagnac}, {Stern}, {Taggart}, {van Santen}, {Ward},
  {Woudt}, \& {Yao}}]{2021NatAs...5..510S}
{Stein}, R., {Velzen}, S.~v., {Kowalski}, M., {et~al.} 2021, Nature Astronomy,
  5, 510, \dodoi{10.1038/s41550-020-01295-8}

\bibitem[{{Tavecchio} {et~al.}(2010){Tavecchio}, {Ghisellini}, {Bonnoli}, \&
  {Ghirlanda}}]{2010MNRAS.405L..94T}
{Tavecchio}, F., {Ghisellini}, G., {Bonnoli}, G., \& {Ghirlanda}, G. 2010,
  \mnras, 405, L94, \dodoi{10.1111/j.1745-3933.2010.00867.x}

\bibitem[{{Ulrich} {et~al.}(1997){Ulrich}, {Maraschi}, \&
  {Urry}}]{1997ARA&A..35..445U}
{Ulrich}, M.-H., {Maraschi}, L., \& {Urry}, C.~M. 1997, \araa, 35, 445,
  \dodoi{10.1146/annurev.astro.35.1.445}

\bibitem[{{van Velzen} {et~al.}(2021){van Velzen}, {Stein}, {Gilfanov},
  {Kowalski}, {Hayasaki}, {Reusch}, {Yao}, {Garrappa}, {Franckowiak}, {Gezari},
  {Nordin}, {Fremling}, {Sharma}, {Yan}, {Kool}, {Sollerman}, {Medvedev},
  {Sunyaev}, {Bellm}, {Dekany}, {Duev}, {Graham}, {Kasliwal}, {Laher},
  {Riddle}, \& {Rusholme}}]{2021arXiv211109391V}
{van Velzen}, S., {Stein}, R., {Gilfanov}, M., {et~al.} 2021, arXiv e-prints,
  arXiv:2111.09391.
\newblock \doarXiv{2111.09391}

\bibitem[{{Wagner} \& {Witzel}(1995)}]{1995ARA&A..33..163W}
{Wagner}, S.~J., \& {Witzel}, A. 1995, \araa, 33, 163,
  \dodoi{10.1146/annurev.aa.33.090195.001115}

\bibitem[{{Waxman}(1995)}]{1995PhRvL..75..386W}
{Waxman}, E. 1995, \prl, 75, 386, \dodoi{10.1103/PhysRevLett.75.386}

\bibitem[{{WISE Team}(2019)}]{allwise}
{WISE Team}. 2019, AllWISE Source Catalog,  IPAC, \dodoi{10.26131/IRSA1}

\bibitem[{{WISE Team}(2020)}]{neowise}
---. 2020, NEOWISE 2-Band Post-Cryo Single Exposure (L1b) Source Table,  IPAC,
  \dodoi{10.26131/IRSA124}

\bibitem[{{Wright} {et~al.}(2010){Wright}, {Eisenhardt}, {Mainzer}, {Ressler},
  {Cutri}, {Jarrett}, {Kirkpatrick}, {Padgett}, {McMillan}, {Skrutskie},
  {Stanford}, {Cohen}, {Walker}, {Mather}, {Leisawitz}, {Gautier}, {McLean},
  {Benford}, {Lonsdale}, {Blain}, {Mendez}, {Irace}, {Duval}, {Liu}, {Royer},
  {Heinrichsen}, {Howard}, {Shannon}, {Kendall}, {Walsh}, {Larsen}, {Cardon},
  {Schick}, {Schwalm}, {Abid}, {Fabinsky}, {Naes}, \&
  {Tsai}}]{2010AJ....140.1868W}
{Wright}, E.~L., {Eisenhardt}, P. R.~M., {Mainzer}, A.~K., {et~al.} 2010, \aj,
  140, 1868, \dodoi{10.1088/0004-6256/140/6/1868}

\bibitem[{{Zauderer} {et~al.}(2011){Zauderer}, {Berger}, {Soderberg}, {Loeb},
  {Narayan}, {Frail}, {Petitpas}, {Brunthaler}, {Chornock}, {Carpenter},
  {Pooley}, {Mooley}, {Kulkarni}, {Margutti}, {Fox}, {Nakar}, {Patel},
  {Volgenau}, {Culverhouse}, {Bietenholz}, {Rupen}, {Max-Moerbeck}, {Readhead},
  {Richards}, {Shepherd}, {Storm}, \& {Hull}}]{2011Natur.476..425Z}
{Zauderer}, B.~A., {Berger}, E., {Soderberg}, A.~M., {et~al.} 2011, \nat, 476,
  425, \dodoi{10.1038/nature10366}

\end{thebibliography}

\end{document}